\documentclass[aps,prd,preprint,showpacs,nofootinbib,12pt]{article}

\title{Full One-Loop Electroweak Corrections to the Charged Higgs Decays into a Chargino and Neutralino in the MSSM}
\author{Zhao Li\thanks{Electronics address: zhli@water.pku.edu.cn}, Chong Sheng Li
\thanks{Electronics address: csli@pku.edu.cn}
, Qiang Li\thanks{Electronics address: qliphy@pku.edu.cn}, 
Li Lin Yang\thanks{Electronics address: llyang@pku.edu.cn}, and Jun Zhao\\
\\
\textit{Department of Physics, Peking University, Beijing 100871, China}}
\date{}
\usepackage{epsfig}
\usepackage{graphicx}
\usepackage{amsmath}
\usepackage{hhline}
\usepackage{amssymb}
\usepackage{times}
\usepackage{cite}
\usepackage{array}
\usepackage{multirow}
\def\rpv{\mbox{$\rlap{\kern0.25em/}R_p$ }}

\begin{document}

\maketitle

\begin{center} \Large Abstract
\end{center}
We calculate the full one-loop electroweak (FEW) corrections to
$H^-\rightarrow\widetilde{\chi}^-_1\widetilde{\chi}^0_i$ in the
minimal supersymmetric standard model (MSSM), and compare with the
leading order (LO) corrections including the loops of the
(s)quarks only in the third generation and the complete leading
order (CLO) corrections including the loops of the (s)quarks in
all the three generations. We find that the magnitudes of the FEW
corrections can be larger than $10\%$ for $\tan\beta>30$.
Moreover, comparing with the FEW corrections, both of the LO and
the CLO corrections are negligible small for the mode 1($H^-\rightarrow\widetilde{\chi}^-_1\widetilde{\chi}^0_1$) when
$\tan\beta<5$, and for the mode 2($H^-\rightarrow\widetilde{\chi}^-_1\widetilde{\chi}^0_2$) when $\tan\beta>45$,
respectively, since there are not enhancements from the Yukawa
couplings. We also calculate the FEW corrections in the minimal
supergravity (mSUGRA) scenario, where the FEW corrections can be
larger than the LO and the CLO corrections by more than $60\%$ and
$50\%$, respectively.
\\
PACS numbers: 12.15.Lk, 12.60.Jv, 14.80.Cp, 14.80.Ly
\newpage
\section{\bf Introduction}
\indent Beyond the standard model(SM), the
supersymmetric(SUSY)\cite{fayet} extensions of the SM
provide a great opportunity to solve some mysterious problems in
the SM. The SUSY partners of the SM particles cancel the quadratic
divergences in the corrections to the Higgs boson mass, and the
hierarchy problem can be solved naturally. If we
consider the R-parity conservation as the essential condition, the
lightest SUSY particles(LSP) will never decay, and the stability
of the LSP provides the most important candidate for the dark
matter\cite{smith}. The most attractive extension of the SM is the
minimal supersymmetric standard model(MSSM)\cite{mssm}. If we set
all the parameters as real in the MSSM, there will be five Higgs
bosons\cite{higgs}: two CP even bosons ($H^0$, $h^0$), one CP odd
boson ($A^0$), and two charged bosons ($H^{\pm}$). When the Higgs
boson of the SM has a mass below 130-140 Gev and the $h^0$ of the
MSSM are in the decoupling limit (which means that $H^\pm$ is too
heavy anyway to be possibly produced), the lightest neutral Higgs
boson may be difficult to be distinguished from the neutral Higgs
boson of the SM. But the charged Higgs bosons carry a distinctive
signature of the Higgs sector in the MSSM. Therefore, the search
for the charged Higgs bosons is very important for probing the Higgs
sector of the MSSM, and will be one of the prime objectives of the
CERN Large Hadron Collider(LHC)\cite{lhc1,lhc2}.
\\
\indent Current bounds on charged Higgs mass can be obtained at
the Tevatron, by studying the top decay $t\to bH^+$, which already
eliminates some region of parameter space \cite{LHCs}, whereas the
combined LEP experiments gives a low bounds approximately
$m_{H^+}>78.6$GeV at $95\%$CL\cite{LEPs}. In the MSSM, we have
$m_{H^{\pm}}\ge 120$ GeV from the mass bounds from LEP--II for the
neutral pseudoscalar $A^0$ of the MSSM ($m_{A^0}\ge 91.9$
GeV)\cite{H+masss}.
\\
\indent If the charged Higgs masses could be large enough, there
will be many SM and SUSY decay modes. In the MSSM the channels of
decay into neutralino and the chargino($
H^{\pm}\rightarrow\widetilde{\chi}^0_i\widetilde{\chi}^{\pm}_j$)
are very important\cite{H+chi}, which have been discussed in the
Ref.\cite{mawengan1}, where only the LO corrections were
calculated. The one loop corrected effective lagrangian for
the charged higgs-neutralino-chargino
couplings is calculated in Ref.\cite{ibrahim}.
In this paper, we present the calculations of the FEW
corrections, which include the contributions of the one-loop
virtual contributions of the (s)leptons and (s)quarks of all the
three generations, and all the possible Higgs and gauge bosons,
charginos and neutralinos and the real corrections i.e. the real
photon emission. In the preparation of this paper, a relevant work
was given in the Ref.\cite{erbelhold} as a conference's short
report, which doesn't show any detail of the calculation and the
numerical results are not complete.
\\
\indent In Sec.2 we define the relevant notations and show the
tree-level result. In Sec.3 we present the virtual corrections,
including the vertex corrections and the counterterms. In Sec.4 we
illustrate the real corrections from the real photon emission
using the phase space slicing(PSS) method\cite{pss}. In Sec.5 we
present the numerical results and conclusion in the low-energy
MSSM and the mSUGRA breaking scenario\cite{msugra}.
\section{\bf Notations and Tree-level Width}
In order to make this paper self-contained, we first present the
relevant interaction Lagrangian\cite{mawengan1} of the MSSM and
the tree level decay width for
$H^+\widetilde{\chi}^-_i\widetilde{\chi}^0_j$
$(i=1,2,\;\;j=1,2,3,4)$. The Lagrangian is
\begin{equation}\label{lagrangian0}
{\mathcal L}_{H^+\widetilde{\chi}^-_j\widetilde{\chi}^0_i}=-H^+\overline{\widetilde{\chi}^0_i}
\left(C^L_{ij}P_L+C^R_{ij}P_R\right)\widetilde{\chi}^-_j+H.c.~,
\end{equation}
where,
$$P_{L,R}=\frac{1}{2}(1\mp\gamma_5),$$
\begin{equation}\label{vcoeff}
\begin{array}{c}
\displaystyle{
C^L_{ij}=\frac{e}{s_W}s_\beta\left(
U_{j1}^*N_{i3}^*-\frac{1}{\sqrt{2}}
U_{j2}^*(N_{i2}^*+t_WN_{i1}^*)\right),
}\\ \\ \displaystyle{
C^R_{ij}=\frac{e}{s_W}c_\beta\left(
V_{j1}N_{i4}+\frac{1}{\sqrt{2}}
V_{j2}(N_{i2}+t_WN_{i1})\right)
},
\end{array}
\end{equation}
for convenience, we take $s_W=\sin\theta_W$, $c_W=\cos\theta_W$, $t_W=\tan\theta_W$,
$s_\beta=\sin\beta$, $c_\beta=\cos\beta$ and $t_\beta=\tan\beta$.
\\
\indent Here the matrixes $U$, $V$ and $N$ are the chargino and neutralino mixing
matrixes, which can diagonalize the corresponding mass
matrixes. The chargino mass matrix is
\begin{equation}\label{chamassmatrix}
\mathbf{X}=
\left(\begin{array}{cc}
M & \sqrt{2}m_Ws_\beta\\
\sqrt{2}m_Wc_\beta & \mu\\
\end{array}
\right)\\
\end{equation}
and the neutralino mass matrix is
\begin{equation}\label{neumassmatrix}
\mathbf{Y}=
\left(\begin{array}{cccc}
M' & 0 & -m_Zs_Wc_\beta & m_Zs_Ws_\beta\\
0 & M & m_Zc_Wc_\beta & -m_Zc_Ws_\beta\\
-m_Zs_Wc_\beta & m_Zc_Wc_\beta & 0 & -\mu\\
m_Zs_Ws_\beta & -m_Zc_Ws_\beta & -\mu & 0\\
\end{array}
\right).
\end{equation}
\\
\indent
The chargino mixing matrixes ($U$, $V$) diagonalize the chargino mass matrix
\begin{equation}\label{diagcha}
U {\mathbf X} V^{\dagger}=
\textrm{diag}\left(
\eta_1 M_{\widetilde{\chi}^-_1},\,
\eta_2 M_{\widetilde{\chi}^-_2}
\right),
\end{equation}
and the neutralino mixing matrix ($N$) diagonalizes the neutralino mass matrix
\begin{equation}\label{diagneu}
N {\mathbf Y} N^{\dagger}=
\textrm{diag}\left(
\epsilon_1 M_{\widetilde{\chi}^0_1},\,
\epsilon_2 M_{\widetilde{\chi}^0_2},\,
\epsilon_3 M_{\widetilde{\chi}^0_3},\,
\epsilon_4 M_{\widetilde{\chi}^0_4}
\right),
\end{equation}
where $\eta_i=\pm1\;(i=1,2)$ and $\epsilon_j=\pm1\;(j=1,2,3,4)$,
these signs depend on the configuration of the mixing matrixes.
The chargino and the neutralino physical masses are
\begin{equation}\label{phymass}
m_{\widetilde{\chi}^-_i}=\left|M_{\widetilde{\chi}^-_i}\right|,\;
m_{\widetilde{\chi}^0_j}=\left|M_{\widetilde{\chi}^0_j}\right|.
\end{equation}
\indent
From the interaction Lagrangian (\ref{lagrangian0}) we can derive the tree-level amplitude as following:
\begin{equation}
{\mathcal M}_0=
\overline{u}(p_{\widetilde{\chi}^-_j})
(C^L_{ij}P_L+C^R_{ij}P_R)
v(p_{\widetilde{\chi}^0_i}).
\end{equation}
\indent
Then the tree-level decay width is thus given by
\begin{equation}\label{born}
\Gamma_0=\frac{1}{8\pi}\frac{p_{out}}{m_{H^-}^2}\left|{\mathcal M}_0\right|^2,
\end{equation}
where the momentum value of the outgoing particle
\begin{equation}
p_{out}=
\frac{1}{2m_{H^-}} \sqrt{(m_{H^-}^2+m_{\widetilde{\chi}^-_j}^2-
m_{\widetilde{\chi}^0_i}^2)^2-4m_{H^-}m_{\widetilde{\chi}^-_j}}.
\end{equation}
\indent For future convenience, we also present here the vertex
$G^+\widetilde{\chi}^-_i\widetilde{\chi}^0_j$ to fix the
renormalization constant of $G^-$ and $H^-$ mixing.
\begin{equation}\label{lagrangian1}
{\mathcal L}_{G^+\widetilde{\chi}^-_j\widetilde{\chi}^0_i}=
-G^+\overline{\widetilde{\chi}^0_i}\left(
D^L_{ij}P_L+D^R_{ij}P_R
\right)\widetilde{\chi}^-_j+H.c.~,
\end{equation}
where $D^L_{ij}=-\cot\beta C^L_{ij},\;D^R_{ij}=\tan\beta C^R_{ij}$.
\section{\bf Virtual Correction}
\begin{figure}[!h]
\centering
\includegraphics[width=340pt]{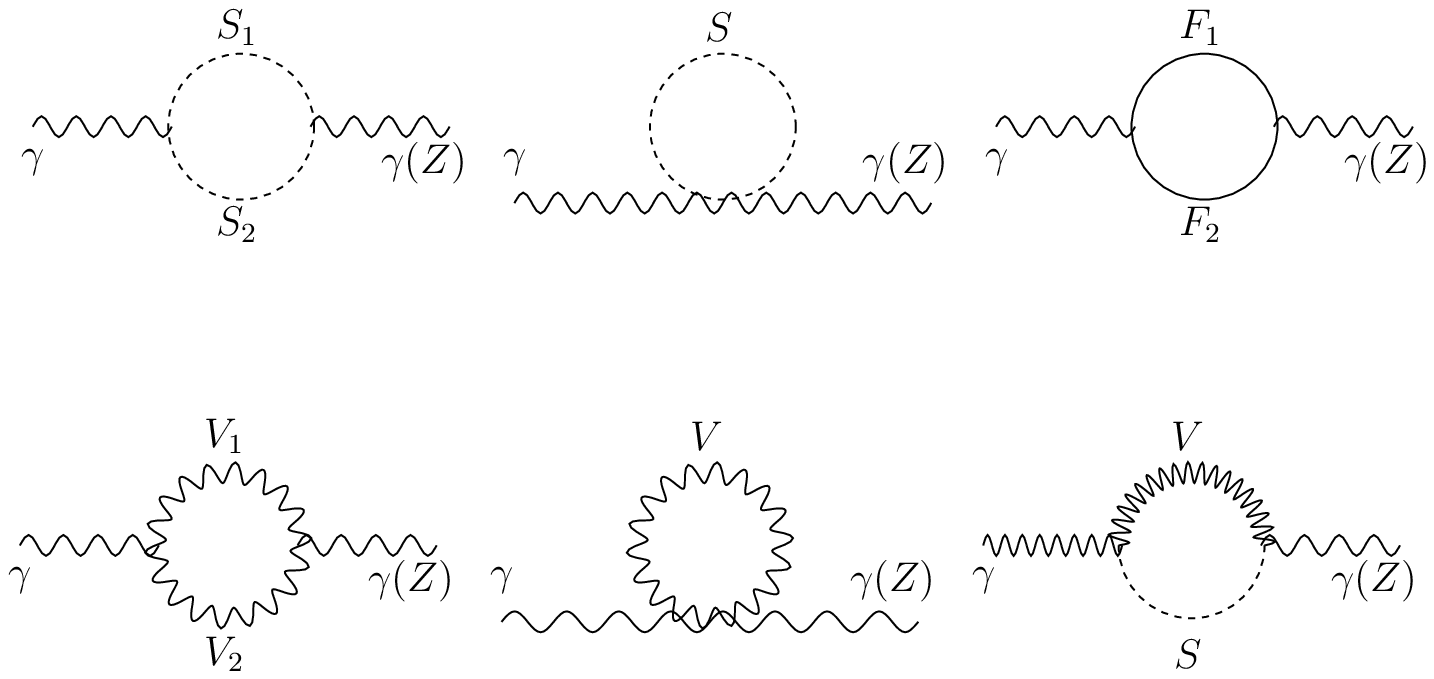}
\centering
\caption{The $\gamma\gamma$ and $\gamma{Z}$ self-energy Feynman diagrams.
$S_1S_2$: $H^-H^+$, $G^-G^+$, $\widetilde{e}\bar{\widetilde{e}}$, $
\widetilde{u}\bar{\widetilde{u}}$, $\widetilde{d}\bar{\widetilde{d}}$; $
S$: $\phi^-$, $\widetilde{e}$, $\widetilde{u}$, $\widetilde{d}$; $
F_1F_2$: $e\bar{e}$, $u\bar{u}$, $d\bar{d}$, $
\widetilde{\chi}^-\widetilde{\chi}^+$; $V_1V_2$: $W^-W^+$; $V$: $W^-$; $
VS$: $W^-G^+$, $W^+G^-$.}
\label{self1}
\end{figure}

\begin{figure}[!h]
\centering
\includegraphics[width=340pt]{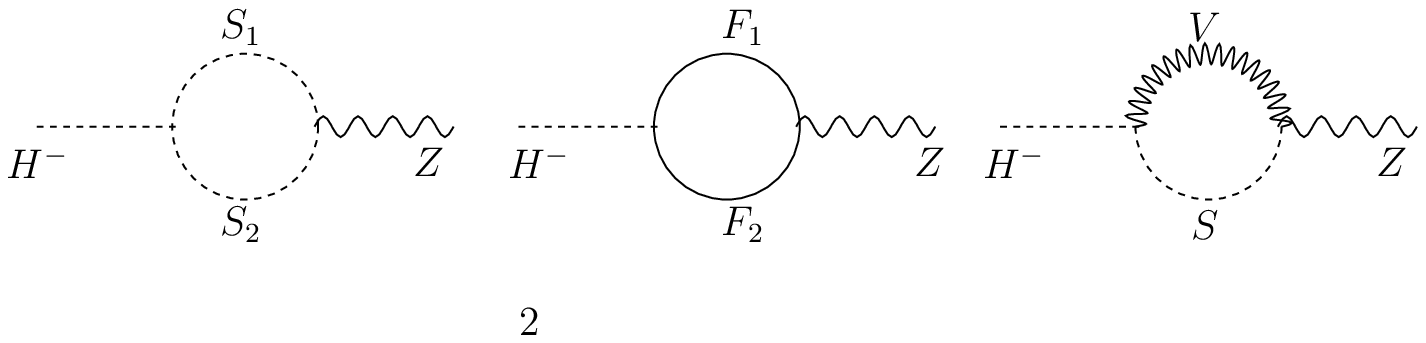}
\centering
\caption{The $H^-W^-$ self-energy Feynman diagrams.
$S_1S_2$: $H^-\phi^0_B$, $G^-\phi^0_B$, $\widetilde{e}\bar{\widetilde{\nu}}$, $
\widetilde{d}\bar{\widetilde{u}}$; $
F_1F_2$: $e\bar{\nu}$, $d\bar{u}$, $\widetilde{\chi}^-\widetilde{\chi}^0$; $
VS$: $W^-\phi^0_B$.}
\label{self2}
\end{figure}

\begin{figure}[!h]
\centering
\includegraphics[width=340pt]{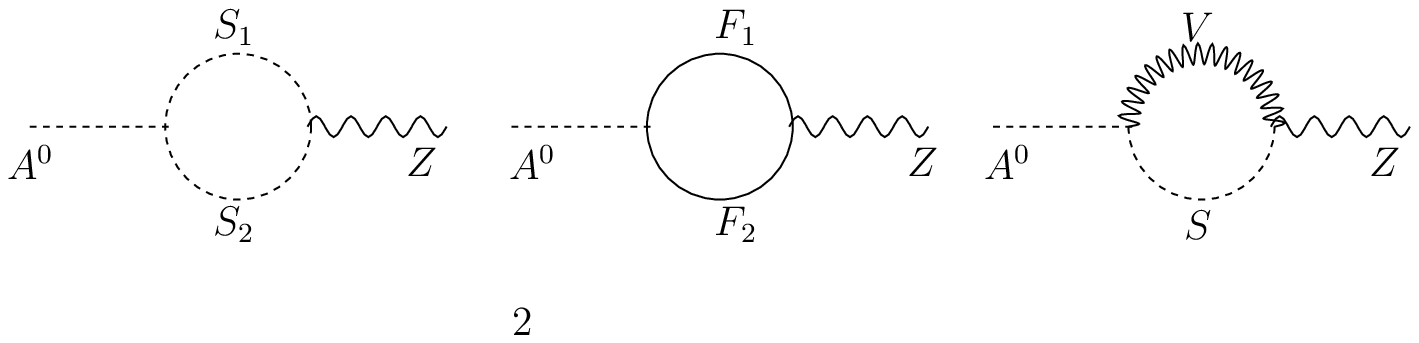}
\centering
\caption{The $A^0Z$ self-energy Feynman diagrams.
$S_1S_2$: $A^0\phi^0_B$, $G^0\phi^0_B$, $\widetilde{e}\bar{\widetilde{e}}$, $
\widetilde{u}\bar{\widetilde{u}}$, $\widetilde{d}\bar{\widetilde{d}}$; $
F_1F_2$: $e\bar{e}$, $u\bar{u},d\bar{d}$, $\widetilde{\chi}^-\widetilde{\chi}^+$, $
\widetilde{\chi}^0\widetilde{\chi}^0$; $
VS$: $Z\phi^0_B$.}
\label{self3}
\end{figure}

\begin{figure}[!h]
\centering
\includegraphics[width=340pt]{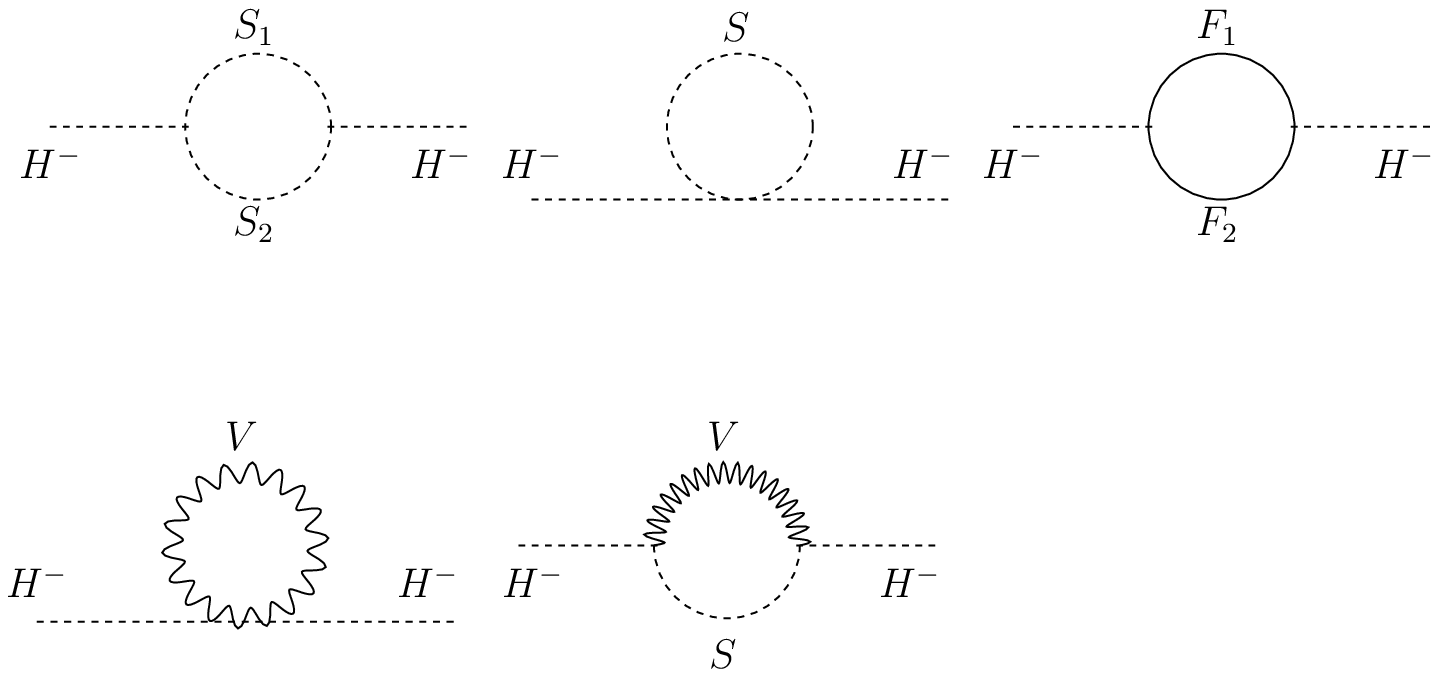}
\centering
\caption{The $H^-H^-$ self-energy Feynman diagrams.
$S_1S_2$: $H^-\phi^0_B$, $G^-\phi^0_T$, $\widetilde{e}\bar{\widetilde{\nu}}$, $
\widetilde{d}\bar{\widetilde{u}}$; $
S$: $\phi^-$, $\phi^0$, $\widetilde{e}$, $\widetilde{\nu}$, $
\widetilde{u}$, $\widetilde{d}$; $
F_1F_2$: $e\bar\nu$, $d\bar{u}$, $\widetilde{\chi}^-\widetilde{\chi}^0$; $
V$: $\gamma$, $Z$, $W^-$; $VS$: $\gamma{H}^-$, $ZH^-$, $W^-\phi^0_T$.}
\label{self4}
\end{figure}

\begin{figure}[!h]
\centering
\includegraphics[width=220pt]{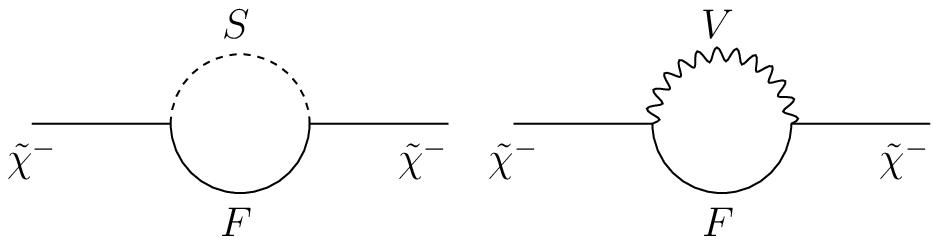}
\centering
\caption{The $\widetilde{\chi}^-\widetilde{\chi}^-$ self-energy Feynman diagrams.
$SF$: $\phi^0\widetilde\chi^-$, $\phi^-\widetilde\chi^0$, $\widetilde\nu{e}$, $
\widetilde{e}\nu$, $\widetilde{u}d$, $\widetilde{d}u$; $
VF$: $\gamma\widetilde\chi^-$, $Z\widetilde\chi^-$, $W^-\widetilde\chi^0$.}
\label{self5}
\end{figure}

\begin{figure}[!h]
\centering
\includegraphics[width=220pt]{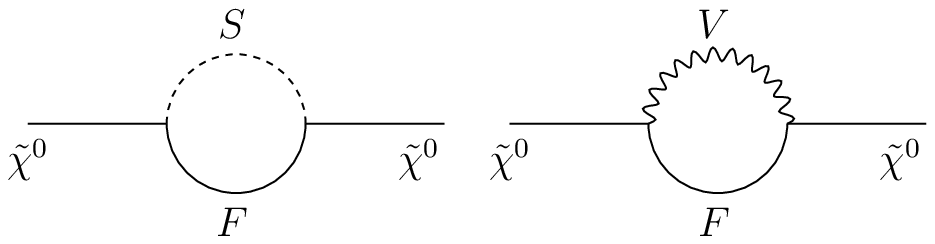}
\centering
\caption{The $\widetilde{\chi}^0\widetilde{\chi}^0$ self-energy Feynman diagrams.
$SF$: $\phi^0\widetilde\chi^0$, $\phi^-\widetilde\chi^+$, $\phi^+\widetilde\chi^-$, 
$\widetilde\nu\bar\nu$, $\widetilde{e}\bar{e}$, $\widetilde{u}\bar{u}$, 
$\widetilde{d}\bar{d}$, 
$\bar{\widetilde\nu}\nu$, $\bar{\widetilde{e}}e$, $\bar{\widetilde{u}}u$, 
$\bar{\widetilde{d}}d$; 
$VF$: $Z\widetilde\chi^0$, $W^-\widetilde\chi^+$, $W^+\widetilde\chi^-$.}
\label{self6}
\end{figure}

The Feynman diagrams, contributing to the virtual corrections to
$H^-\rightarrow\widetilde{\chi}^0_i\widetilde{\chi}^-_j$
are shown in Figs.\ref{oneloopfd}-\ref{self6}.
In the calculation we use the 't Hooft-Feynman gauge,
the dimensional regularization $(D=4-2\epsilon)$ to regularize
the ultraviolet (UV) and infrared (IR) divergences in the virtual loop corrections,
and the on-mass-shell scheme for the renormalization\cite{mawengan1}.
We use the FormCalc program\cite{formcalc}
to calculate the amplitudes of the one-loop vertex amplitudes
and the self-energy diagrams. In order to keep
supersymmetry the corrections with the vector bosons are performed
by the dimensional reduction.
\\
\indent
The relevant renormalization constants for the wave function and the fields mixing
have the following definitions:
\begin{equation}
\begin{array}{c}
\widetilde{\chi}^-_{i0}=\Big(\delta_{ik}+
\frac{1}{2}\delta Z^L_{-,ik}P_L+\frac{1}{2}\delta Z^R_{-,ik}P_R\Big)
\widetilde{\chi}^-_{k},
\\ \\
\widetilde{\chi}^0_{i0}=\Big(\delta_{ik}+
\frac{1}{2}\delta Z^L_{0,ik}P_L+\frac{1}{2}\delta Z^R_{0,ik}P_R\Big)
\widetilde{\chi}^0_{k},
\\ \\
\left(
\begin{array}{c}
H^-\\ \\
G^-
\end{array}
\right)_0
=
\left(
\begin{array}{cc}
\displaystyle{
1+\frac{1}{2}\delta Z_{H^-}
}&\displaystyle{ \frac{1}{2}\delta Z_{H^-G^-}
}\\ \\ \displaystyle{
\frac{1}{2}\delta Z_{G^-H^-}
}&\displaystyle{ 1+\frac{1}{2}\delta Z_{H^-}
}
\end{array}
\right)
\left(
\begin{array}{c}
H^-\\ \\
G^-
\end{array}
\right).
\end{array}
\end{equation}
\indent
The renormalization constants for the vertex parameters are defined as
\begin{equation}
\begin{array}{c}
e_0=\Big(1+\delta Z_e\Big)e, \;
s_{W0}= s_W+\delta s_W, \;
t_{\beta0}= t_\beta+\delta t_\beta,
\\ \\
U_0=U+\delta U, \;
V_0=V+\delta V, \;
N_0=N+\delta N,
\\ \\
\displaystyle{
\delta U=\frac{1}{4}
\Big(\delta Z_-^L-\delta Z_-^{L\dagger}\Big)U,
}\\ \\ \displaystyle{
\delta V=\frac{1}{4}
\Big(\delta Z_-^R-\delta Z_-^{R\dagger}\Big)V,
}\\ \\ \displaystyle{
\delta N=\frac{1}{4}
\Big(\delta Z_0^L-\delta Z_0^{L\dagger}\Big)N.
}
\end{array}
\end{equation}
With the above rotation matrixes renormalization
counterterm definitions, the chargino and neutralino mass matrixes
get radiative corrections\cite{eberlkincel}. The corrections are UV finite shifts
on the tree-level matrixes X and Y. The shift $\Delta X$ is
\begin{eqnarray}
\Delta X_{11} &=& 0\,\\
\Delta X_{12} &=& \left(\frac{\delta m_W}{m_W} + c_\beta^2\,
   \frac{\delta t_\beta}{t_\beta} \right)\,
   X_{12} - \delta X_{12}\,\\
\Delta X_{21} &=& \left(\frac{\delta m_W}{m_W} - s_\beta^2\,
   \frac{\delta t_\beta}{t_\beta} \right)\,
   X_{21} - \delta X_{21}\,\\
\Delta X_{22} &=& 0\, ,
\end{eqnarray}
where
\begin{eqnarray}
 \left(\delta X\right)_{ij} & = &
  \sum_{k=1}^2 \left[ m_{\widetilde\chi^+_k} \left(
  \delta U_{ki} V_{kj} +  U_{ki} \delta V_{kj}\right) + \delta
  m_{\widetilde\chi^+_k} U_{ki} V_{kj}\right]\,.
  \end{eqnarray}
The shift $\Delta Y$ is
\begin{eqnarray}
\Delta Y_{11} &=& 0\,\\
\Delta Y_{12} &=& - \delta Y_{12}\,\\
\Delta Y_{13} &=& \left(\frac{\delta m_Z}{m_Z}+ \frac{\delta
s_W}{s_W} - s_\beta^2\, \frac{\delta t_\beta}{t_\beta}
\right)\, Y_{13} - \delta Y_{13}\,\\
\Delta Y_{14} &=& \left(\frac{\delta m_Z}{m_Z}+ \frac{\delta
s_W}{s_W} + c_\beta^2\, \frac{\delta t_\beta}{t_\beta}
\right)\, Y_{14} - \delta Y_{14}\,\\
\Delta Y_{22} &=& \delta M - \delta Y_{22}\, \,\\
\Delta Y_{23} &=& \left(\frac{\delta m_Z}{m_Z} - t_W^2\,
\frac{\delta s_W}{s_W} - s_\beta^2\, \frac{\delta
t_\beta}{t_\beta}\right)\, Y_{23} - \delta Y_{23}\,\\
\Delta Y_{24} &=& \left(\frac{\delta m_Z}{m_Z} - t_W^2\,
\frac{\delta s_W}{s_W} + c_\beta^2\, \frac{\delta
t_\beta}{t_\beta}\right)\, Y_{24} - \delta Y_{24}\,\\
\Delta Y_{33} &=&  - \delta Y_{33}\,\\
\Delta Y_{34} &=& -\delta\mu - \delta Y_{34}\,\\
\Delta Y_{44} &=& - \delta Y_{44}\, .
\end{eqnarray}
where
\begin{eqnarray}
 (\delta Y)_{ij} & = &
  \sum_{k=1}^4 \left[
  \delta m_{\widetilde \chi^0_k} Z_{ki} Z_{kj} + m_{\widetilde \chi^0_k}
  \delta Z_{ki} Z_{kj} + m_{\widetilde \chi^0_k} Z_{ki} \delta
  Z_{kj}\right]\, .
  \end{eqnarray}
Then the corrected mixing matrixes are $X+\Delta X$ and $Y+\Delta Y$.
Through the diagonalization (5) and (6), the corrected pole masses and rotation matrixes
can be extracted. Using the corrected couplings, the tree-level decay widths
are also changed into the improved tree-level widths\cite{oeller}.
\\
\indent
The chargino and the neutralino mixing matrixes renormalization constants cancel
the antisymmetric parts of their wave function renormalization constants. Consequently,
the chargino and the neutralino wave function renormalization constants are
shifted as,
\begin{equation}
\delta Z_-\rightarrow\frac{1}{2}(\delta Z_-+\delta Z_-^{\dagger}), \;
\delta Z_0\rightarrow\frac{1}{2}(\delta Z_0+\delta Z_0^{\dagger}).
\end{equation}
\indent
Meanwhile, the renormalization for $\tan\beta$ cancels half of
the $G^--H^-$ renormalization\cite{mawengan1} as
\begin{equation}
\delta Z_{G^-H^-}\rightarrow\frac{1}{2}\delta Z_{G^-H^-}.
\end{equation}
\indent
The renormalized virtual amplitudes can be written as
\begin{equation}
{\mathcal M}^V_1={\mathcal M}^{(v)}_1+{\mathcal M}^{(c)}_1
\end{equation}
including the vertex one-loop contribution ${\mathcal M}^{(v)}_1$
and the corresponding counterterm ${\mathcal M}^{(c)}_1$.
The vertex part can be derived from the vertex one-loop
Feynman diagrams in Fig.\ref{oneloopfd}. We list their
analytic expressions in Appendix.
\\
\indent
With the previous definitions of the renormalization constants,
the counterterm Lagrangian for the vertex is
\begin{equation}
\begin{array}{l}
\displaystyle{
\delta{\mathcal L}_{H^+\widetilde{\chi}^-_j\widetilde{\chi}^0_i}=
-H^+\overline{\widetilde{\chi}^0_i}\Big\{\Big[C^L_{ij}
\Big(\delta Z_e-\frac{\delta s_W}{s_W}
+\frac{1}{2}\delta Z_{H^-}
-\frac{1}{4}\cot\beta\delta Z_{G^-H^-}\Big)
}\\ \\ \displaystyle{
+\frac{1}{4}\sum^2_{k=1}
C^L_{ik}\Big(\delta Z^L_{-,kj}+\delta Z^{L*}_{-,jk}\Big)
+\frac{1}{4}\sum^4_{k=1}
C^L_{kj}\Big(\delta Z^{L*}_{0,ki}+\delta Z^L_{0,ik}\Big)
-\frac{e s_\beta}{\sqrt{2}s_W}
 U_{j2}^*\delta t_W N_{i1}^*
\Big]P_L
}\\ \\ \displaystyle{
+\Big[
C^R_{ij}\Big(\delta Z_e-\frac{\delta s_W}{s_W}
+\frac{1}{2}\delta Z_{H^-}
+\frac{1}{4}\tan\beta\delta Z_{G^-H^-}\Big)
+\frac{1}{4}\sum^2_{k=1}
C^R_{ik}\Big(\delta Z^R_{-,kj}+\delta Z^{R*}_{-,jk}\Big)
}\\ \\ \displaystyle{
+\frac{1}{4}\sum^4_{k=1}
C^L_{kj}\Big(\delta Z^{L*}_{0,ki}+\delta Z^L_{0,ik}\Big)
+\frac{e c_\beta}{\sqrt{2}s_W}
V_{i2}\delta t_WN_{j1}
\Big]P_R\Big\}\widetilde{\chi}^-_j
}.
\end{array}
\end{equation}
\indent
The counterterm amplitude ${\mathcal M}^{(c)}_1$
can be explicitly derived from the above Lagrangian.
The renormalization of the input parameters $e$, $\theta_W$, $m_Z$ and $m_W$
follows the conventional on-mass-shell scheme\cite{mawengan1}.
The other renormalization constants with the on-mass-shell scheme
are defined as follows.
\\
\indent
The charged Higgs wave function renormalization constant is
\begin{equation}
\begin{array}{l}
\displaystyle{
\delta Z_{H^-}=-\widetilde{Re}\frac{\partial\Sigma_{H^-H^-}}{\partial p^2}
(m^2_{H^-}).
}
\end{array}
\end{equation}
\indent
The fermion wave function renormalization\cite{mawengan1} constants are
\begin{equation}
\begin{array}{l}
\displaystyle{
\delta Z^L_{ii}=-\widetilde{Re}\Big\{\Sigma^L_{ii}(m^2_i)+
m^2_i\Big[\frac{\partial \Sigma^L_{ii}}{\partial p^2}(m^2_i)+
\frac{\partial \Sigma^R_{ii}}{\partial p^2}(m^2_i)\Big]+
m_i\Big[\frac{\partial\Sigma^{SL}_{ii}}{\partial p^2}(m^2_i)+
\frac{\partial\Sigma^{SR}_{ii}}{\partial p^2}(m^2_i)\Big]\Big\},
}\\ \\ \displaystyle{
\delta Z^R_{ii}=-\widetilde{Re}\Big\{\Sigma^R_{ii}(m^2_i)+
m^2_i\Big[\frac{\partial \Sigma^L_{ii}}{\partial p^2}(m^2_i)+
\frac{\partial \Sigma^R_{ii}}{\partial p^2}(m^2_i)\Big]+
m_i\Big[\frac{\partial\Sigma^{SL}_{ii}}{\partial p^2}(m^2_i)+
\frac{\partial\Sigma^{SR}_{ii}}{\partial p^2}(m^2_i)\Big]\Big\},
}\\ \\ \displaystyle{
\delta Z^L_{ij}=\frac{2}{m^2_i-m^2_j}\widetilde{Re}
\Big[m_j^2\Sigma^L_{ij}(m^2_j)+m_im_j\Sigma^R_{ij}(m^2_j)+
m_i\Sigma^{SL}_{ij}(m_j^2)+m_j\Sigma^{SR}_{ij}(m_j^2)\Big],
\hspace{10pt}(i\not=j),
}\\ \\ \displaystyle{
\delta Z^R_{ij}=\frac{2}{m^2_i-m^2_j}\widetilde{Re}
\Big[m_j^2\Sigma^R_{ij}(m^2_j)+m_im_j\Sigma^L_{ij}(m^2_j)+
m_i\Sigma^{SR}_{ij}(m_j^2)+m_j\Sigma^{SL}_{ij}(m_j^2)\Big],
\hspace{10pt}(i\not=j).
}
\end{array}
\end{equation}
\indent
We use the scheme in Ref.\cite{mawengan1} to fix $G^--H^-$ mixing
renormalization constant,
\begin{equation}
\begin{array}{l}
\displaystyle{
\delta Z_{G^-H^-}=-\frac{2}{m_W}\widetilde{Re}\Sigma_{H^-W^-}(m_{H^-}^2).
}
\end{array}
\end{equation}
\indent Then the renormalization constants could be derived from
the self-energy Feynman diagrams shown in
Figs.\ref{self1}-\ref{self6}. Through the calculation, the UV
divergences of the one-loop vertex and the counterterm are
\begin{equation}
\begin{array}{l}
\displaystyle{
{\mathcal M}_1^{(v)}\Big|_{UV}=\Big(-\frac{1}{\epsilon}\Big)\Big\{
\Big[\frac{\alpha}{16{\pi}m_W^2s^2_{\beta}s^2_W}
[s^2_{\beta}(m^2_Z+18m^2_W)+\sum_{g=1}^3 6m^2_{u_g}]+
\frac{{\alpha}ec_{\beta}}{\sqrt{2}{\pi}s^2_Wc_W}N_{i1}V_{j2}\Big]C^R_{ij}P_R
} \\ \\ \displaystyle{
+\Big[\frac{\alpha}{16{\pi}m_W^2c^2_{\beta}s^2_W}
[c^2_{\beta}(m^2_Z+18m^2_W)+\sum_{g=1}^3(6m^2_{d_g}+2m^2_{e_g})]-
\frac{{\alpha}es_{\beta}}{\sqrt{2}{\pi}s^2_Wc_W}N_{i1}U_{j2}\Big]C^L_{ij}P_L
\Big\},
}
\end{array}
\end{equation}
\begin{equation}
\begin{array}{l}
\displaystyle{
{\mathcal M}_1^{(c)}\Big|_{UV}=\frac{1}{\epsilon}\Big\{
\Big[\frac{\alpha}{16{\pi}m_W^2s^2_{\beta}s^2_W}
[s^2_{\beta}(m^2_Z+18m^2_W)+\sum_{g=1}^3 6m^2_{u_g}]+
\frac{{\alpha}ec_{\beta}}{\sqrt{2}{\pi}s^2_Wc_W}N_{i1}V_{j2}\Big]C^R_{ij}P_R
} \\ \\ \displaystyle{
+\Big[\frac{\alpha}{16{\pi}m_W^2c^2_{\beta}s^2_W}
[c^2_{\beta}(m^2_Z+18m^2_W)+\sum_{g=1}^3(6m^2_{d_g}+2m^2_{l_g})]-
\frac{{\alpha}es_{\beta}}{\sqrt{2}{\pi}s^2_Wc_W}N_{i1}U_{j2}\Big]C^L_{ij}P_L
\Big\},
}
\end{array}
\end{equation}
where, $m_{u_g}, m_{d_g}$ and $m_{l_g}$ represent the mass of the u-type quarks,
the d-type quarks and the leptons, and g is the generation index.
\\
\indent Obviously, the UV divergences of ${\mathcal M}^{(v)}_1$
and ${\mathcal M}^{(c)}_1$ can cancel each other, as they must.
Then the renormalized amplitude at one-loop order is UV convergent
\begin{equation}
{\mathcal M}^V_1\Big|_{UV}=0.
\end{equation}
\indent
Thus the full one-loop virtual correction for the decay width is
\begin{equation}\label{vcorr}
\Gamma_V=\frac{1}{8\pi}\frac{p_{out}}{m^2_{H^-}}
2Re({\mathcal M}_0{\mathcal M}^{V*}_1),
\end{equation}
where the renormalized amplitude ${\mathcal M}^V_1$ is UV finite,
but it still contains infrared (IR) divergences, which can be
written as:
\begin{equation}\label{virtualIR}
\begin{array}{l}
\displaystyle{ {\mathcal M}_1^{V}\Big|_{IR}
=\frac{\alpha}{2\pi}\frac{1}{\epsilon}\Big[-2+\frac{2x_1x_2-x_1-x_2}{x_1-x_2}\ln\Big(
\frac{x_1x_2-x_2}{x_1x_2-x_1}\Big)\Big]{\mathcal M}_0, }
\end{array}
\end{equation}
where,
\begin{equation}
\begin{array}{l}
\displaystyle{ x_{1,2}=\frac{
m_{H^-}^2-m_{\widetilde{\chi}^-_j}^2+m_{\widetilde{\chi}^0_i}^2\pm2m_{H^-}p_{out}}
{2m_{\widetilde{\chi}^0_i}^2}
}.
\end{array}
\end{equation}
\indent
The IR divergences can be cancelled after adding the
contributions from the emission of real photons, which will be
described in detail in the following section.
\section{\bf Real Correction}
\begin{figure}[h]
\begin{center}
\includegraphics[width=200pt]{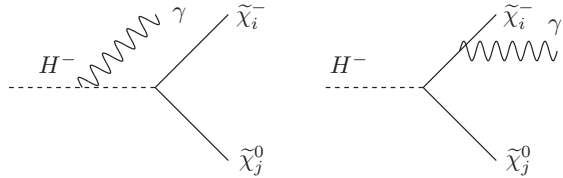}
\caption{The photon emission Feynman diagrams.\label{realcorrfd}}
\end{center}
\end{figure}

The Feynman diagrams for the real corrections
are shown in Fig.\ref{realcorrfd}.
\\
\indent
The relevant three-body decay width is
\begin{equation}
\Gamma_R=\frac{1}{2\Phi}\int\overline{\sum}|{\cal M}_3|^2dPS^{(3)},
\end{equation}
where $\Phi=m_{H^-}$ is the usual flux factor for the one particle initial state.
$dPS^{(3)}$ is the three-body phase space.
$\overline{\sum}|{\cal M}_3|^2$ is the squared amplitude averaged
over the initial degrees of freedom,
\\ $ \displaystyle{
\overline{\sum}|{\mathcal M}_3|^2=
\frac{2e^2}{(p_{\widetilde{\chi}^-}\cdot~p_{\gamma})^2}[-m_{\widetilde{\chi}^-}^2
(p_{\widetilde{\chi}^0}\cdot~p_{\gamma})-m_{\widetilde{\chi}^-}^2(p_{\widetilde{\chi}^-}
\cdot~p_{\widetilde{\chi}^0})+(p_{\widetilde{\chi}^-}\cdot~p_{\gamma})(p_{\widetilde{\chi}^0}
\cdot~p_{\gamma})](C_LC_R^{\dagger}+C_RC_L^{\dagger})
} $ \\ \\ $ \displaystyle{
\hspace{24pt}
+\frac{2e^2}{(p_{\widetilde{\chi}^-}\cdot~p_{\gamma})^2}
m_{\widetilde{\chi}^-}m_{\widetilde{\chi}^0}
[m_{\widetilde{\chi}^-}^2+(p_{\widetilde{\chi}^-}\cdot~p_{\gamma})]
(C_LC_L^{\dagger}+C_RC_R^{\dagger})
+\frac{2e^2}{[(p_{\widetilde{\chi}^-}\cdot~p_{\gamma})+
(p_{\widetilde{\chi}^0}\cdot~p_{\gamma})]^2}
} $ \\ \\ $ \displaystyle{
\hspace{24pt}
\times[m_{\widetilde{\chi}^-}^2+m_{\widetilde{\chi}^0}^2+2(p_{\widetilde{\chi}^-}
\cdot~p_{\widetilde{\chi}^0})+(p_{\widetilde{\chi}^-}\cdot~p_{\gamma})+
(p_{\widetilde{\chi}^0}\cdot~p_{\gamma})]
[-(p_{\widetilde{\chi}^-}\cdot~p_{\widetilde{\chi}^0})(C_LC_R^{\dagger}
+C_RC_L^{\dagger})
} $ \\ \\ $ \displaystyle{
\hspace{24pt}
+m_{\widetilde{\chi}^-}m_{\widetilde{\chi}^0}
(C_LC_L^{\dagger}+C_RC_R^{\dagger})]
+\frac{2e^2}{(p_{\widetilde{\chi}^-}\cdot~p_{\gamma})
[(p_{\widetilde{\chi}^-}\cdot~p_{\gamma})+(p_{\widetilde{\chi}^0}\cdot~p_{\gamma})]}
[2m_{\widetilde{\chi}^-}^2(p_{\widetilde{\chi}^-}\cdot~p_{\widetilde{\chi}^0})
} $ \\ \\ $ \displaystyle{
\hspace{24pt}
+2(p_{\widetilde{\chi}^-}\cdot~p_{\widetilde{\chi}^0})^2+2(p_{\widetilde{\chi}^-}
\cdot~p_{\widetilde{\chi}^0})(p_{\widetilde{\chi}^0}\cdot~p_{\gamma})
+(p_{\widetilde{\chi}^-}\cdot~p_{\widetilde{\chi}^0})(p_{\widetilde{\chi}^0}\cdot~
p_{\gamma})+m_{\widetilde{\chi}^-}^2(p_{\widetilde{\chi}^0}\cdot~p_{\gamma})
} $ \\ \\ $ \displaystyle{
\hspace{24pt}
-4m_{\widetilde{\chi}^0}^2(p_{\widetilde{\chi}^-}\cdot~p_{\gamma})]
(C_LC_R^{\dagger}+C_RC_L^{\dagger})
+\frac{2e^2}{(p_{\widetilde{\chi}^-}\cdot~p_{\gamma})
[(p_{\widetilde{\chi}^-}\cdot~p_{\gamma})+(p_{\widetilde{\chi}^0}\cdot~p_{\gamma})]}
(-m_{\widetilde{\chi}^-}m_{\widetilde{\chi}^0})
} $ \\ \\ $ \displaystyle{
\hspace{24pt}
\times[2m_{\widetilde{\chi}^-}^2+
2(p_{\widetilde{\chi}^-}\cdot~p_{\gamma})
+(p_{\widetilde{\chi}^0}\cdot~p_{\gamma})
+2(p_{\widetilde{\chi}^-}\cdot~p_{\widetilde{\chi}^0})](C_LC_L^{\dagger}+C_RC_R^{\dagger})
} $ \\
where $p_{\widetilde{\chi}^-}$, $p_{\widetilde{\chi}^0}$ and $p_{\gamma}$ is the relevant four
dimensional momentums.
\\
\indent
The IR singularities arise from the phase space integration
for the real soft photon emission, which can be conveniently isolated
by slicing the space into two regions defined by suitable cut-off
$\delta_s$, according to whether the energy of the emitted photon is soft,
i.e. $E_\gamma\le\delta_s m_{H^-}/2$, or not. Correspondingly,
the three-body decay width can be written
into two parts as following.
\begin{equation}
\frac{1}{2\Phi}\int\overline{\sum}|{\cal M}_3|^2dPS^{(3)}
=\frac{1}{2\Phi}\int_{soft}\overline{\sum}|{\cal M}_3|^2dPS^{(3)}
+\frac{1}{2\Phi}\int_{hard}\overline{\sum}|{\cal M}_3|^2dPS^{(3)},
\end{equation}
where the corresponding parts are $\Gamma_{soft}$ and $\Gamma_{hard}$ respectively.
\\
\indent The hard part $\Gamma_{hard}$ is IR finite and can be
numerically calculated using the Cuba program\cite{cuba}. The IR
divergences only live in the soft part $\Gamma_{soft}$. Using the
eikonal approximation, the soft part can be factorized into an IR
factor, multiplied by the tree-level decay width.
\begin{equation}
\Gamma_{soft}=\delta_{IR}\Gamma_0,
\end{equation}
where
\begin{equation}\label{realIR}
\begin{array}{l}
\displaystyle{
\delta_{IR}=\frac{\alpha}{2\pi}\Big\{\frac{1}{\epsilon}\Big[
2-\frac{2x_1x_2-x_1-x_2}{x_1-x_2}\ln\Big(\frac{x_1x_2-x_2}{x_1x_2-x_1}\Big)\Big]
+\Big[\ln\Big(\frac{4\pi\mu^2}{\delta_s^2m^2_{H^-}}\Big)-\gamma_E\Big]
} \\ \\ \displaystyle{
\hspace{24pt}
\times\Big[2-\frac{2x_1x_2-x_1-x_2}{x_1-x_2}
\ln\Big(\frac{x_1x_2-x_2}{x_1x_2-x_1}\Big)\Big]
+2+\frac{2x_1x_2-x_1-x_2}{x_1-x_2}
} \\ \\ \displaystyle{
\hspace{24pt}
\times\Big[\ln\Big(\frac{x_1x_2-x_2}{x_1x_2-x_1}\Big)
-2Li_2\Big(\frac{2(x_1-x_2)}{x_1x_2-x_2}\Big)
-\frac{1}{2}\ln^2\Big(\frac{x_1x_2-x_2}{x_1x_2-x_1}\Big)
\Big]\Big\}.
}
\end{array}
\end{equation}
\indent From Eq.(\ref{virtualIR}) and Eq.(\ref{realIR}) we can see
that the IR divergences in $\Gamma_V$ and $\Gamma_R$ can be
cancelled. Finally, summing up the tree-level, the virtual and the
real corrections, the decay width of
$H^-\rightarrow\widetilde{\chi}^-\widetilde{\chi}^0$, including
the FEW corrections, is
\begin{equation}
\Gamma=\Gamma_0+\Gamma_V+\Gamma_R.
\end{equation}
\section{\bf Numerical Results}
\begin{figure}[!h]
\includegraphics[width=400pt]{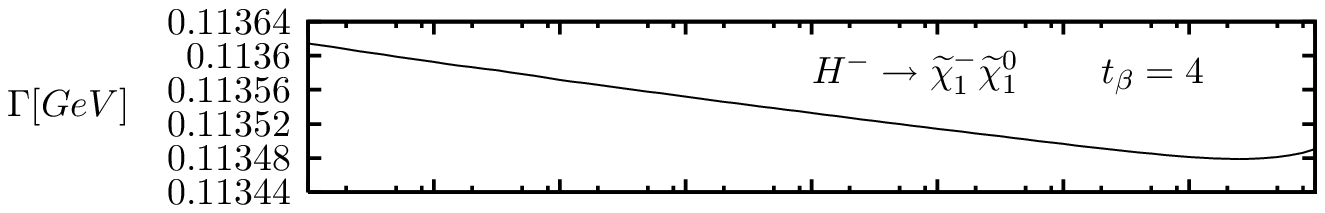}
\includegraphics[width=400pt]{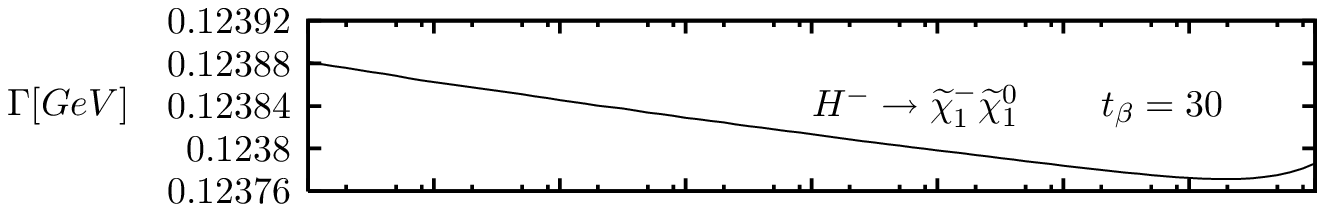}
\includegraphics[width=400pt]{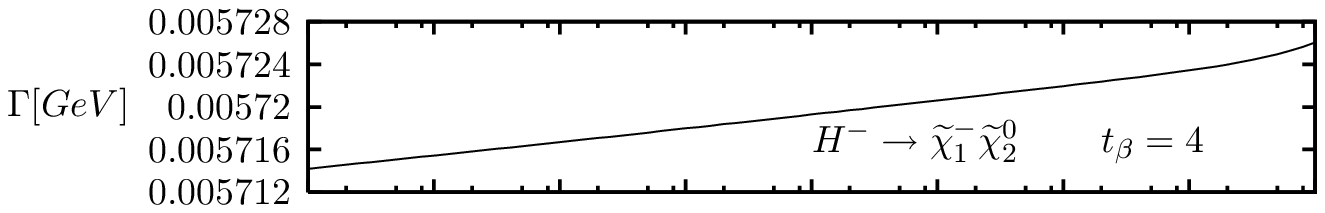}
\includegraphics[width=400pt]{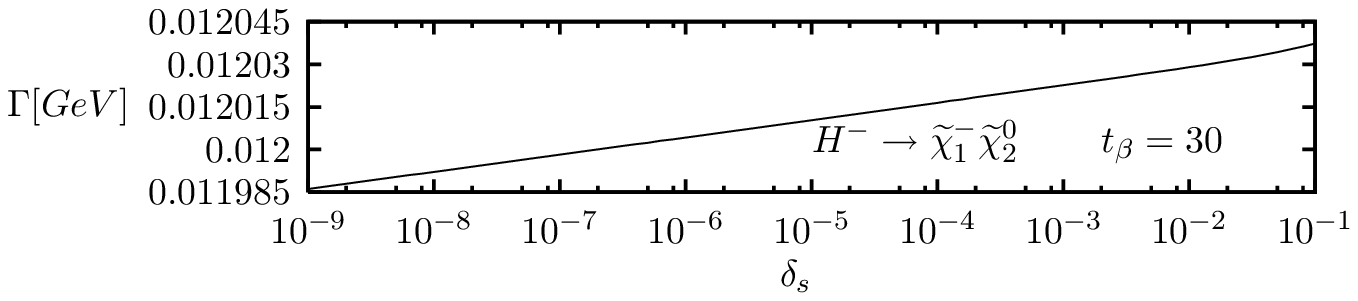}
\caption{The dependence of the decay width on
the soft photon cut-off scale.
\label{abssoft}}
\end{figure}

\begin{figure}[!h]
\includegraphics[width=400pt,height=280pt]{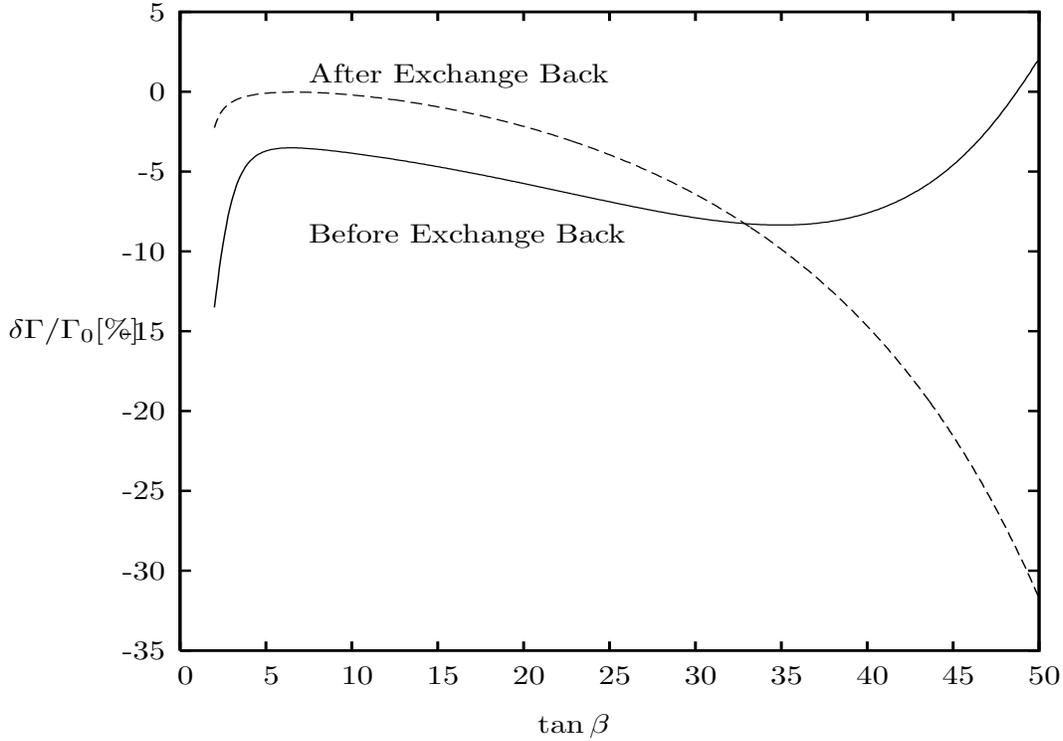}
\caption{The LO corrections to
$H^-\rightarrow\widetilde{\chi}^-_1\widetilde{\chi}^0_2$
before and after exchanging back
\label{exchangingback}}
\end{figure}

\begin{figure}[!h]
\includegraphics[width=400pt,height=135pt]{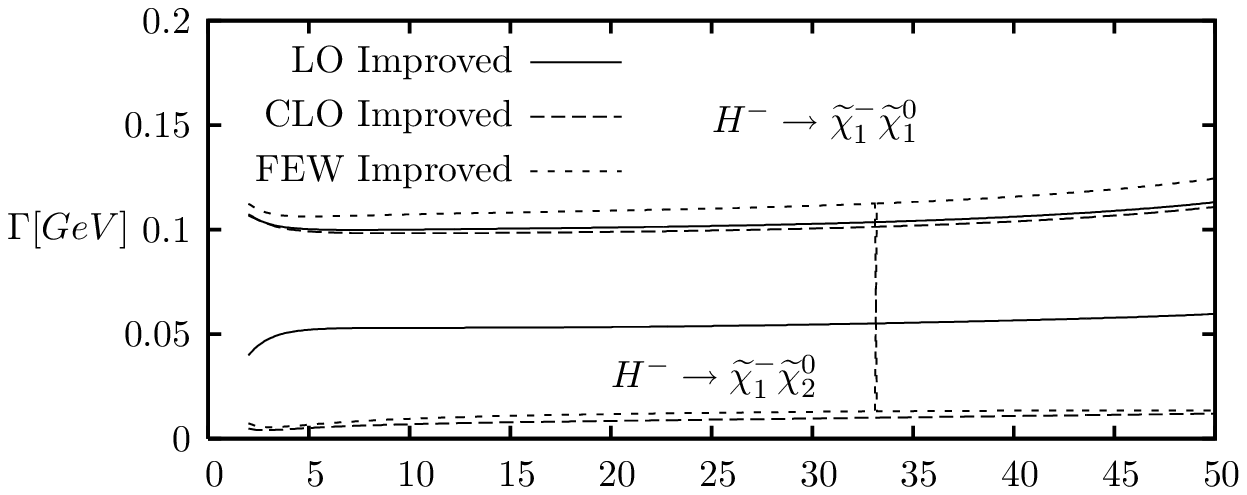}
\includegraphics[width=400pt,height=145pt]{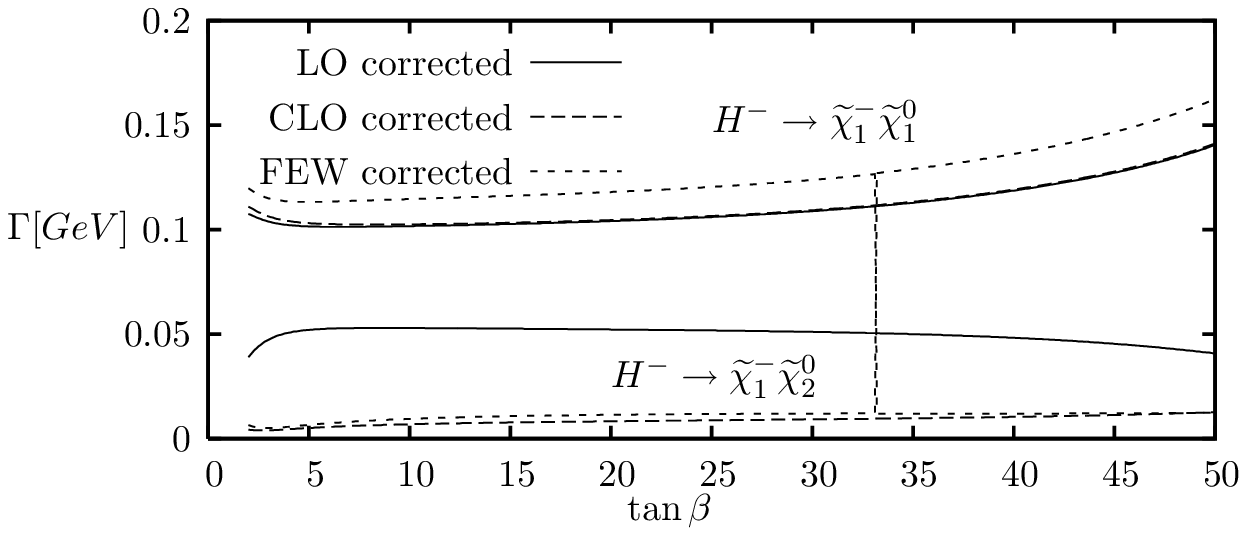}
\caption{The improved tree-level decay width (above) and
the corrected decay width (below) as the functions of
$\tan\beta$.
\label{abstanbeta}}
\end{figure}

\begin{figure}[!h]
\includegraphics[width=400pt,height=280pt]{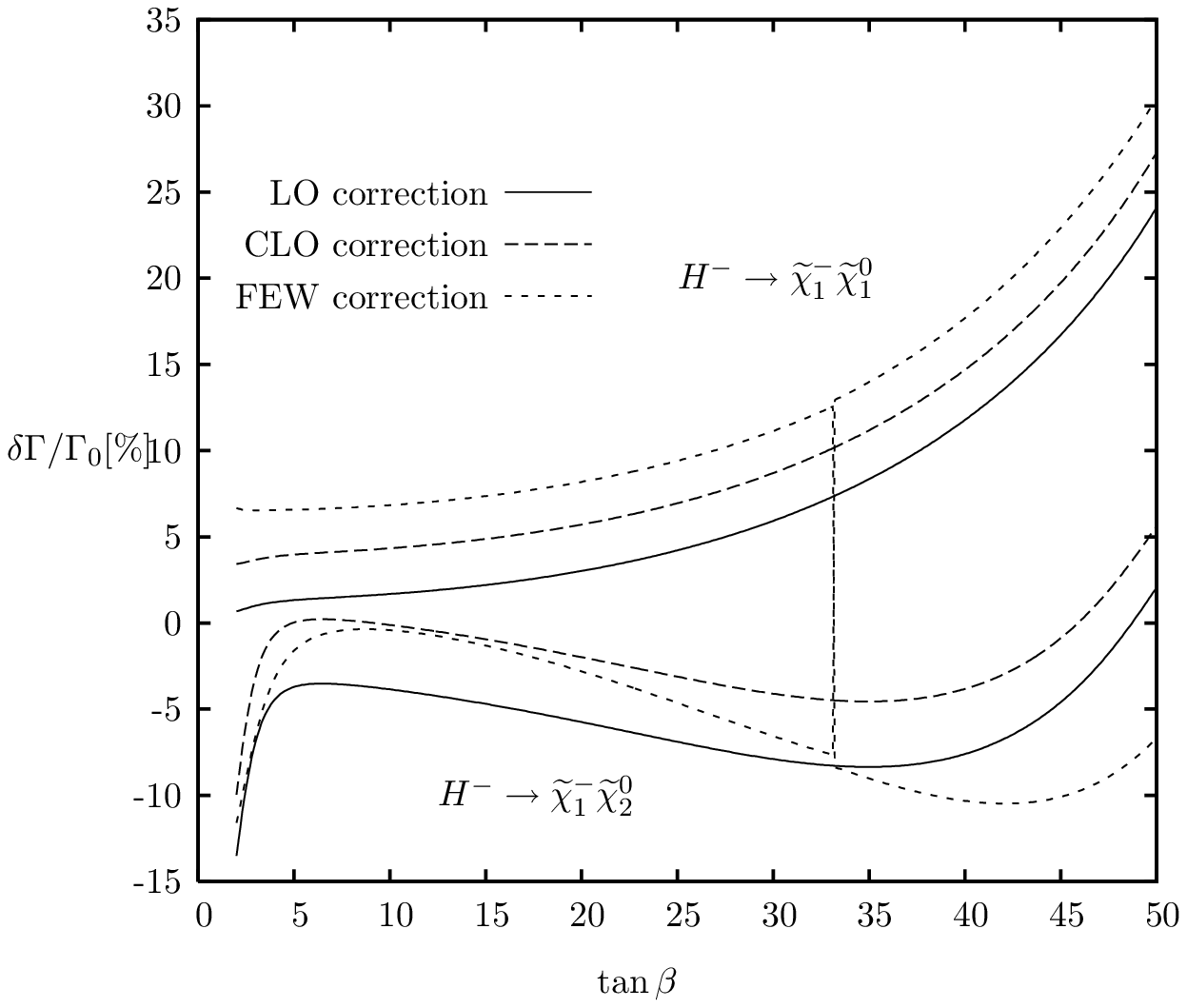}
\caption{The relative correction as the functions of
$\tan\beta$.
\label{retanbeta}}
\end{figure}

\begin{figure}[!h]
\includegraphics[width=400pt,height=140pt]{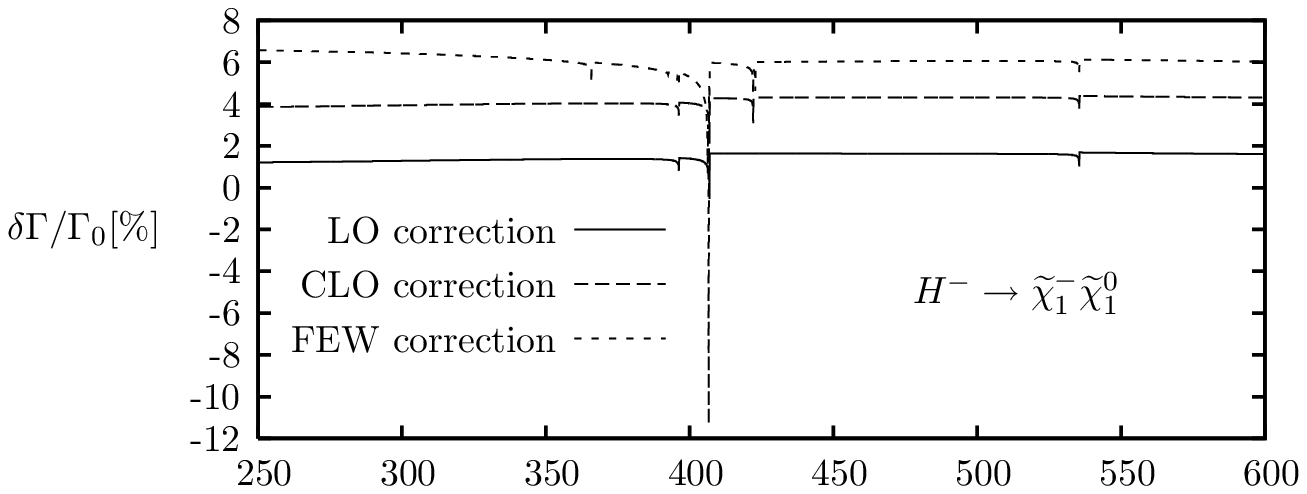}
\includegraphics[width=400pt,height=155pt]{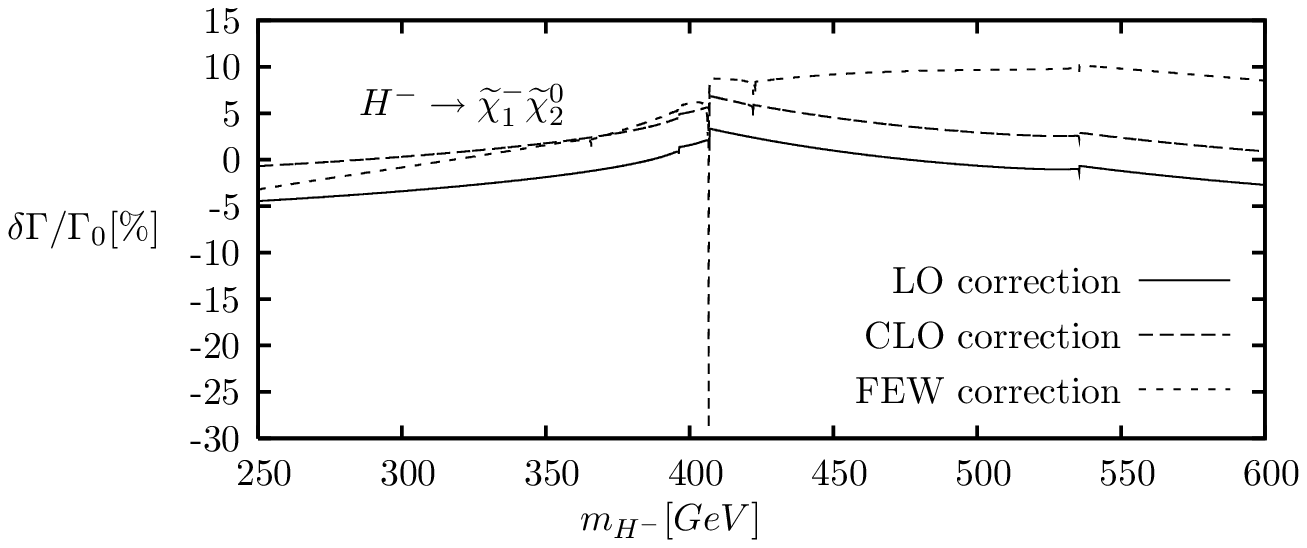}
\caption{The relative correction
as the functions of $m_{H^-}$
for $\tan\beta=4$.
\label{remhp1}}
\end{figure}

\begin{figure}[!h]
\includegraphics[width=400pt,height=140pt]{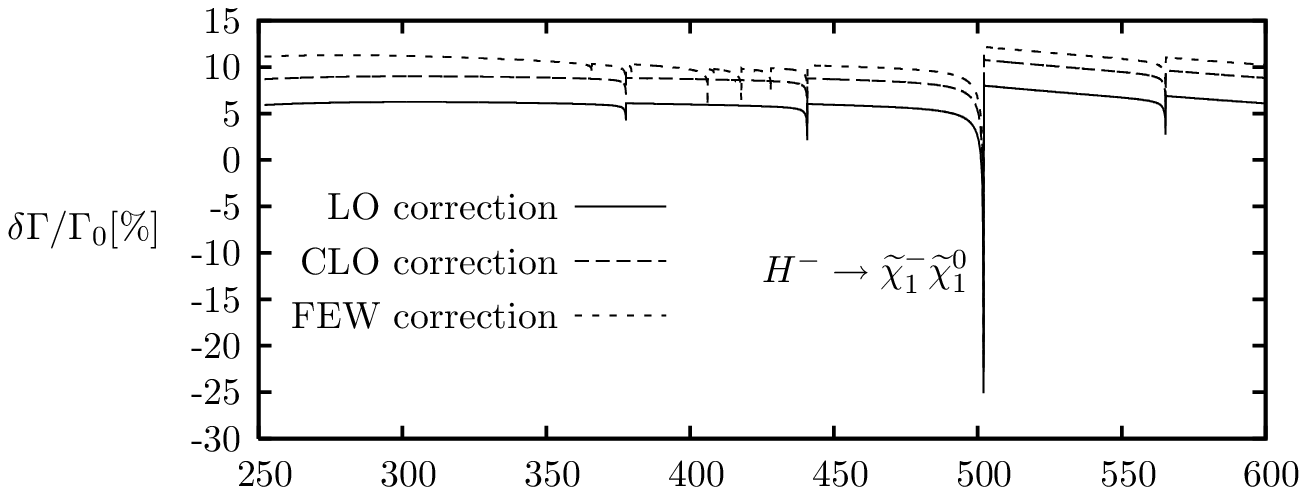}
\includegraphics[width=400pt,height=155pt]{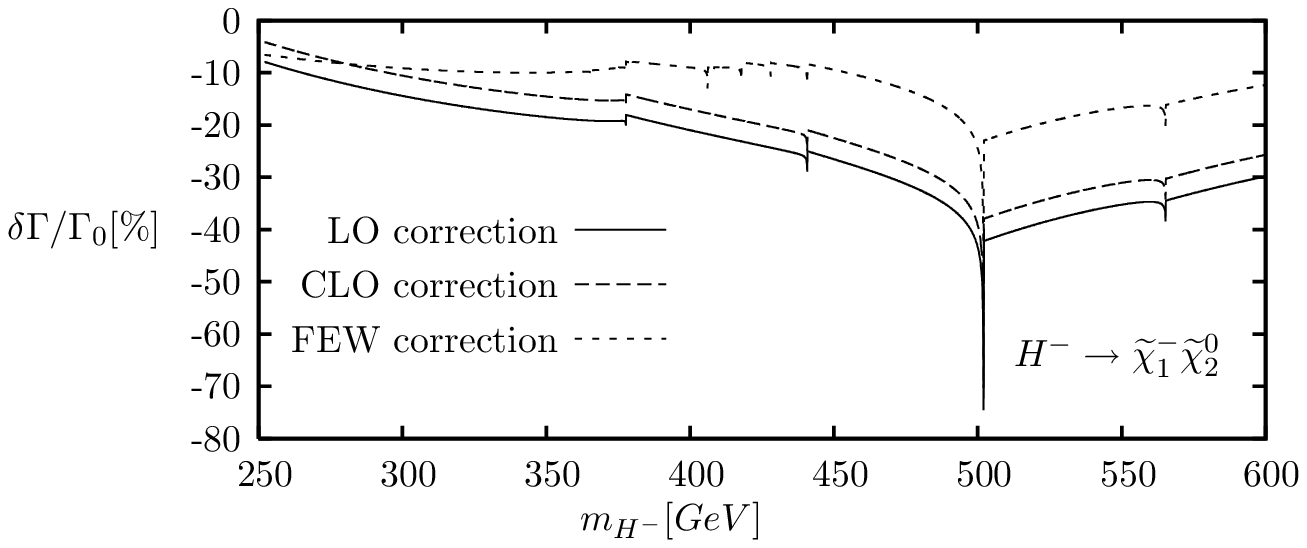}
\caption{The relative correction
as the functions of $m_{H^-}$
for $\tan\beta=30$.
\label{remhp2}}
\end{figure}

\begin{figure}[!h]
\includegraphics[width=400pt,height=140pt]{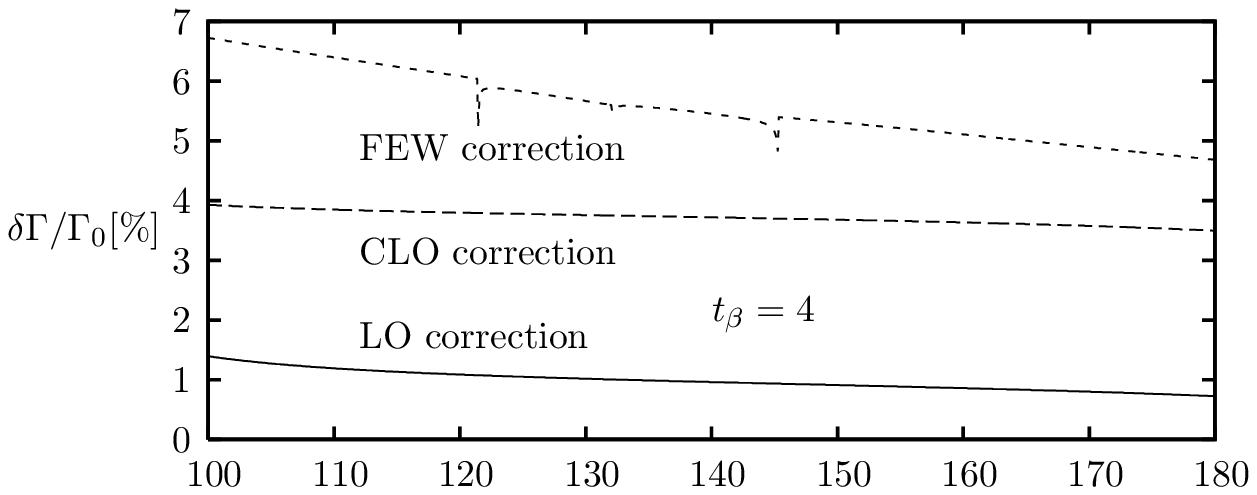}
\includegraphics[width=400pt,height=155pt]{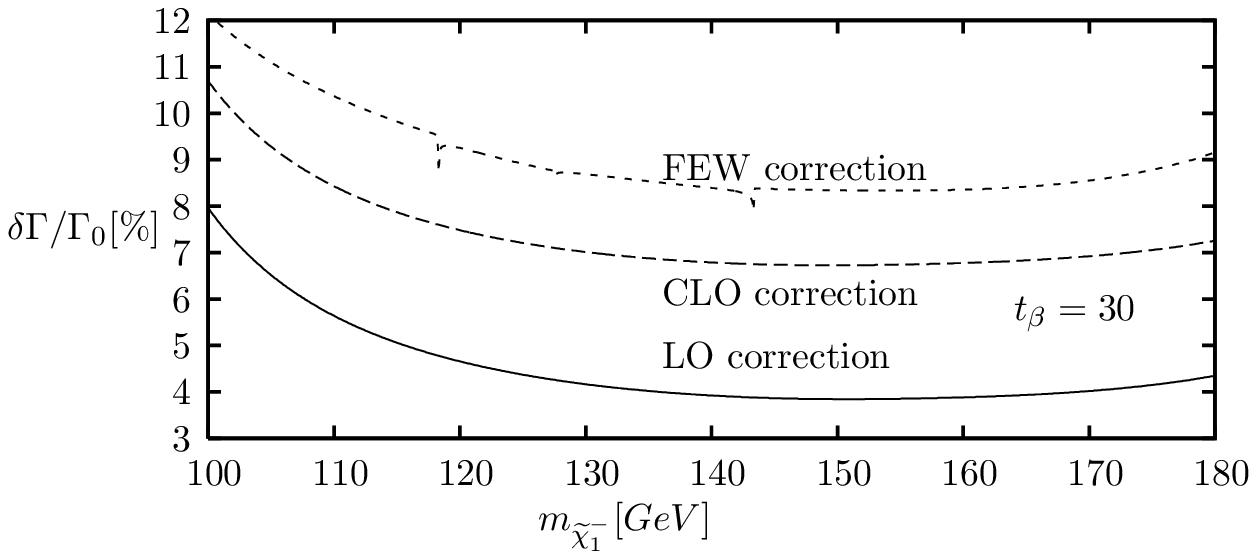}
\caption{The relative correction
as the functions of $m_{\widetilde{\chi}^-_1}$
for $H^-\rightarrow\widetilde{\chi}^-_1\widetilde{\chi}^0_1$.
\label{remcha}}
\end{figure}

\begin{figure}[!h]
\includegraphics[width=400pt,height=140pt]{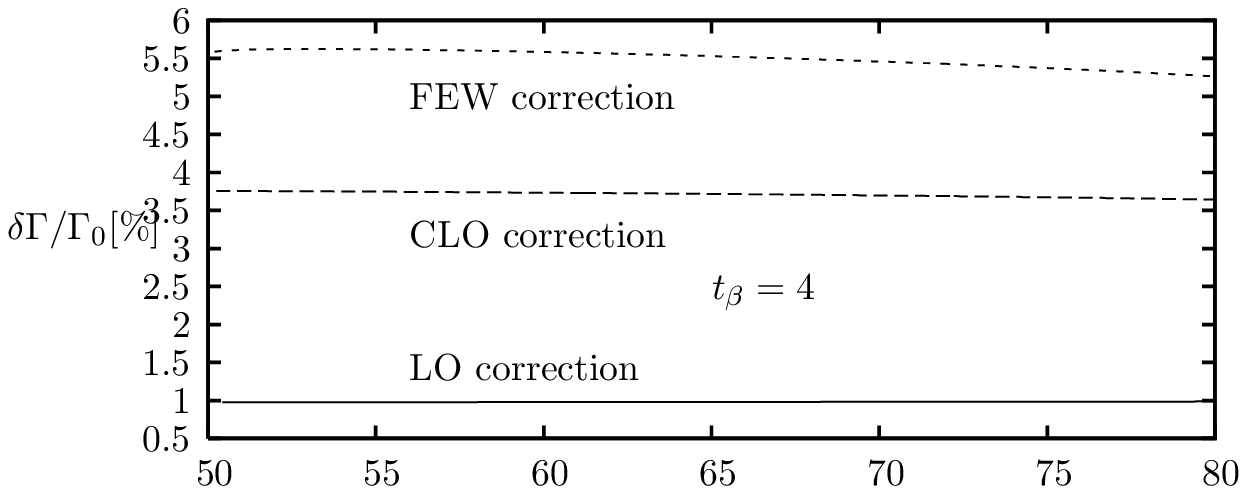}
\includegraphics[width=400pt,height=155pt]{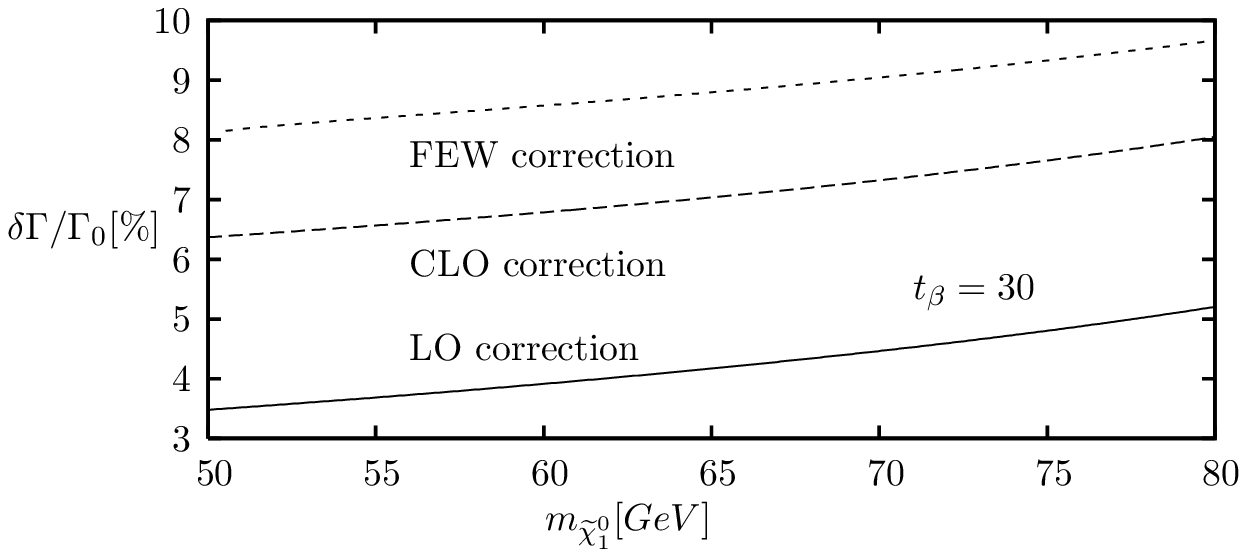}
\caption{The relative correction
as the functions of $m_{\widetilde{\chi}^0_1}$
for $H^-\rightarrow\widetilde{\chi}^-_1\widetilde{\chi}^0_1$.
\label{remneu}}
\end{figure}

\begin{figure}[!h]
\includegraphics[width=400pt,height=140pt]{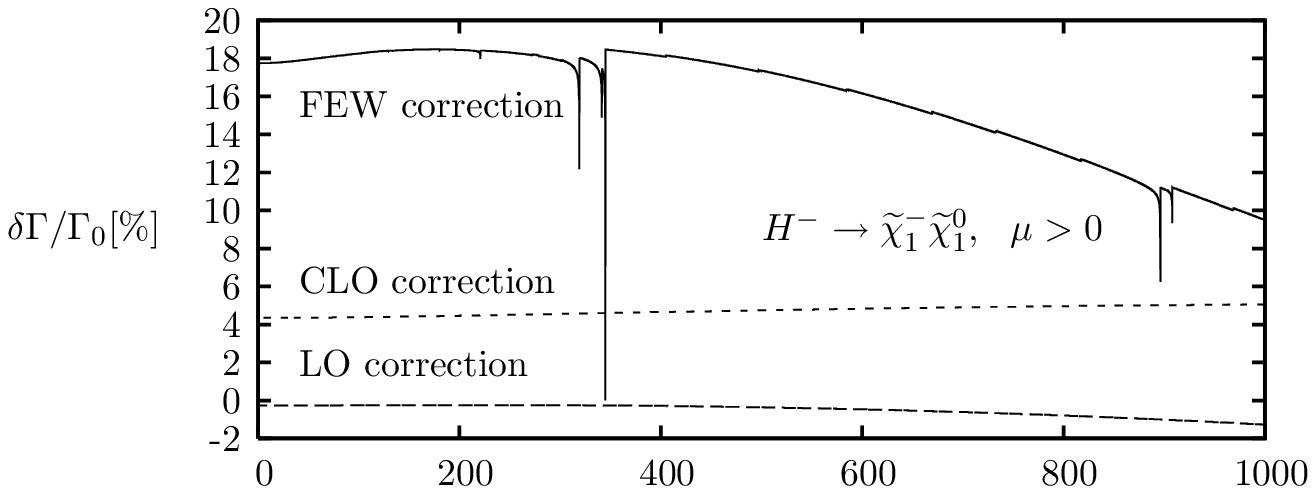}
\includegraphics[width=400pt,height=140pt]{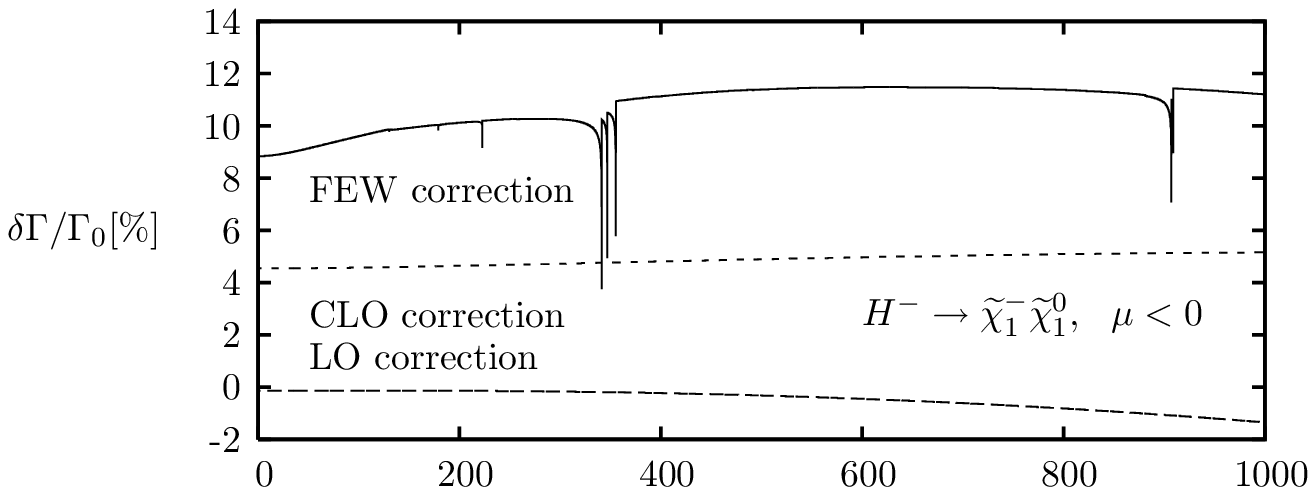}
\includegraphics[width=400pt,height=140pt]{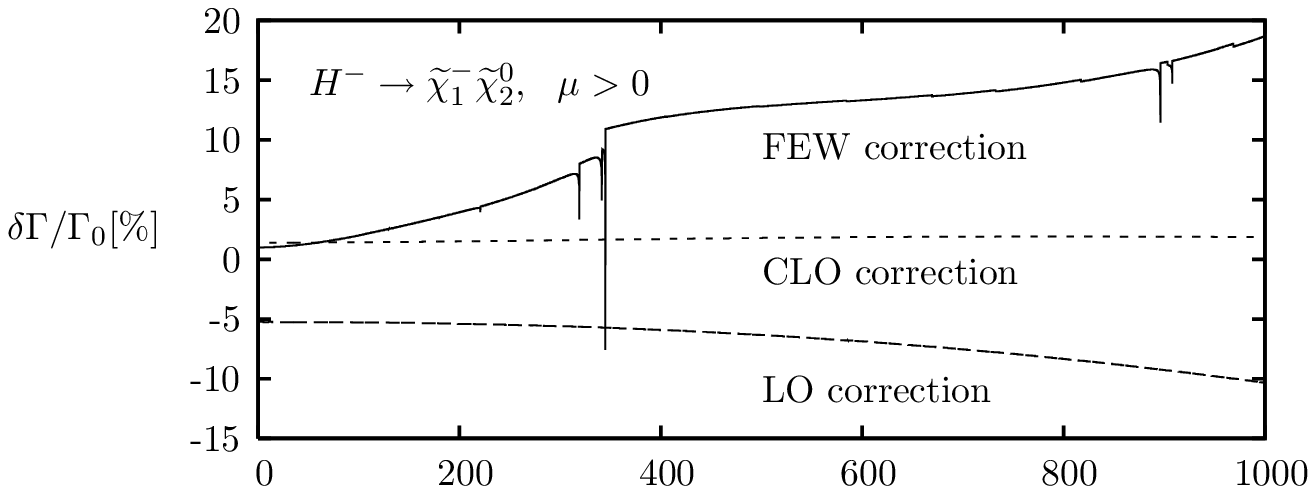}
\includegraphics[width=400pt,height=155pt]{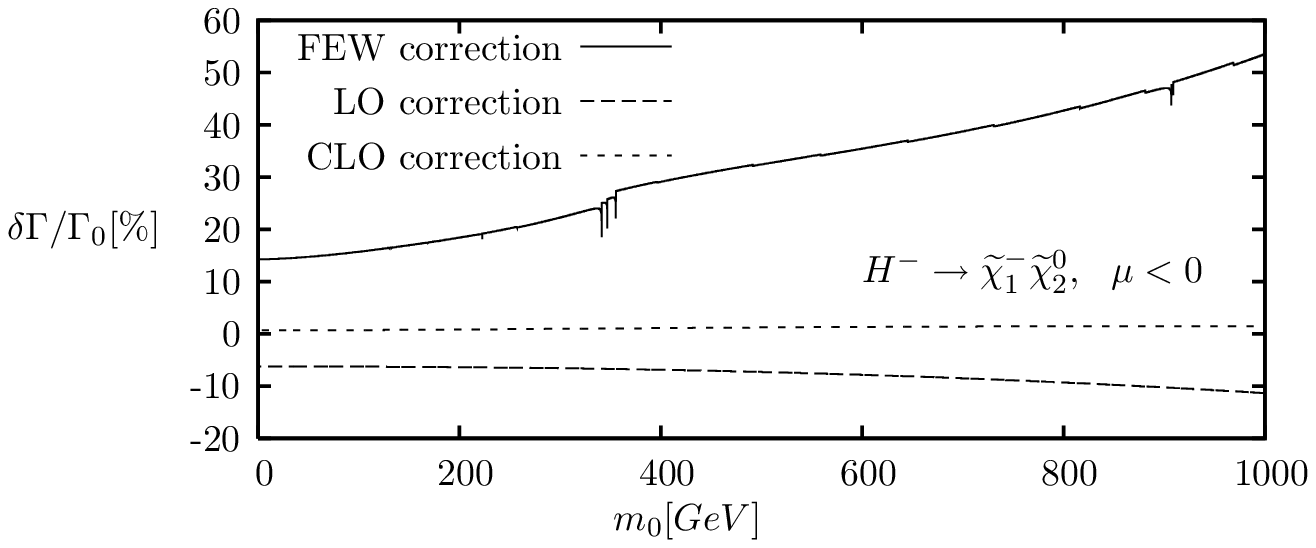}
\caption{The relative correction
as the functions of $m_0$.
\label{rem0}}
\end{figure}

\begin{figure}[!h]
\includegraphics[width=400pt,height=140pt]{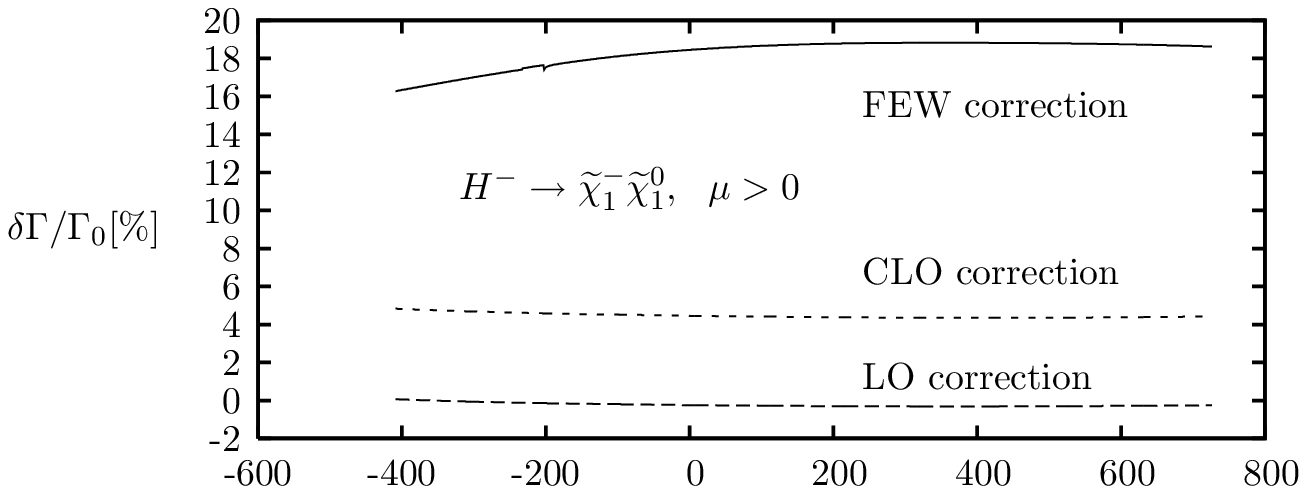}
\includegraphics[width=400pt,height=140pt]{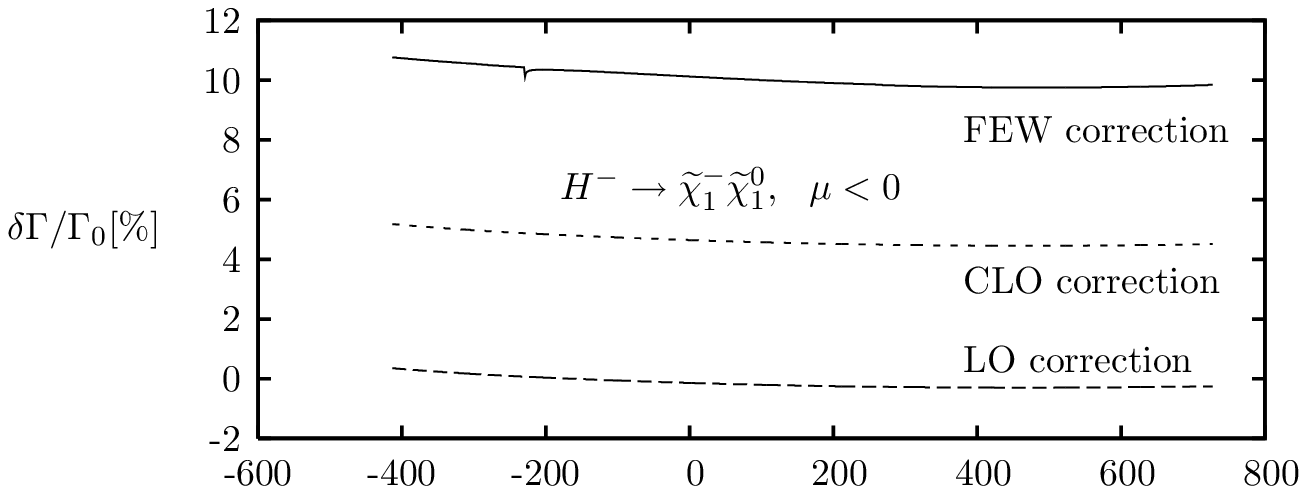}
\includegraphics[width=400pt,height=140pt]{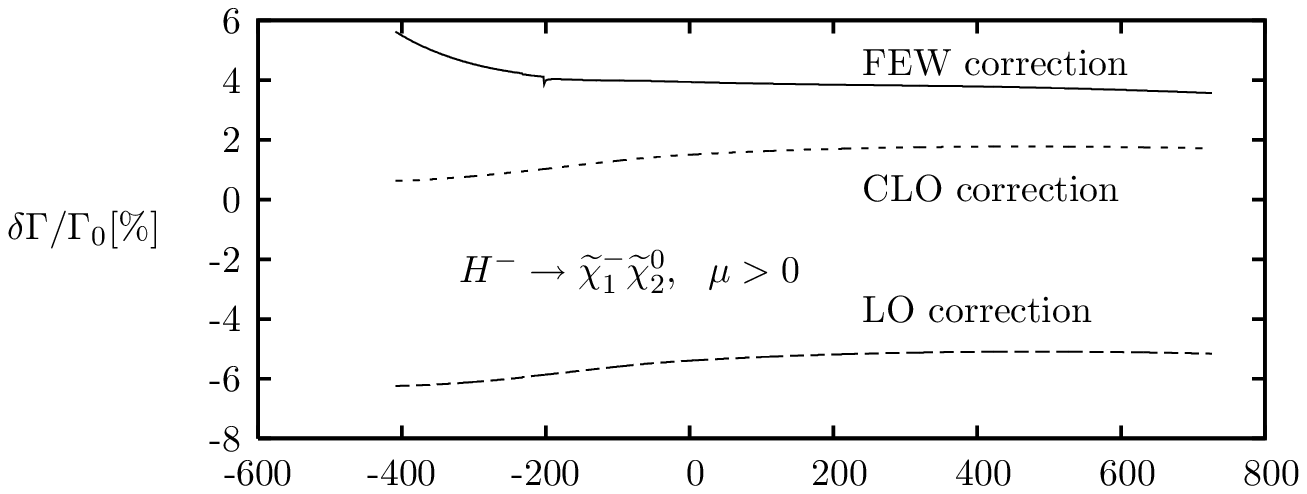}
\includegraphics[width=400pt,height=155pt]{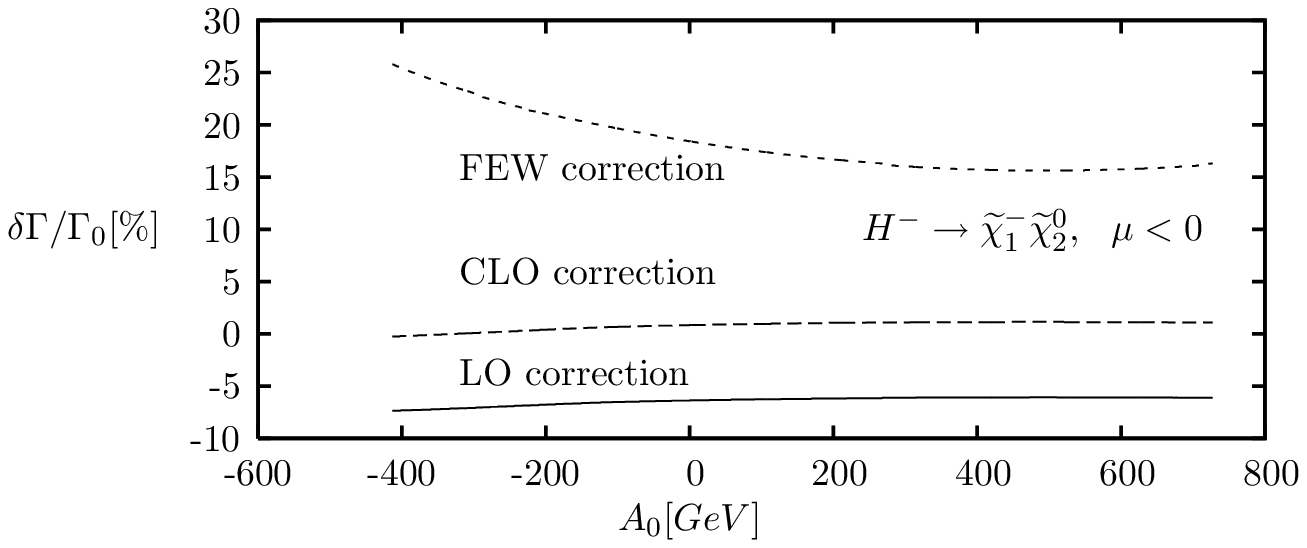}
\caption{The relative correction
as the functions of $A_0$.
\label{rea0}}
\end{figure}

We now present some numerical results of two charged Higgs decay
modes:
$H^-\rightarrow\widetilde{\chi}^-_1\widetilde{\chi}^0_1$ (mode 1)
and $H^-\rightarrow\widetilde{\chi}^-_1\widetilde{\chi}^0_2$ (mode
2), which are dominant decay modes allowed by kinetics. The SM input parameters are chosen as follows\cite{PDG},
$$
\begin{array}{lll}
m_Z=91.1875GeV, &
m_W=80.45GeV, &
\alpha_{EW}=1/137,
\\
m_e=0.51MeV, &
m_\mu=105.658MeV, &
m_\tau=1.777GeV,
\\
m_u=53.8MeV, &
m_c=1.5GeV, &
m_t=178GeV,
\\
m_d=53.8MeV, &
m_s=150MeV, &
m_b=4.7GeV.
\end{array}
$$
As mentioned in Ref.\cite{suspect}, the masses of the up and down quarks
are effective parameters which are
adjusted such that the five-flavor hadronic contribution to $\Delta\alpha$
is 0.02788 \cite{Je99}, ie
\begin{equation*}
\Delta\alpha_{\text{had}}^{(5)}(s = M_Z^2)
= \frac{\alpha}{\pi}\sum_{f = u,c,d,s,b}
 q_f^2 \Bigl(\log\frac{M_Z^2}{m_f^2} - \frac 53\Bigr)
  \overset{!}{=} 0.02778\,.
   \end{equation*}
\indent
For the phenomenological MSSM parameters, we choose all
the parameters are real and use the suspect program\cite{suspect}
to calculate the particle spectrum. Our calculations are mainly
based on the relevant inputs as following unless specified:
$$
m_{H^-}=250GeV,
\hspace{20pt} m_{\widetilde{\chi}^-_1}=100GeV,
\hspace{20pt} m_{\widetilde{\chi}^-_2}=300GeV,
\hspace{20pt} m_{\widetilde{\chi}^0_1}=60GeV,
$$ $$
m_{\widetilde{Q}}=m_{\widetilde{U}}=m_{\widetilde{D}}=
m_{\widetilde{E}}=m_{\widetilde{L}}=A_t=A_b=A_\tau=M_{SUSY}=200GeV.
$$
\indent
With the above chargino and neutralino masses, the fundamental parameters M, M' and $\mu$
are extracted from the tree-level mass matrixes (3) and (4), assuming $\mu<0$ and 
the magnitude of M is always larger than that of $\mu$.
The above input parameters are consistent with all the
existing experiment data\cite{PDG}. With the these mass parameters,
we calculate out the basic phenomenological MSSM parameters.
Note that the inputs are the
same as the ones in Ref.\cite{mawengan1}. But we vary $\tan\beta$,
$m_{H^-}$, $m_{\widetilde{\chi}^-_1}$ and
$m_{\widetilde{\chi}^0_1}$ to examine their effects on the decay
widths.
\\
\indent Fig.\ref{abssoft} presents the dependence of the FEW
corrected decay width on the arbitrary soft cut-off scale
$\delta_s$, introduced in the PSS method. As $\delta_s$ varies
from $10^{-1}$ to $10^{-9}$, the uncertainty of the decay width is
below $\pm0.1\%$. Therefore we set the soft cut-off scale
$\delta_s$ as $10^{-5}$ through our numerical calculations.
\\
\indent When we include the corrections as shown in Eqs.(14)-(29)
from mixing matrixes , the sequence of the masses of the
neutralino 2 and 3 will be exchanged. Consequently, the second and
the third row in the rotation matrix N will be exchanged. This is
so-called level crossings\cite{bartl}. To prove this viewpoint, we
force to exchange back between neutralino 2 and 3.
Fig.\ref{exchangingback} shows the LO corrections before and after
the exchanging. We can see the corrections after exchanging are
almost the same as the corresponding corrections in
Ref.\cite{mawengan1}.
\\
\indent Fig.\ref{abstanbeta} shows that the improved tree-level
decay width and the LO, the CLO and the FEW corrected decay width
as the functions of $\tan\beta$, respectively. As $\tan\beta\ge4$,
the LO corrections increases the tree-level decay width for the
decay mode 1 and slightly decreases it for the decay mode 2.
\\
\indent Fig.\ref{retanbeta} shows the LO, the CLO and the FEW
relative corrections as the functions of $\tan\beta$,
respectively. As $\tan\beta$ ranges between 2 and 50, all the
corrections keep increasing with the increasing of $\tan\beta$ for
the decay mode 1, which can reach $30\%$, and vary between $5\%$
and $-15\%$ for the decay mode 2. Comparing with the LO
corrections, the FEW and the CLO corrections for the mode 1 are in
general larger by almost $6\%$ and $3\%$, respectively. From
Fig.\ref{retanbeta}, we can also see the changes of FEW and CLO
corrections for the mode 2 are not negligible, and the LO
corrections are no longer important as $\tan\beta<5$ for the mode
1 and $\tan\beta>45$ for the mode 2, since in these conditions the
quark mass-dependent terms in the
$\widetilde{\chi}\widetilde{q}\bar{q}$ vertexes are small and the
mass-independent terms are important, thus the contributions from
the first and second generation quarks should also be significant.
Moreover, we find that the curves for the LO corrections of the
mode 1 are almost the same as that in Ref.\cite{mawengan1}.
However, the curve for LO corrections of the mode 2 is different
due to the level crossings as discussed above. For the same
reason, the FEW corrections for the mode 1 and 2 are changed to
each other for $\tan\beta=33.1$.
\\
\indent Fig.\ref{remhp1} shows the LO, the CLO and the FEW
corrections for $\tan\beta=4$ as the functions of $m_{H^-}$,
respectively. These corrections are not very sensitive to
$m_{H^-}$ for the decay mode 1, and have a little dependence of
$m_{H^-}$ for the decay mode 2. As $m_{H^-}$ ranges between 250GeV
and 600GeV, the corrections do not change too much. Comparing with
the LO and the CLO corrections, in general, the FEW corrections
for the mode 1 increase about $6\%$ and $2\%$, respectively, and
the magnitude of the FEW corrections for the mode 2 can increase
about $8\%$ and $6\%$, respectively. There are many dips on the
curves, which come from the singularities at the threshold points,
for example, respective ones of which on the LO and the CLO
corrections curves are shown as following:
\begin{center}
$m_{H^-}(396.2GeV)=m_{\widetilde{t}_1}(186.6GeV)+m_{\widetilde{b}_1}(209.6GeV),$
\\
$m_{H^-}(407.0GeV)=m_{\widetilde{t}_1}(186.6GeV)+m_{\widetilde{b}_2}(220.4GeV),$
\\
$m_{H^-}(535.4GeV)=m_{\widetilde{t}_2}(325.8GeV)+m_{\widetilde{b}_1}(209.6GeV).$
\end{center}
Moreover, there are also more little dips appearing on the curves
of the FEW corrections that come from the singularities of other
loop Feynman diagrams.
\\
\indent Fig.\ref{remhp2} gives almost the same case as
Fig.\ref{remhp1} except $\tan\beta=30$. Comparing with the LO and
the CLO corrections, in general, the FEW corrections for the mode
1 increase about $5\%$ and $2\%$, respectively, and the magnitude
of the FEW corrections for the mode 2 can increase about $14\%$
and $10\%$, respectively. The respective dips on the LO and the
CLO corrections curves, arising from the singularities at the
threshold points, are
\begin{center}
$m_{H^-}(377.6GeV)=m_{\widetilde{t}_1}(195.8GeV)+m_{\widetilde{b}_1}(181.8GeV),$
\\
$m_{H^-}(439.9GeV)=m_{\widetilde{t}_1}(195.8GeV)+m_{\widetilde{b}_2}(244.1GeV),$
\\
$m_{H^-}(501.0GeV)=m_{\widetilde{t}_2}(319.2GeV)+m_{\widetilde{b}_1}(181.8GeV),$
\\
$m_{H^-}(563.3GeV)=m_{\widetilde{t}_2}(319.2GeV)+m_{\widetilde{b}_2}(244.1GeV).$
\end{center}
\indent Moreover, there are also more little dips on the curves of
the FEW corrections. In comparison, the results of the LO
corrections shown in Figs.\ref{remhp1} and \ref{remhp2} agree with
the ones in Ref.\cite{mawengan1}.
\\
\indent Fig.\ref{remcha} gives the LO, the CLO and the FEW
corrections to
$H^-\rightarrow\widetilde{\chi}^-_1\widetilde{\chi}^0_1$ as the
functions of $m_{\widetilde{\chi}^-_1}$ for $\tan\beta=4$ and
$30$, respectively. The LO corrections are about $1\%$ for
$\tan\beta=4$ and generally vary from $4\%$ to $8\%$ for
$\tan\beta=30$. Comparing with the LO corrections, in general, for
both above cases the CLO corrections increase about $3\%$, and the
FEW corrections increase about $5\%$, respectively.
\\
\indent Fig.\ref{remneu} presents the LO, the CLO and the FEW
corrections to
$H^-\rightarrow\widetilde{\chi}^-_1\widetilde{\chi}^0_1$ as the
functions of $m_{\widetilde{\chi}^0_1}$, respectively. Here, we
choose the same parameters as above except
$m_{\widetilde{\chi}^-_1}=128$GeV. When $\tan\beta=4$, the LO and
the CLO corrections almost do not change, and the FEW corrections
slightly decrease with the increasing of
$m_{\widetilde{\chi}^0_1}$. When $\tan\beta=30$, all three
corrections increase with the increasing of
$m_{\widetilde{\chi}^0_1}$. Comparing with the LO corrections, in
general, the CLO corrections increase about $3\%$, and the FEW
corrections increase about $5\%$, respectively.
\\
\indent In the following calculations, the MSSM parameters are
constrained within the mSUGRA\cite{msugra}, in which there are
only five free input parameters, i.e. $m_{1/2},m_0,A_0,\tan\beta$
and sign of $\mu$, where $m_{1/2},m_0,A_0$ are the universal
gaugino mass, scalar mass at GUT scale and the trilinear soft
breaking parameter in the superpotential terms, respectively.
\\
\indent Fig.\ref{rem0} shows the LO, the CLO and the FEW
corrections as the functions of $m_0$ ranging between 0 and
$1000$GeV, respectively, assuming $m_{1/2}=200$GeV, $A_0=0$ and
$\tan\beta=10$. We find that for both decay modes the LO and the
CLO corrections only slightly change as $m_0$ varies. For the
decay mode 1, the FEW corrections can be larger than the LO and
the CLO corrections by $18\%$ and $14\%$, respectively. For the
decay mode 2, the FEW corrections can be larger than the LO
corrections by $50\%$, and than the CLO corrections by $60\%$.
\\
\indent Fig.\ref{rea0} gives the LO, the CLO and the FEW
corrections as the functions of $A_0$, respectively, assuming
$m_{1/2}=200$GeV, $m_0=200$GeV and $\tan\beta=10$. The LO and the
CLO corrections almost do not change with varying of $A_0$, but
the FEW corrections change much. The FEW corrections can be larger
than the LO and the CLO corrections by about $18\%$ and $13\%$ for
the decay mode 1, respectively. For the decay mode 2, the FEW
corrections tend to decrease and can be larger than the LO and the
CLO corrections by about $30\%$ and $25\%$, respectively.
\\
\indent In conclusion, we have calculated the FEW corrections to
the charged Higgs decays into a neutralino and chargino in the
MSSM, and compared with the LO and the CLO corrections. Our
results show that the magnitudes of the FEW corrections can be
larger than $10\%$ for both decay modes for $\tan\beta>30$.
Moreover, comparing with the FEW corrections, both of the LO and
the CLO corrections are negligible small for the mode 1 when
$\tan\beta<5$, and for the mode 2 when $\tan\beta>45$,
respectively, since there are not enhancements from the Yukawa
couplings. We have also calculated the FEW corrections in the
mSUGRA scenario, where the FEW corrections can be larger than the
LO and the CLO corrections by more than $60\%$ and $50\%$,
respectively. Thus the FEW corrections are significant, which
might be observable in the future high precision experiments for
Higgs physics.

\section*{\bf Acknowledgment}
This work was supported in part by the National Natural Science
Foundation of China and Specialized Research Fund for the Doctoral
Program of Higher Education.

\section*{\bf Appendix}
\begin{figure}[!h]
\centering
\includegraphics[width=340pt]{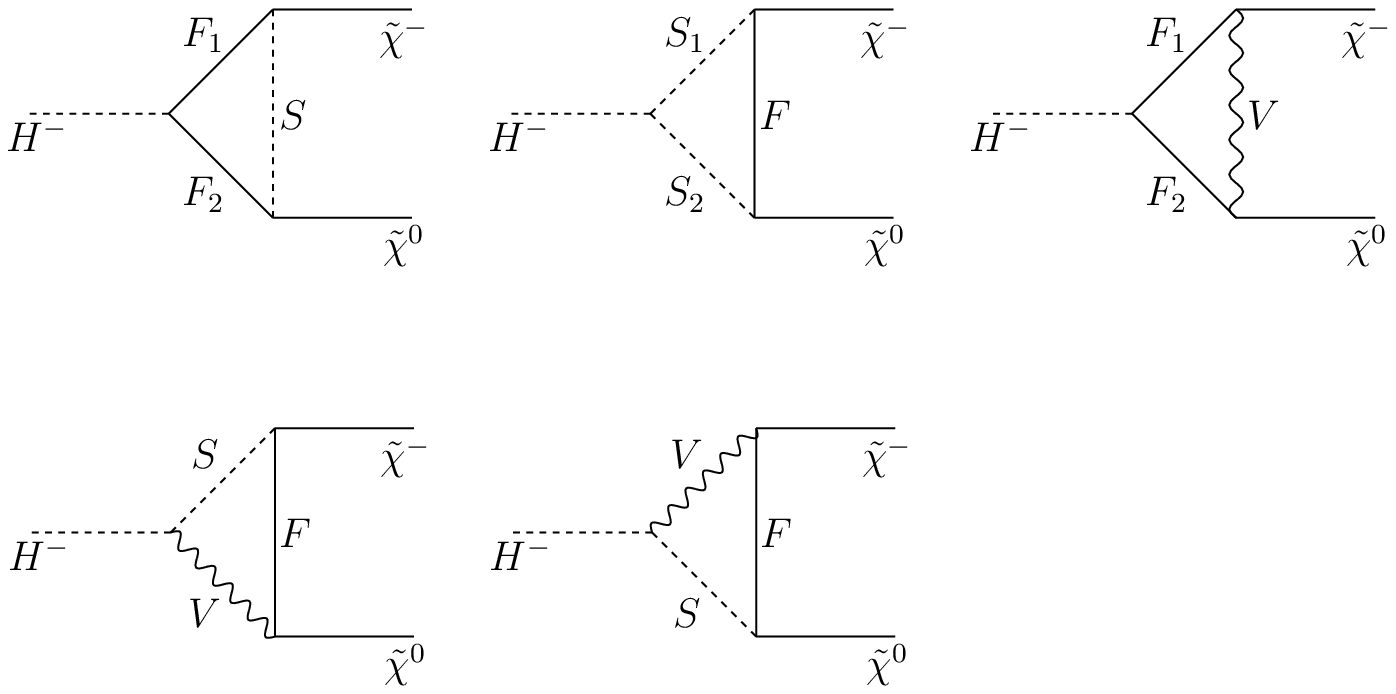}
\centering
\caption{The vertex one-loop Feynman diagrams for
$H^-\to\widetilde\chi^-\widetilde\chi^0$.
$F_1F_2S$: $\widetilde{\chi}^0\widetilde{\chi}^-\phi^0$, $
\widetilde{\chi}^-\widetilde{\chi}^0\phi^-$, $f'f\bar{f}$; $
S_1S_2F$: $\phi^0_BH^-\widetilde{\chi}^-$, $\phi^0_TG^-\widetilde{\chi}^-$, $
H^-\phi^0_B\widetilde{\chi}^0$, $G^-\phi^0_T\widetilde{\chi}^0$, $
\widetilde{f}'\widetilde{f}f$; $
F_1F_2V$: $\widetilde{\chi}^-\widetilde{\chi}^0Z$, $
\widetilde{\chi}^0\widetilde{\chi}^-W^-$; $
SVF$: $H^-Z\widetilde{\chi}^0$, $\phi^0_TW^-\widetilde{\chi}^-$; $
VSF$: ${\gamma}H^-\widetilde{\chi}^-$, $ZH^-\widetilde{\chi}^-$, $
W^-\phi^0_T\widetilde{\chi}^0$.
$
\phi^0=\{h^0,H^0,A^0,G^0\}$, $
\phi^0_B=\{h^0,H^0\}$, $
\phi^0_T=\{h^0,H^0,A^0\}$, $
\phi^-=\{H^-,G^-\}$
\label{oneloopfd}}
\end{figure}

In this appendix, we list the explicit expressions for the vertex one-loop
amplitude.
The vertex one-loop amplitudes are expanded with two Dirac
matrix elements\cite{denner} with 35 coefficients for each of them,
corresponding to the 35 Feynman diagrams in Fig.\ref{oneloopfd}.
\begin{equation}
\displaystyle{
\mathcal{M}^{(v)}_1=\sum^{35}_{i=1}(f^i_1 F_1+f^i_2 F_2)
},
\end{equation}
where,
\begin{equation}
F_1=\overline{u}(p_{\widetilde{\chi}^-_j})P_Rv(p_{\widetilde{\chi}^0_i}),~~
F_2=\overline{u}(p_{\widetilde{\chi}^-_j})P_Lv(p_{\widetilde{\chi}^0_i}).
\end{equation}
\indent In our paper, we use the Passarino-Veltman integrals,
which are defined in Ref.\cite{denner}. For simplicity, we define
the notations as following:
\\
 $
\displaystyle{
c^1_{(0,1,2)}=
\mathbf{C_{(0,1,2)}}(
    \mathbf{m}_{\widetilde{\chi}^-_j}^2,
    \mathbf{m}_{\widetilde{\chi}^0_i}^2,
    \mathbf{m}_{H^-}^2,
    \mathbf{m}_{\widetilde{\chi}^-_{c_1}}^2,
    \mathbf{m}_{h^0}^2,
    \mathbf{m}_{\widetilde{\chi}^0_{n_1}}^2
    )
,~~
b^1_0=
\mathbf{B_0}(
  \mathbf{m}_{\widetilde{\chi}^0_i}^2,
  \mathbf{m}_{h^0}^2,
  \mathbf{m}_{\widetilde{\chi}^0_{n_1}}^2
  )
} $ \\ \\ $ \displaystyle{
c^2_{(0,1,2)}=
\mathbf{C_{(0,1,2)}}(
    \mathbf{m}_{\widetilde{\chi}^-_j}^2,
    \mathbf{m}_{\widetilde{\chi}^0_i}^2,
    \mathbf{m}_{H^-}^2,
    \mathbf{m}_{\widetilde{\chi}^-_{c_1}}^2,
    \mathbf{m}_{H^0}^2,
    \mathbf{m}_{\widetilde{\chi}^0_{n_1}}^2
    )
,~~
b^2_0=
\mathbf{B_0}(
  \mathbf{m}_{\widetilde{\chi}^0_i}^2,
  \mathbf{m}_{H^0}^2,
  \mathbf{m}_{\widetilde{\chi}^0_{n_1}}^2
  )
} $ \\ \\ $ \displaystyle{
c^3_{(0,1,2)}=
\mathbf{C_{(0,1,2)}}(
    \mathbf{m}_{\widetilde{\chi}^-_j}^2,
    \mathbf{m}_{\widetilde{\chi}^0_i}^2,
    \mathbf{m}_{H^-}^2,
    \mathbf{m}_{\widetilde{\chi}^-_{c_1}}^2,
    \mathbf{m}_{A^0}^2,
    \mathbf{m}_{\widetilde{\chi}^0_{n_1}}^2
    )
,~~
b^3_0=
\mathbf{B_0}(
  \mathbf{m}_{\widetilde{\chi}^0_i}^2,
  \mathbf{m}_{A^0}^2,
  \mathbf{m}_{\widetilde{\chi}^0_{n_1}}^2
  )
} $ \\ \\ $ \displaystyle{
c^4_{(0,1,2)}=
\mathbf{C_{(0,1,2)}}(
    \mathbf{m}_{\widetilde{\chi}^-_j}^2,
    \mathbf{m}_{\widetilde{\chi}^0_i}^2,
    \mathbf{m}_{H^-}^2,
    \mathbf{m}_{\widetilde{\chi}^-_{c_1}}^2,
    \mathbf{m}_Z^2,
    \mathbf{m}_{\widetilde{\chi}^0_{n_1}}^2
    )
,~~
b^4_0=
\mathbf{B_0}(
  \mathbf{m}_{\widetilde{\chi}^0_i}^2,
  \mathbf{m}_Z^2,
  \mathbf{m}_{\widetilde{\chi}^0_{n_1}}^2
  )
} $ \\ \\ $ \displaystyle{
c^5_{(0,1,2)}=
\mathbf{C_{(0,1,2)}}(
    \mathbf{m}_{\widetilde{\chi}^-_j}^2,
    \mathbf{m}_{\widetilde{\chi}^0_i}^2,
    \mathbf{m}_{H^-}^2,
    \mathbf{m}_{\widetilde{\chi}^-_{n_1}}^2,
    \mathbf{m}_{H^-}^2,
    \mathbf{m}_{\widetilde{\chi}^0_{c_1}}^2
    )
,~~
b^5_0=
\mathbf{B_0}(
  \mathbf{m}_{\widetilde{\chi}^0_i}^2,
  \mathbf{m}_{H^-}^2,
  \mathbf{m}_{\widetilde{\chi}^0_{c_1}}^2
  )
} $ \\ \\ $ \displaystyle{
c^6_{(0,1,2)}=
\mathbf{C_{(0,1,2)}}(
    \mathbf{m}_{\widetilde{\chi}^-_j}^2,
    \mathbf{m}_{\widetilde{\chi}^0_i}^2,
    \mathbf{m}_{H^-}^2,
    \mathbf{m}_{\widetilde{\chi}^-_{n_1}}^2,
    \mathbf{m}_W^2,
    \mathbf{m}_{\widetilde{\chi}^0_{c_1}}^2
    )
,~~
b^6_0=
\mathbf{B_0}(
  \mathbf{m}_{\widetilde{\chi}^0_i}^2,
  \mathbf{m}_W^2,
  \mathbf{m}_{\widetilde{\chi}^0_{c_1}}^2
  )
} $ \\ \\ $ \displaystyle{
c^7_{(0,1,2)}=
\mathbf{C_{(0,1,2)}}(
    \mathbf{m}_{\widetilde{\chi}^-_j}^2,
    \mathbf{m}_{\widetilde{\chi}^0_i}^2,
    \mathbf{m}_{H^-}^2,
    \mathbf{m}_{e_{g_1}}^2,
    \mathbf{m}_{\widetilde{\nu}_{g_1}}^2,
    0
    )
,~~
b^7_0=
\mathbf{B_0}(
  \mathbf{m}_{\widetilde{\chi}^0_i}^2,
  0,
  \mathbf{m}_{\widetilde{\nu}_{g_1}}^2
  )
} $ \\ \\ $ \displaystyle{
c^8_{(0,1,2)}=
\mathbf{C_{(0,1,2)}}(
    \mathbf{m}_{H^-}^2,
    \mathbf{m}_{\widetilde{\chi}^0_i}^2,
    \mathbf{m}_{\widetilde{\chi}^-_j}^2,
    0,
    \mathbf{m}_{e_{g_1}}^2,
    \mathbf{m}_{\widetilde{e}_{s_1g_1}}^2
    )
,~~
b^8_0=
\mathbf{B_0}(
  \mathbf{m}_{\widetilde{\chi}^0_i}^2,
  \mathbf{m}_{e_{g_1}}^2,
  \mathbf{m}_{\widetilde{e}_{s_1g_1}}^2
  )
} $ \\ \\ $ \displaystyle{
c^9_{(0,1,2)}=
\mathbf{C_{(0,1,2)}}(
    \mathbf{m}_{H^-}^2,
    \mathbf{m}_{\widetilde{\chi}^-_j}^2,
    \mathbf{m}_{\widetilde{\chi}^0_i}^2,
    \mathbf{m}_{u_{g_2}}^2,
    \mathbf{m}_{d_{g_1}}^2,
    \mathbf{m}_{\widetilde{u}_{s_1g_1}}^2
    )
,~~
b^9_0=
\mathbf{B_0}(
  \mathbf{m}_{\widetilde{\chi}^-_j}^2,
  \mathbf{m}^2_{d_{g_1}},
  \mathbf{m}_{\widetilde{u}_{s_1g_2}}^2
  )
} $ \\ \\ $ \displaystyle{
c^{10}_{(0,1,2)}=
\mathbf{C_{(0,1,2)}}(
    \mathbf{m}_{H^-}^2,
    \mathbf{m}_{\widetilde{\chi}^0_i}^2,
    \mathbf{m}_{\widetilde{\chi}^-_j}^2,
    \mathbf{m}_{u_{g_1}}^2,
    \mathbf{m}_{d_{g_2}}^2,
    \mathbf{m}_{\widetilde{d}_{s_1g_2}}^2
    )
,~~
b^{10}_0=
\mathbf{B_0}(
  \mathbf{m}_{\widetilde{\chi}^0_i}^2,
  \mathbf{m}^2_{d_{g_2}},
  \mathbf{m}_{\widetilde{u}_{s_1g_2}}^2
  )
} $ \\ \\ $ \displaystyle{
c^{11}_{(0,1,2)}=
\mathbf{C_{(0,1,2)}}(
    \mathbf{m}_{\widetilde{\chi}^-_j}^2,
    \mathbf{m}_{\widetilde{\chi}^0_i}^2,
    \mathbf{m}_{H^-}^2,
    \mathbf{m}_{H^-}^2,
    \mathbf{m}_{\widetilde{\chi}^0_{n_1}}^2,
    \mathbf{m}_{h^0}^2
    )
} $ \\ \\ $ \displaystyle{
c^{12}_{(0,1,2)}=
\mathbf{C_{(0,1,2)}}(
    \mathbf{m}_{\widetilde{\chi}^-_j}^2,
    \mathbf{m}_{\widetilde{\chi}^0_i}^2,
    \mathbf{m}_{H^-}^2,
    \mathbf{m}_{h^0}^2,
    \mathbf{m}_{\widetilde{\chi}^-_{c_1}}^2,
    \mathbf{m}_{H^-}^2
    )
} $ \\ \\ $ \displaystyle{
c^{13}_{(0,1,2)}=
\mathbf{C_{(0,1,2)}}(
    \mathbf{m}_{\widetilde{\chi}^-_j}^2,
    \mathbf{m}_{\widetilde{\chi}^0_i}^2,
    \mathbf{m}_{H^-}^2,
    \mathbf{m}_{H^-}^2,
    \mathbf{m}_{\widetilde{\chi}^0_{n_1}}^2,
    \mathbf{m}_{H^0}^2
    )
} $ \\ \\ $ \displaystyle{
c^{14}_{(0,1,2)}=
\mathbf{C_{(0,1,2)}}(
    \mathbf{m}_{\widetilde{\chi}^-_j}^2,
    \mathbf{m}_{\widetilde{\chi}^0_i}^2,
    \mathbf{m}_{H^-}^2,
    \mathbf{m}_{H^0}^2,
    \mathbf{m}_{\widetilde{\chi}^-_{c_1}}^2,
    \mathbf{m}_{H^-}^2
    )
} $ \\ \\ $ \displaystyle{
c^{15}_{(0,1,2)}=
\mathbf{C_{(0,1,2)}}(
    \mathbf{m}_{\widetilde{\chi}^-_j}^2,
    \mathbf{m}_{\widetilde{\chi}^0_i}^2,
    \mathbf{m}_{H^-}^2,
    \mathbf{m}_{W}^2,
    \mathbf{m}_{\widetilde{\chi}^0_{n_1}}^2,
    \mathbf{m}_{h^0}^2
    )
} $ \\ \\ $ \displaystyle{
c^{16}_{(0,1,2)}=
\mathbf{C_{(0,1,2)}}(
    \mathbf{m}_{\widetilde{\chi}^-_j}^2,
    \mathbf{m}_{\widetilde{\chi}^0_i}^2,
    \mathbf{m}_{H^-}^2,
    \mathbf{m}_{h^0}^2,
    \mathbf{m}_{\widetilde{\chi}^-_{c_1}}^2,
    \mathbf{m}_{W}^2
    )
} $ \\ \\ $ \displaystyle{
c^{17}_{(0,1,2)}=
\mathbf{C_{(0,1,2)}}(
    \mathbf{m}_{\widetilde{\chi}^-_j}^2,
    \mathbf{m}_{\widetilde{\chi}^0_i}^2,
    \mathbf{m}_{H^-}^2,
    \mathbf{m}_{W}^2,
    \mathbf{m}_{\widetilde{\chi}^0_{n_1}}^2,
    \mathbf{m}_{H^0}^2
    )
} $ \\ \\ $ \displaystyle{
c^{18}_{(0,1,2)}=
\mathbf{C_{(0,1,2)}}(
    \mathbf{m}_{\widetilde{\chi}^-_j}^2,
    \mathbf{m}_{\widetilde{\chi}^0_i}^2,
    \mathbf{m}_{H^-}^2,
    \mathbf{m}_{H^0}^2,
    \mathbf{m}_{\widetilde{\chi}^-_{c_1}}^2,
    \mathbf{m}_{W}^2
    )
} $ \\ \\ $ \displaystyle{
c^{19}_{(0,1,2)}=
\mathbf{C_{(0,1,2)}}(
    \mathbf{m}_{\widetilde{\chi}^-_j}^2,
    \mathbf{m}_{\widetilde{\chi}^0_i}^2,
    \mathbf{m}_{H^-}^2,
    \mathbf{m}_{W}^2,
    \mathbf{m}_{\widetilde{\chi}^0_{n_1}}^2,
    \mathbf{m}_{A^0}^2
    )
} $ \\ \\ $ \displaystyle{
c^{20}_{(0,1,2)}=
\mathbf{C_{(0,1,2)}}(
    \mathbf{m}_{\widetilde{\chi}^-_j}^2,
    \mathbf{m}_{\widetilde{\chi}^0_i}^2,
    \mathbf{m}_{H^-}^2,
    \mathbf{m}_{A^0}^2,
    \mathbf{m}_{\widetilde{\chi}^-_{c_1}}^2,
    \mathbf{m}_{W}^2
    )
} $ \\ \\ $ \displaystyle{
c^{21}_{(0,1,2)}=
\mathbf{C_{(0,1,2)}}(
    \mathbf{m}_{\widetilde{\chi}^-_j}^2,
    \mathbf{m}_{\widetilde{\chi}^0_i}^2,
    \mathbf{m}_{H^-}^2,
    \mathbf{m}_{\widetilde{e}_{s_1g_1}}^2,
    0,
    \mathbf{m}_{\widetilde{\nu}_{g_1}}^2
    )
} $ \\ \\ $ \displaystyle{
c^{22}_{(0,1,2)}=
\mathbf{C_{(0,1,2)}}(
    \mathbf{m}_{\widetilde{\chi}^-_j}^2,
    \mathbf{m}_{H^-}^2,
    \mathbf{m}_{\widetilde{\chi}^0_i}^2,
    \mathbf{m}_{e_{g_1}}^2,
    \mathbf{m}_{\widetilde{\nu}_{g_1}}^2,
    \mathbf{m}_{\widetilde{e}_{s_1g_1}}^2
    )
} $ \\ \\ $ \displaystyle{
c^{23}_{(0,1,2)}=
\mathbf{C_{(0,1,2)}}(
    \mathbf{m}_{\widetilde{\chi}^-_j}^2,
    \mathbf{m}_{H^-}^2,
    \mathbf{m}_{\widetilde{\chi}^0_i}^2,
    \mathbf{m}_{u_{g_2}}^2,
    \mathbf{m}_{\widetilde{d}_{s_1g_1}}^2
    \mathbf{m}_{\widetilde{u}_{s_2g_2}}^2
    )
} $ \\ \\ $ \displaystyle{
c^{24}_{(0,1,2)}=
\mathbf{C_{(0,1,2)}}(
    \mathbf{m}_{\widetilde{\chi}^-_j}^2,
    \mathbf{m}_{H^-}^2,
    \mathbf{m}_{\widetilde{\chi}^0_i}^2,
    \mathbf{m}_{d_{g_2}}^2,
    \mathbf{m}_{\widetilde{u}_{s_1g_1}}^2,
    \mathbf{m}_{\widetilde{d}_{s_2g_2}}^2
    )
} $ \\ \\ $ \displaystyle{
c^{25}_{(0,1,2)}=
\mathbf{C_{(0,1,2)}}(
    \mathbf{m}_{\widetilde{\chi}^-_j}^2,
    \mathbf{m}_{\widetilde{\chi}^0_i}^2,
    \mathbf{m}_{H^-}^2,
    \mathbf{m}_{\widetilde{\chi}^-_{c_1}}^2,
    \mathbf{m}_{Z}^2,
    \mathbf{m}_{\widetilde{\chi}^0_{n_1}}^2
    )
,~~
b^{25}_0=
\mathbf{B_0}(
  \mathbf{m}_{\widetilde{\chi}^0_i}^2,
  \mathbf{m}_{Z}^2,
  \mathbf{m}_{\widetilde{\chi}^0_{n_1}}^2
  )
} $ \\ \\ $ \displaystyle{
c^{26}_{(0,1,2)}=
\mathbf{C_{(0,1,2)}}(
    \mathbf{m}_{\widetilde{\chi}^-_j}^2,
    \mathbf{m}_{\widetilde{\chi}^0_i}^2,
    \mathbf{m}_{H^-}^2,
    \mathbf{m}_{\widetilde{\chi}^-_{n_1}}^2,
    \mathbf{m}_{W}^2,
    \mathbf{m}_{\widetilde{\chi}^0_{c_1}}^2
    )
,~~
b^{26}_0=
\mathbf{B_0}(
  \mathbf{m}_{\widetilde{\chi}^0_i}^2,
  \mathbf{m}_{W}^2,
  \mathbf{m}_{\widetilde{\chi}^-_{c_1}}^2
  )
} $ \\ \\ $ \displaystyle{
c^{27}_{(0,1,2)}=
\mathbf{C_{(0,1,2)}}(
    \mathbf{m}_{\widetilde{\chi}^-_j}^2,
    \mathbf{m}_{\widetilde{\chi}^0_i}^2,
    \mathbf{m}_{H^-}^2,
    \mathbf{m}_{H^-}^2,
    \mathbf{m}_{\widetilde{\chi}^0_{n_1}}^2,
    \mathbf{m}_{Z}^2
    )
,~~
b^{27}_0=
\mathbf{B_0}(
  \mathbf{m}_{\widetilde{\chi}^0_i}^2,
  \mathbf{m}_{Z}^2,
  \mathbf{m}_{\widetilde{\chi}^-_{n_1}}^2
  )
} $ \\ \\ $ \displaystyle{
c^{28}_{(0,1,2)}=
\mathbf{C_{(0,1,2)}}(
    \mathbf{m}_{\widetilde{\chi}^-_j}^2,
    \mathbf{m}_{\widetilde{\chi}^0_i}^2,
    \mathbf{m}_{H^-}^2,
    \mathbf{m}_{h^0}^2,
    \mathbf{m}_{\widetilde{\chi}^-_{c_1}}^2,
    \mathbf{m}_{W}^2
    )
,~~
b^{28}_0=
\mathbf{B_0}(
  \mathbf{m}_{\widetilde{\chi}^0_i}^2,
  \mathbf{m}_{W}^2,
  \mathbf{m}_{\widetilde{\chi}^-_{c_1}}^2
  )
} $ \\ \\ $ \displaystyle{
c^{29}_{(0,1,2)}=
\mathbf{C_{(0,1,2)}}(
    \mathbf{m}_{\widetilde{\chi}^-_j}^2,
    \mathbf{m}_{\widetilde{\chi}^0_i}^2,
    \mathbf{m}_{H^-}^2,
    \mathbf{m}_{H^0}^2,
    \mathbf{m}_{\widetilde{\chi}^-_{c_1}}^2,
    \mathbf{m}_{W}^2
    )
,~~
b^{29}_0=
\mathbf{B_0}(
  \mathbf{m}_{\widetilde{\chi}^0_i}^2,
  \mathbf{m}_{W}^2,
  \mathbf{m}_{\widetilde{\chi}^-_{c_1}}^2
  )
} $ \\ \\ $ \displaystyle{
c^{30}_{(0,1,2)}=
\mathbf{C_{(0,1,2)}}(
    \mathbf{m}_{\widetilde{\chi}^-_j}^2,
    \mathbf{m}_{\widetilde{\chi}^0_i}^2,
    \mathbf{m}_{H^-}^2,
    \mathbf{m}_{A^0}^2,
    \mathbf{m}_{\widetilde{\chi}^-_{c_1}}^2,
    \mathbf{m}_{W}^2
    )
,~~
b^{30}_0=
\mathbf{B_0}(
  \mathbf{m}_{\widetilde{\chi}^0_i}^2,
  \mathbf{m}_{W}^2,
  \mathbf{m}_{\widetilde{\chi}^-_{c_1}}^2
  )
} $ \\ \\ $ \displaystyle{
c^{31}_{(0,1,2)}=
\mathbf{C_{(0,1,2)}}(
    \mathbf{m}_{\widetilde{\chi}^-_j}^2,
    \mathbf{m}_{\widetilde{\chi}^0_i}^2,
    \mathbf{m}_{H^-}^2,
    0,
    \mathbf{m}_{\widetilde{\chi}^-_j}^2,
    \mathbf{m}_{H^-}^2
    )
,~~
b^{31}_0=
\mathbf{B_0}(
  \mathbf{m}_{\widetilde{\chi}^0_i}^2,
  \mathbf{m}_{H^-}^2,
  \mathbf{m}_{\widetilde{\chi}^-_j}^2
  )
} $ \\ \\ $ \displaystyle{
c^{32}_{(0,1,2)}=
\mathbf{C_{(0,1,2)}}(
    \mathbf{m}_{\widetilde{\chi}^-_j}^2,
    \mathbf{m}_{\widetilde{\chi}^0_i}^2,
    \mathbf{m}_{H^-}^2,
    \mathbf{m}_Z^2,
    \mathbf{m}_{\widetilde{\chi}^-_{c_1}}^2,
    \mathbf{m}_{H^-}^2
    )
,~~
b^{32}_0=
\mathbf{B_0}(
  \mathbf{m}_{\widetilde{\chi}^0_i}^2,
  \mathbf{m}_{H^-}^2,
  \mathbf{m}_{\widetilde{\chi}^-_{c_1}}^2
  )
} $ \\ \\ $ \displaystyle{
c^{33}_{(0,1,2)}=
\mathbf{C_{(0,1,2)}}(
    \mathbf{m}_{\widetilde{\chi}^-_j}^2,
    \mathbf{m}_{\widetilde{\chi}^0_i}^2,
    \mathbf{m}_{H^-}^2,
    \mathbf{m}_W^2,
    \mathbf{m}_{\widetilde{\chi}^0_{n_1}}^2,
    \mathbf{m}_{h^0}^2
    )
,~~
b^{33}_0=
\mathbf{B_0}(
  \mathbf{m}_{\widetilde{\chi}^0_i}^2,
  \mathbf{m}_{h^0}^2,
  \mathbf{m}_{\widetilde{\chi}^0_{n_1}}^2
  )
} $ \\ \\ $ \displaystyle{
c^{34}_{(0,1,2)}=
\mathbf{C_{(0,1,2)}}(
    \mathbf{m}_{\widetilde{\chi}^-_j}^2,
    \mathbf{m}_{\widetilde{\chi}^0_i}^2,
    \mathbf{m}_{H^-}^2,
    \mathbf{m}_W^2,
    \mathbf{m}_{\widetilde{\chi}^0_{n_1}}^2,
    \mathbf{m}_{H^0}^2
    )
,~~
b^{34}_0=
\mathbf{B_0}(
  \mathbf{m}_{\widetilde{\chi}^0_i}^2,
  \mathbf{m}_{H^0}^2,
  \mathbf{m}_{\widetilde{\chi}^0_{n_1}}^2
  )
} $ \\ \\ $ \displaystyle{
c^{35}_{(0,1,2)}=
\mathbf{C_{(0,1,2)}}(
    \mathbf{m}_{\widetilde{\chi}^-_j}^2,
    \mathbf{m}_{\widetilde{\chi}^0_i}^2,
    \mathbf{m}_{H^-}^2,
    \mathbf{m}_W^2,
    \mathbf{m}_{\widetilde{\chi}^0_{n_1}}^2,
    \mathbf{m}_{A^0}^2
    )
,~~
b^{35}_0=
\mathbf{B_0}(
  \mathbf{m}_{\widetilde{\chi}^0_i}^2,
  \mathbf{m}_{A^0}^2,
  \mathbf{m}_{\widetilde{\chi}^0_{n_1}}^2
  )
} $ \\ \\
\indent
We also define some abbreviations for the frequent combinations as following:
\\ \\ $ \displaystyle{
\mathcal{X}^1_{ab}=
(c_W N_{a2}-s_W N_{a1})(s_{\alpha} N_{b3} +c_{\alpha} N_{b4})
+(c_W N_{b2}-s_W N_{b1})(s_{\alpha} N_{a3} +c_{\alpha} N_{a4})
} $ \\ \\ $ \displaystyle{
\mathcal{X}^2_{ab}=
(c_W N_{a2}- s_W N_{a1})(c_\alpha N_{b3} -s_\alpha N_{b4})+
(c_W N_{b2}- s_W N_{b1})(c_\alpha N_{a3} -s_\alpha N_{a4})
} $ \\ \\ $ \displaystyle{
\mathcal{X}^3_{ab}=
(c_W N_{a2}- s_W N_{a1})(c_\beta N_{b4} -s_\beta N_{b3})+
(c_W N_{b2}- s_W N_{b1})(c_\beta N_{a4} -s_\beta N_{a3})
} $ \\ \\ $ \displaystyle{
\mathcal{X}^4_{ab}=
(c_W N_{a2}- s_W N_{a1})(c_\beta N_{b3} +s_\beta N_{b4})+
(c_W N_{b2}- s_W N_{b1})(c_\beta N_{a3} +s_\beta N_{a4})
} $ \\ \\ $ \displaystyle{
\mathcal{X}^8_{gs}=
  c_\beta \mathbf{m}_W R^{\widetilde{e}_g*}_{s1}
  (s_W N_{i1} + c_W N_{i2})- c_W \mathbf{m}_{e_g} R^{\widetilde{e}_g*}_{s2} N_{i3}
} $ \\ \\ $ \displaystyle{
\mathcal{X}^9_{gs}=2\mathbf{m}_W s_\beta R^{\widetilde{u}_g*}_{s1} V_{j1}-\sqrt{2}\mathbf{m}_{u_g}
        R^{\widetilde{u}_g*}_{s2}V_{j2}
} $ \\ \\ $ \displaystyle{
\mathcal{X}^{10}_{gs}=2 c_\beta \mathbf{m}_W U^*_{j1}
   R^{\widetilde{d}_g}_{s1} - \sqrt{2} \mathbf{m}_{d_g} U^*_{j2} R^{\widetilde{d}_g}_{s2}
} $ \\ \\ $ \displaystyle{
\mathcal{X}^{21}_{gs}=
 (\mathbf{m}_W^2 s_{2\beta}-
t_\beta  \mathbf{m}_{e_g}^2)
R^{\widetilde{e}_g*}_{s1}-
 (\mu+ t_\beta \mathbf{A}_{e_g}^*)
\mathbf{m}_{e_g}
R^{\widetilde{e}_g*}_{s2}
} $ \\ \\ $ \displaystyle{
\mathcal{X}^{25}_{jc}=
\mathit{s_W^2}\delta_{c,j}-
\mathbf{U}_{c1}\mathbf{U}^*_{j1}-
\frac{1}{2}\mathbf{U}_{c2}\mathbf{U}^*_{j2}
} $ \\ \\ $ \displaystyle{
\mathcal{X}^{26}_{cn}=
U^*_{c1} N_{n2}+ U^*_{c2} N_{n3}/\sqrt{2}
} $ \\ \\ $ \displaystyle{
\mathcal{Y}^1_{ab}=s_{\alpha} U_{a2} V_{b1} - c_{\alpha} U_{a1} V_{b2}
,\hspace{24pt}
\mathcal{Y}^2_{ab}=c_\alpha U_{a2} V_{b1} +s_\alpha U_{b1} V_{a2}
} $ \\ \\ $ \displaystyle{
\mathcal{Y}^3_{ab}=s_\beta U_{a2} V_{b1} + c_\beta U_{a1} V_{b2}
,\hspace{24pt}
\mathcal{Y}^4_{ab}=s_\beta U_{a1} V_{b2} - c_\beta U_{a2} V_{b1}
} $ \\ \\ $ \displaystyle{
\mathcal{Y}^8_{gs}= -2 U^*_{j1} R^{\widetilde{e}_g}_{s1}+
  \sqrt{2} \mathbf{m}_{e_g} U^*_{j2} R^{\widetilde{e}_g}_{s2}/(c_\beta \mathbf{m}_W)
} $ \\ \\ $ \displaystyle{
\mathcal{Y}^9_{gs}= 4\mathbf{m}_W s_\beta s_W R^{\widetilde{u}_g}_{s2} N_{i1}-3 c_W
        \mathbf{m}_{u_g} R^{\widetilde{u}_g}_{s1}N_{i4}
} $ \\ \\ $ \displaystyle{
\mathcal{Y}^{10}_{gs}= (2 c_\beta \mathbf{m}_W s_W R^{\widetilde{d}_g*}_{s2} N^*_{i1}+ 3 c_W
   \mathbf{m}_{d_g} R^{\widetilde{d}_g*}_{s1} N^*_{i3})\mathbf{m}_{d_g}
} $ \\ \\ $ \displaystyle{
\mathcal{Y}^{25}_{cj}=\mathit{s_W^2}\delta_{c,j}-\mathbf{V}_{j1}\mathbf{V}^*_{c1}-
\frac{1}{2}\mathbf{V}_{j2}\mathbf{V}^*_{c2}
} $ \\ \\ $ \displaystyle{
\mathcal{Y}^{26}_{cn}=V^*_{c1}N_{n2}-V^*_{c2} N_{n4}/\sqrt{2}
} $ \\ \\ $ \displaystyle{
\mathcal{Z}^8_{gs}= 2 c_\beta \mathbf{m}_W s_W R^{\widetilde{e}_g*}_{s2} N^*_{i1}+
  c_W \mathbf{m}_{e_g} R^{\widetilde{e}_g*}_{s1} N^*_{i3}
} $ \\ \\ $ \displaystyle{
\mathcal{Z}^9_{gs}= \mathbf{m}_W s_\beta R^{\widetilde{u}_g}_{s1}(s_W  N^*_{i1} + 3c_W  N^*_{i2})+
   3 c_W  \mathbf{m}_{u_g} R^{\widetilde{u}_g}_{s2} N^*_{i4}
} $ \\ \\ $ \displaystyle{
\mathcal{Z}^{10}_{gs}= c_\beta \mathbf{m}_W R^{\widetilde{d}_g*}_{s1}
  (-s_W N_{i1}+ 3 c_W N_{i2})- 3 c_W \mathbf{m}_{d_g} R^{\widetilde{d}_g*}_{s2} N_{i3}
} $ \\ \\ $ \displaystyle{
\mathcal{Z}^{25}_{ab}=
\mathbf{N}_{a3}\mathbf{N}^*_{b3}-\mathbf{N}_{a4}\mathbf{N}^*_{b4}
} $ \\ \\ $ \displaystyle{
\mathcal{W}^1_{nc}=
(\mathbf{m}_{\widetilde{\chi}^-_{c}}C^L_{nc}+
\mathbf{m}_{\widetilde{\chi}^0_{n}}C^R_{nc})
,\hspace{24pt}
\mathcal{W}^2_{nc}=
(\mathbf{m}_{\widetilde{\chi}^0_{n}}C^L_{nc}+
\mathbf{m}_{\widetilde{\chi}^-_{c}}C^R_{nc})
} $ \\ \\ $ \displaystyle{
\mathcal{W}^3_{nc}=
(\mathbf{m}_{\widetilde{\chi}^-_{c}}D^L_{nc}+
\mathbf{m}_{\widetilde{\chi}^0_{n}}D^R_{nc})
,\hspace{24pt}
\mathcal{W}^4_{nc}=
(\mathbf{m}_{\widetilde{\chi}^0_{n}}D^L_{nc}+
\mathbf{m}_{\widetilde{\chi}^-_{c}}D^R_{nc})
} $ \\ \\ $ \displaystyle{
\mathcal{W}^5_{c_1c_2}=
(\mathbf{m}_{\widetilde{\chi}^-_{c_1}}\mathcal{Y}^1+
\mathbf{m}_{\widetilde{\chi}^-_{c_2}}\mathcal{Y}^{1*})
,\hspace{24pt}
\mathcal{W}^6_{c_1c_2}=
(\mathbf{m}_{\widetilde{\chi}^-_{c_1}}\mathcal{Y}^2+
\mathbf{m}_{\widetilde{\chi}^-_{c_2}}\mathcal{Y}^{2*})
} $ \\ \\ $ \displaystyle{
\mathcal{W}^7_{c_1c_2}=
(\mathbf{m}_{\widetilde{\chi}^-_{c_1}}\mathcal{Y}^3-
\mathbf{m}_{\widetilde{\chi}^-_{c_2}}\mathcal{Y}^{3*})
,\hspace{24pt}
\mathcal{W}^8_{n_1n_2}=
(\mathbf{m}_{\widetilde{\chi}^0_{n_1}}\mathcal{X}^1-
\mathbf{m}_{\widetilde{\chi}^0_{n_2}}\mathcal{X}^{1*})
} $ \\ \\ $ \displaystyle{
\mathcal{W}^9_{g_1g_2s_1}=\sqrt{2} t_\beta \mathbf{m}^2_{d_{g_1}}
U^*_{j2} R^{\widetilde{u}_{g_2}*}_{s_11}
,\hspace{24pt}
\mathcal{W}^{10}_{g_1g_2s_1}=\sqrt{2} V_{j2}
\mathbf{m}_{u_{g_1}}^2 R^{\widetilde{d}_{g_2}}_{s_11}
} $ \\ \\ $ \displaystyle{
\mathcal{T}^1=\mathcal{T}^2=\mathcal{T}^3=\mathcal{T}^4=
\alpha/(8 \sqrt{2} c_W \pi s_W^2)
,\hspace{24pt}
\mathcal{T}^5=\mathcal{T}^6=\alpha/(4 \pi e^2)
} $ \\ \\ $ \displaystyle{
\mathcal{T}^7=\alpha e t_\beta(c_W N^*_{i2} - s_W N^*_{i1})/(8 \mathbf{m}_W c_W \pi s_W^3)
,\hspace{24pt}
\mathcal{T}^8=\alpha e t_\beta/(16 c_\beta c_W \mathbf{m}_W^2 \pi s_W^3)
} $ \\ \\ $ \displaystyle{
\mathcal{T}^9=\alpha e \mathcal{K}_{g_2g_1} \mathcal{K}^*_{g_2g_1}
/(16 c_W \mathbf{m}^3_W \pi s_\beta^2 s_W^3 t_\beta)
,\hspace{24pt}
\mathcal{T}^{10}= \alpha e \mathcal{K}_{g_1g_2} \mathcal{K}^*_{g_1g_2}
/(16 c_W \mathbf{m}_W^3 \pi s_\beta^2 s_W^3)
} $ \\ \\ $ \displaystyle{
\mathcal{T}^{11}=-\alpha \mathbf{m}_W[c_{2\beta} s_{\alpha+\beta}/(2c_W^2)+s_{\beta-\alpha}]
/(8 c_W \pi s_W^2)
} $ \\ \\ $ \displaystyle{
\mathcal{T}^{12}=-\alpha \mathbf{m}_W[c_{2\beta} s_{\alpha+\beta}/(2c_W^2)+s_{\beta-\alpha}]
/(4 \sqrt{2} \pi s_W^2)
} $ \\ \\ $ \displaystyle{
\mathcal{T}^{13}=\alpha \mathbf{m}_W[c_{\beta-\alpha}-c_{2\beta} c_{\alpha+\beta}/(2c_W^2)]
/(8 c_W \pi s_W^2)
} $ \\ \\ $ \displaystyle{
\mathcal{T}^{14}=\alpha \mathbf{m}_W[c_{\beta-\alpha}-c_{2\beta} c_{\alpha+\beta}/(2c_W^2)]
/(4 \sqrt{2} \pi s_W^2)
} $ \\ \\ $ \displaystyle{
\mathcal{T}^{15}=\alpha \mathbf{m}_W[c_{\beta-\alpha}-s_{2\beta} s_{\alpha+\beta}/(2c_W^2)]
/(16 c_W \pi s_W^2)
} $ \\ \\ $ \displaystyle{
\mathcal{T}^{16}=\alpha \mathbf{m}_W(c_{\beta-\alpha}-s_{2\beta} s_{\alpha+\beta}/c_W^2)
/(8 \sqrt{2} \pi s_W^2)
} $ \\ \\ $ \displaystyle{
\mathcal{T}^{17}=\alpha \mathbf{m}_W[s_{\beta-\alpha}-s_{2\beta}c_{\alpha+\beta}/c_W^2]
/(16 c_W \pi s_W^2)
} $ \\ \\ $ \displaystyle{
\mathcal{T}^{18}=\alpha \mathbf{m}_W[s_{\beta-\alpha}-s_{2\beta}c_{\alpha+\beta}/c_W^2]
/(8 \sqrt{2} \pi s_W^2)
} $ \\ \\ $ \displaystyle{
\mathcal{T}^{19}=\alpha\mathbf{m}_W/(16 c_W \pi s_W^2)
,\hspace{24pt}
\mathcal{T}^{20}=\alpha \mathbf{m}_W/(8 \sqrt{2} \pi s_W^2)
} $ \\ \\ $ \displaystyle{
\mathcal{T}^{21}=\alpha e/(16 c_W \mathbf{m}_W \pi s_W^3)
,\hspace{24pt}
\mathcal{T}^{22}=\alpha e/(8 \sqrt{2} c_\beta^2 c_W \mathbf{m}_W^3 \pi s_W^3)
} $ \\ \\ $ \displaystyle{
\mathcal{T}^{23}=\alpha e \mathcal{K}_{g_2g_1} \mathcal{K}^*_{g_2g_1}
/(16 c_W \mathbf{m}_W^3 \pi s_\beta^2 s_W^3 t_\beta)
,\hspace{24pt}
\mathcal{T}^{24}=\alpha e \mathcal{K}_{g_1g_2} \mathcal{K}^*_{g_1g_2}/
(16 c_W \mathbf{m}_W^3\pi s_\beta^2 s_W^3)
} $ \\ \\ $ \displaystyle{
\mathcal{T}^{25}=\alpha/(4 c_W^2 \pi s_W^2)
,\hspace{24pt}
\mathcal{T}^{26}=\alpha/(2\pi s_W^2)
} $ \\ \\ $ \displaystyle{
\mathcal{T}^{27}=\alpha c_{2W}/(16 \pi c_W^2 s_W^2)
,\hspace{24pt}
\mathcal{T}^{28}=\alpha e c_{\beta-\alpha}/(8 \sqrt{2} \pi s_W^3)
} $ \\ \\ $ \displaystyle{
\mathcal{T}^{29}=\alpha e s_{\beta-\alpha}/(8\sqrt{2}\pi s_W^3)
,\hspace{24pt}
\mathcal{T}^{30}=\alpha e/(8\sqrt{2}\pi s_W^3)
} $ \\ \\ $ \displaystyle{
\mathcal{T}^{31}=-\alpha/(4\pi)
,\hspace{24pt}
\mathcal{T}^{32}=\alpha c_{2W}/(8\pi c_W^2 s_W^2)
} $ \\ \\ $ \displaystyle{
\mathcal{T}^{33}=\alpha e c_{\beta-\alpha}/(16 c_W \pi s_W^2)
,\hspace{24pt}
\mathcal{T}^{34}=\alpha e s_{\beta-\alpha}/(16 c_W \pi s_W^2)
} $ \\ \\ $ \displaystyle{
\mathcal{T}^{35}=\alpha e/(16 c_W \pi s_W^2)
}
$
\\ \\
\indent With the above abbreviations, we present the coefficients
as following:
\\
\\
$
\displaystyle{
f^1_1=
\mathcal{T}^1
\sum^2_{c_1=1}
\sum^4_{n_1=1}
[
b^1_0
\mathcal{Y}^1_{c_1j}
\mathcal{X}^1_{in_1}
C^L_{n_1c_1}
 +
  c^1_2
  (
    \mathbf{m}_{\widetilde{\chi}^0_i}
\mathcal{W}^2_{n_1c_1}
  \mathcal{Y}^1_{c_1j}
      \mathcal{X}^{1*}_{in_1}
  -
       \mathbf{m}_{\widetilde{\chi}^-_j}
       \mathcal{W}^1_{n_1c_1}
  \mathcal{Y}^{1*}_{jc_1}
      \mathcal{X}^1_{in_1}
       +
    \mathbf{m}_{H^-}^2
\mathcal{Y}^1_{c_1j}
 \mathcal{X}^1_{in_1}
C^L_{n_1c_1}
)
  } $ \\ \\ $ \displaystyle{
  \indent
+
c^1_0
  (
     \mathbf{m}_{\widetilde{\chi}^-_{c_1}}
  \mathcal{W}^1_{n_1c_1}
  \mathcal{Y}^1_{c_1j}
     \mathcal{X}^1_{in_1}
      -
    \mathbf{m}_{\widetilde{\chi}^-_{c_1}}
    \mathbf{m}_{\widetilde{\chi}^-_j}
  \mathcal{Y}^{1*}_{jc_1}
  \mathcal{X}^1_{in_1}
C^L_{n_1c_1}
      +
    \mathbf{m}_{\widetilde{\chi}^-_{c_1}}
    \mathbf{m}_{\widetilde{\chi}^0_i}
      \mathcal{Y}^1_{c_1j}
     C^R_{n_1c_1}
     \mathcal{X}^{1*}_{in_1}
     )
  } $ \\ \\ $ \displaystyle{
  \indent
     +
  c^1_1
  (
   \mathbf{m}_{\widetilde{\chi}^-_j}^2
\mathcal{Y}^1_{c_1j}
\mathcal{X}^1_{in_1}
C^L_{n_1c_1}
       -
     \mathbf{m}_{\widetilde{\chi}^-_j}
    \mathbf{m}_{\widetilde{\chi}^0_i}
\mathcal{Y}^{1*}_{jc_1}
      C^R_{n_1c_1}
     \mathcal{X}^{1*}_{in_1}
  -
       \mathbf{m}_{\widetilde{\chi}^-_j}
  \mathcal{W}^1_{n_1c_1}
  \mathcal{Y}^{1*}_{jc_1}
  \mathcal{X}^1_{in_1}
)
]
} $ \\ \\ $ \displaystyle{
f^1_2=f^1_1(
\mathcal{Y}^1_{c_1j}\leftrightarrow\mathcal{Y}^{1*}_{jc_1},~
\mathcal{X}^1_{in_1}\leftrightarrow\mathcal{X}^{1*}_{in_1},~
C^L_{n_1c_1}\leftrightarrow C^R_{n_1c_1},~
\mathcal{W}^1_{n_1c_1}\leftrightarrow\mathcal{W}^2_{n_1c_1}
)
} $ \\ \\ $ \displaystyle{
f^2_1=f^1_1(
\mathcal{T}^1\rightarrow\mathcal{T}^2,~
b^1_0\rightarrow b^2_0,~
c^1_{(0,1,2)}\rightarrow c^2_{(0,1,2)},~
\mathcal{X}^1_{in_1}\rightarrow\mathcal{X}^2_{in_1},~
\mathcal{X}^{1*}_{in_1}\rightarrow\mathcal{X}^{2*}_{in_1},~
  } $ \\ \\ $ \displaystyle{
  \indent
\mathcal{Y}^1_{c_1j}\rightarrow\mathcal{Y}^2_{c_1j},~
\mathcal{Y}^{1*}_{jc_1}\rightarrow\mathcal{Y}^{2*}_{jc_1}
)
} $ \\ \\ $ \displaystyle{
f^2_2=f^2_1(
\mathcal{Y}^2_{c_1j}\leftrightarrow\mathcal{Y}^{2*}_{jc_1},~
\mathcal{X}^2_{in_1}\leftrightarrow\mathcal{X}^{2*}_{in_1},~
C^L_{n_1c_1}\leftrightarrow C^R_{n_1c_1},~
\mathcal{W}^1_{n_1c_1}\leftrightarrow\mathcal{W}^2_{n_1c_1}
)
} $ \\ \\ $ \displaystyle{
f^3_1=f^1_1(
\mathcal{T}^1\rightarrow\mathcal{T}^3,~
b^1_0\rightarrow b^3_0,~
c^1_{(0,1,2)}\rightarrow c^3_{(0,1,2)},~
\mathcal{X}^1_{in_1}\rightarrow\mathcal{X}^3_{in_1},~
\mathcal{X}^{1*}_{in_1}\rightarrow-\mathcal{X}^{3*}_{in_1},~
  } $ \\ \\ $ \displaystyle{
  \indent
\mathcal{Y}^1_{c_1j}\rightarrow\mathcal{Y}^3_{c_1j},~
\mathcal{Y}^{1*}_{jc_1}\rightarrow-\mathcal{Y}^{3*}_{jc_1}
)
} $ \\ \\ $ \displaystyle{
f^3_2=f^3_1(
\mathcal{Y}^3_{c_1j}\leftrightarrow\mathcal{Y}^{3*}_{jc_1},~
\mathcal{X}^3_{in_1}\leftrightarrow\mathcal{X}^{3*}_{in_1},~
C^L_{n_1c_1}\leftrightarrow C^R_{n_1c_1},~
\mathcal{W}^1_{n_1c_1}\leftrightarrow\mathcal{W}^2_{n_1c_1}
)
} $ \\ \\ $ \displaystyle{
f^4_1=f^3_1(
\mathcal{T}^3\rightarrow\mathcal{T}^4,~
b^3_0\rightarrow b^4_0,~
c^3_{(0,1,2)}\rightarrow c^4_{(0,1,2)},~
\mathcal{X}^3_{in_1}\rightarrow\mathcal{X}^4_{in_1},~
\mathcal{X}^{3*}_{in_1}\rightarrow\mathcal{X}^{4*}_{in_1},~
  } $ \\ \\ $ \displaystyle{
  \indent
\mathcal{Y}^3_{c_1j}\rightarrow\mathcal{Y}^4_{c_1j},~
\mathcal{Y}^{3*}_{jc_1}\rightarrow\mathcal{Y}^{4*}_{jc_1}
)
} $ \\ \\ $ \displaystyle{
f^4_2=f^4_1(
\mathcal{Y}^4_{c_1j}\leftrightarrow\mathcal{Y}^{4*}_{jc_1},~
\mathcal{X}^4_{in_1}\leftrightarrow\mathcal{X}^{4*}_{in_1},~
C^L_{n_1c_1}\leftrightarrow C^R_{n_1c_1},~
\mathcal{W}^1_{n_1c_1}\leftrightarrow\mathcal{W}^2_{n_1c_1}
)
} $ \\ \\ $ \displaystyle{
f^5_1=f^4_1(
T^4\rightarrow T^5,~
b^4_0\rightarrow b^5_0,~
c^4_{(0,1,2)}\rightarrow c^5_{(0,1,2)},~
\mathbf{m}_{\widetilde{\chi}^0_{n_1}}\rightarrow m_{\widetilde{\chi}^-_{c_1}},~
\mathcal{X}^4_{in_1}\rightarrow C^{L*}_{ic_1},~
\mathcal{X}^{4*}_{in_1}\rightarrow -C^{R*}_{ic_1},~
} $ \\ \\ $ \displaystyle{
\indent
\mathcal{Y}^4_{c_1j}\rightarrow C^R_{n_1j},~
\mathcal{Y}^{4*}_{jc_1}\rightarrow -C^L_{n_1j},~
\mathcal{W}^1_{n_1c_1}\leftrightarrow\mathcal{W}^2_{n_1c_1}
)
} $ \\ \\ $ \displaystyle{
f^5_2=f^5_1(
L\leftrightarrow R,~
\mathcal{W}^1_{n_1c_1}\leftrightarrow\mathcal{W}^2_{n_1c_1}
)
} $ \\ \\ $ \displaystyle{
f^6_1=f^5_1(
\mathcal{T}^5\rightarrow \mathcal{T}^6,~
b^5_0\rightarrow b^6_0,~
c^5_{(0,1,2)}\rightarrow c^6_{(0,1,2)},~
C^{L*}_{ic_1}\rightarrow -C^{L*}_{ic_1},~
C^L_{ic_1}\rightarrow -C^L_{ic_1}
)
} $ \\ \\ $ \displaystyle{
f^6_2=f^6_1(L\leftrightarrow R)
} $ \\ \\ $ \displaystyle{
f^7_1=
-\mathcal{T}^7\sum^3_{g_1=1}
\mathbf{m}^2_{e_{g_1}}\mathbf{m}_{\widetilde{\chi}^0_i}
\Bigg[V_{j1}(c^7_0+c^7_2)+
c^7_1 \mathbf{m}_{\widetilde{\chi}^-_j} U^*_{j2}
/(\sqrt{2} c_\beta \mathbf{m}_W)
\Bigg]
} $ \\ \\ $ \displaystyle{
f^7_2==\mathcal{T}^7\sum^3_{g_1=1}
\mathbf{m}^2_{e_{g_1}}\Bigg(
\frac{U^*_{j2}}{\sqrt{2} c_\beta \mathbf{m}_W}
(b_0^7 + m^2_{H^-} c^7_2 + m^2_{\widetilde{\chi}^-_j} c^7_1
+c_0^7\mathbf{m}^2_{e_{g_1}}) +
\mathbf{m}_{\widetilde{\chi}^-_j} V_{j1}(c^7_0 + c^7_1 +c^7_2)
\Bigg)
} $ \\ \\ $ \displaystyle{
f^8_1=\mathcal{T}^8
\sum^3_{g_1=1}\sum^2_{s_1=1}
(c_1^8+c_2^8)\mathbf{m}_{\widetilde{\chi}^-_j}m^2_{e_{g_1}}
\mathcal{Y}^8_{g_1s_1}\mathcal{X}^8_{g_1s_1}
-\mathbf{m}_{\widetilde{\chi}^-_j} m_{e_{g_1}} m_{\widetilde{\chi}^0_i}
\mathcal{Y}^8_{g_1s_1}\mathcal{Z}^8_{g_1s_1}
} $ \\ \\ $ \displaystyle{
f^8_2=\mathcal{T}^8
\sum^3_{g_1=1}\sum^2_{s_1=1}
(b_0^8+c^8_2 m^2_{\widetilde{\chi}^-_j} + m^2_{H^-})\mathbf{m}_{e_{g_1}}
\mathcal{Y}^8_{g_1s_1}\mathcal{Z}^8_{g_1s_1}
-c_1^8 m^2_{e_{g_1}} \mathbf{m}_{\widetilde{\chi}^0_i}
\mathcal{Y}^8_{g_1s_1}\mathcal{X}^8_{g_1s_1}
} $ \\ \\ $ \displaystyle{
f^9_1=\sum^3_{g_1=1}\sum^3_{g_2=1}\sum^2_{s_1=1}\mathcal{T}^9[
\mathbf{m}_{u_{g_2}}\mathcal{X}^9_{g_2s_1}\mathcal{Y}^9_{g_2s_1}
(b^9_0+c^9_0(\mathbf{m}^2_{u_{g_2}}+t^2_\beta\mathbf{m}^2_{d_{g_1}})
+c^9_1\mathbf{m}^2_{H^-}+c^9_2\mathbf{m}^2_{\widetilde{\chi}^0_i})
} $ \\ \\ $ \displaystyle{
\indent
-t^2_\beta\mathcal{W}^9_{g_1g_2s_1}\mathbf{m}_{\widetilde{\chi}^-_j}
(\mathbf{m}_{u_{g_2}}\mathcal{Y}^9_{g_2s_1}(c^9_0+c^9_1/s^2_\beta)
+c^9_2\mathbf{m}_{\widetilde{\chi}^0_i}\mathcal{Z}^9_{g_2s_1})
} $ \\ \\ $ \displaystyle{
\indent
+\mathbf{m}_{\widetilde{\chi}^0_i}\mathcal{X}^9_{g_2s_1}\mathcal{Z}^9_{g_2s_1}
(c^9_0\mathbf{m}^2_{u_{g_2}}+(c^9_1+c^9_2)(\mathbf{m}^2_{u_{g_2}}+
t^2_\beta \mathbf{m}^2_{d_{g_1}}))
]
} $ \\ \\ $ \displaystyle{
f^9_2=\sum^3_{g_1=1}\sum^3_{g_2=1}\sum^2_{s_1=1}\mathcal{T}^9[
\mathcal{W}^9_{g_1g_2s_1}\mathcal{Z}^9_{g_2s_1}t^2_\beta
(b^9_0+c^9_0\mathbf{m}^2_{u_{g_2}}/s^2_\beta
+c^9_2\mathbf{m}^2_{\widetilde{\chi}^0_i}
+c^9_1\mathbf{m}^2_{H^-})
} $ \\ \\ $ \displaystyle{
\indent
+\mathcal{W}^9_{g_1g_2s_1}\mathcal{Y}^9_{g_2s_1}
\mathbf{m}_{\widetilde{\chi}^0_i}\mathbf{m}_{u_{g_2}}
((c^9_2+c^9_1)/c^2_\beta+c^9_0t^2_\beta)
-\mathcal{X}^9_{g_2s_1}\mathbf{m}_{\widetilde{\chi}^-_j}
(c^9_2\mathbf{m}_{\widetilde{\chi}^0_i}\mathbf{m}_{u_{g_2}}\mathcal{Y}^9_{g_2s_1}
} $ \\ \\ $ \displaystyle{
\indent
+((c^9_0+c^9_1)\mathbf{m}^2_{u_{g_2}}+c^9_1\mathbf{m}^2_{d_{g_1}}t^2_\beta)\mathcal{Z}^9_{g_2s_1})]
} $ \\ \\ $ \displaystyle{
f^{10}_1=\sum^3_{g_1=1}\sum^3_{g_2=1}\sum^2_{s_1=1}\mathcal{T}^{10}[
\mathcal{W}^{10}_{g_1g_2s_1}\mathcal{Z}^{10}_{g_2s_1}(b^{10}_0+
(\mathbf{m}_{u_{g_1}}^2+t^2_\beta\mathbf{m}_{d_{g_2}}^2)c^{10}_0+
\mathbf{m}_{\widetilde{\chi}^-_j}^2c^{10}_2+\mathbf{m}^2_{H^-}c^{10}_1)
} $ \\ \\ $ \displaystyle{
\indent
-\mathcal{X}^{10}_{g_2s_1}\mathcal{Z}^{10}_{g_2s_1}\mathbf{m}_{\widetilde{\chi}^-_j}^2
((c^{10}_2+c^{10}_1)t_\beta(\mathbf{m}_{u_{g_1}}^2+t^2_\beta\mathbf{m}_{d_{g_2}}^2)+
c^{10}_0\mathbf{m}_{u_{g_1}}^2)
} $ \\ \\ $ \displaystyle{
\indent
+\mathcal{Y}^{10}_{g_2s_1}\mathbf{m}_{\widetilde{\chi}^0_i}^2((c^{10}_0t^2_\beta
+c^{10}_1(1+t^2_\beta))\mathcal{W}^{10}_{g_1g_2s_1}-\mathcal{X}^{10}_{g_2s_1}c^{10}_2
\mathbf{m}_{\widetilde{\chi}^-_j}t^3_\beta)]
} $ \\ \\ $ \displaystyle{
f^{10}_2=\sum^3_{g_1=1}\sum^3_{g_2=1}\sum^2_{s_1=1}\mathcal{T}^{10}[
\mathcal{X}^{10}_{g_2s_1}\mathcal{Y}^{10}_{g_2s_1}(t^3_\beta(b^{10}_0+c^{10}_1
\mathbf{m}^2_{H^-}+c^{10}_2\mathbf{m}_{\widetilde{\chi}^-_j}^2)+(1+t^2_\beta)
\mathbf{m}_{u_{g_1}}^2c^{10}_0)
} $ \\ \\ $ \displaystyle{
\indent
-\mathcal{W}^{10}_{g_1g_2s_1}\mathcal{Y}^{10}_{g_2s_1}\mathbf{m}_{\widetilde{\chi}^-_j}
((1+t^2_\beta)(c^{10}_1+c^{10}_2)+c^{10}_0t^2_\beta)
\mathcal{Z}^{10}_{g_2s_1}\mathbf{m}_{\widetilde{\chi}^0_i}(\mathcal{X}^{10}_{g_2s_1}t_\beta(c^{10}_0
\mathbf{m}_{u_{g_1}}^2
} $ \\ \\ $ \displaystyle{
\indent
+c^{10}_1(\mathbf{m}_{u_{g_1}}^2+t^2_\beta\mathbf{m}_{d_{g_2}}^2))-c^{10}_2
\mathbf{m}_{\widetilde{\chi}^-_j}\mathcal{W}^{10}_{g_1g_2s_1})]
} $ \\ \\ $ \displaystyle{
f^{11}_1=\mathcal{T}^{11}\sum^4_{n_1=1}[
c^{11}_1\mathbf{m}_{\widetilde{\chi}^-_j}\mathcal{X}^1_{in_1}C^L_{n_1j}
+c^{11}_0\mathcal{X}^1_{in_1}\mathcal{W}^2_{n_1j}
+c^{11}_2(\mathbf{m}_{\widetilde{\chi}^-_j}\mathcal{X}^1_{in_1}C^L_{n_1j}
-\mathbf{m}_{\widetilde{\chi}^0_i}\mathcal{X}^{1*}_{in_1}C^R_{n_1j})]
} $ \\ \\ $ \displaystyle{
f^{11}_2=f^{11}_1(C^L_{n_1j}\leftrightarrow C^R_{n_1j},~
\mathcal{X}^1_{in_1}\leftrightarrow \mathcal{X}^{1*}_{in_1},~
\mathcal{W}^2_{n_1j}\rightarrow \mathcal{W}^1_{n_1j})
} $ \\ \\ $ \displaystyle{
f^{12}_1=\mathcal{T}^{12}\sum^2_{c_1=1}[
c^{12}_1\mathbf{m}_{\widetilde{\chi}^-_j}\mathcal{Y}^{1*}_{jc_1}C^R_{ic_1}
+c^{12}_0\mathcal{W}^5_{c_1j}C^R_{ic_1}
+c^{12}_2(\mathbf{m}_{\widetilde{\chi}^-_j}\mathcal{Y}^{1*}_{jc_1}C^R_{ic_1}
-\mathbf{m}_{\widetilde{\chi}^0_i}\mathcal{Y}^1_{c_1j}C^L_{ic_1})]
} $ \\ \\ $ \displaystyle{
f^{12}_2=f^{12}_1(C^L_{ic_1}\leftrightarrow C^R_{ic_1},~
\mathcal{Y}^1_{c_1j}\leftrightarrow \mathcal{Y}^{1*}_{jc_1},~
\mathcal{W}^5_{c_1j}\rightarrow \mathcal{W}^5_{jc_1})
} $ \\ \\ $ \displaystyle{
f^{13}_1=\mathcal{T}^{13}\sum^4_{n_1=1}[
c^{13}_1\mathbf{m}_{\widetilde{\chi}^-_j}\mathcal{X}^2_{in_1}C^L_{n_1j}
+c^{13}_0\mathcal{X}^2_{in_1}\mathcal{W}^2_{n_1j}
+c^{13}_2(\mathbf{m}_{\widetilde{\chi}^-_j}\mathcal{X}^2_{in_1}C^L_{n_1j}
+\mathbf{m}_{\widetilde{\chi}^0_i}\mathcal{X}^{2*}_{in_1}C^R_{n_1j})]
} $ \\ \\ $ \displaystyle{
f^{13}_2=f^{13}_1(C^L_{n_1j}\leftrightarrow C^R_{n_1j},~
\mathcal{X}^2_{in_1}\leftrightarrow \mathcal{X}^{2*}_{in_1},~
\mathcal{W}^2_{n_1j}\rightarrow \mathcal{W}^1_{n_1j})
} $ \\ \\ $ \displaystyle{
f^{14}_1=\mathcal{T}^{14}\sum^2_{c_1=1}[
c^{14}_1\mathbf{m}_{\widetilde{\chi}^-_j}\mathcal{Y}^{2*}_{jc_1}C^R_{ic_1}
+c^{14}_0\mathcal{W}^5_{c_1j}C^R_{ic_1}
+c^{14}_2(\mathbf{m}_{\widetilde{\chi}^-_j}\mathcal{Y}^{2*}_{jc_1}C^R_{ic_1}
-\mathbf{m}_{\widetilde{\chi}^0_i}\mathcal{Y}^2_{c_1j}C^L_{ic_1})]
} $ \\ \\ $ \displaystyle{
f^{14}_2=f^{14}_1(C^L_{ic_1}\leftrightarrow C^R_{ic_1},~
\mathcal{Y}^2_{c_1j}\leftrightarrow \mathcal{Y}^{2*}_{jc_1},~
\mathcal{W}^6_{c_1j}\rightarrow \mathcal{W}^6_{jc_1})
} $ \\ \\ $ \displaystyle{
f^{15}_1=\mathcal{T}^{15}\sum^4_{n_1=1}[
c^{15}_1\mathbf{m}_{\widetilde{\chi}^-_j}\mathcal{X}^1_{in_1}D^L_{n_1j}
+c^{15}_0\mathcal{X}^1_{in_1}\mathcal{W}^3_{n_1j}
+c^{15}_2(\mathbf{m}_{\widetilde{\chi}^-_j}\mathcal{X}^1_{in_1}D^L_{n_1j}
-\mathbf{m}_{\widetilde{\chi}^0_i}\mathcal{X}^{1*}_{in_1}D^R_{n_1j})]
} $ \\ \\ $ \displaystyle{
f^{15}_2=f^{15}_1(D^L_{n_1j}\leftrightarrow D^R_{n_1j},~
\mathcal{X}^1_{in_1}\leftrightarrow \mathcal{X}^{1*}_{in_1},~
\mathcal{W}^3_{n_1j}\rightarrow \mathcal{W}^4_{n_1j})
} $ \\ \\ $ \displaystyle{
f^{16}_1=\mathcal{T}^{16}\sum^2_{c_1=1}[
c^{16}_1\mathbf{m}_{\widetilde{\chi}^-_j}\mathcal{Y}^{1*}_{jc_1}D^R_{ic_1}
+c^{16}_0\mathcal{W}^5_{c_1j}D^R_{ic_1}
+c^{16}_2(\mathbf{m}_{\widetilde{\chi}^-_j}\mathcal{Y}^{1*}_{jc_1}D^R_{ic_1}
-\mathbf{m}_{\widetilde{\chi}^0_i}\mathcal{Y}^1_{c_1j}D^L_{ic_1})]
} $ \\ \\ $ \displaystyle{
f^{16}_2=f^{16}_1(D^L_{ic_1}\leftrightarrow D^R_{ic_1},~
\mathcal{Y}^1_{c_1j}\leftrightarrow \mathcal{Y}^{1*}_{jc_1},~
\mathcal{W}^5_{c_1j}\rightarrow \mathcal{W}^5_{jc_1})
} $ \\ \\ $ \displaystyle{
f^{17}_1=\mathcal{T}^{17}\sum^4_{n_1=1}[
c^{17}_1\mathbf{m}_{\widetilde{\chi}^-_j}\mathcal{X}^2_{in_1}D^L_{n_1j}
+c^{17}_0\mathcal{X}^2_{in_1}\mathcal{W}^3_{n_1j}
+c^{17}_2(\mathbf{m}_{\widetilde{\chi}^-_j}\mathcal{X}^2_{in_1}D^L_{n_1j}
-\mathbf{m}_{\widetilde{\chi}^0_i}\mathcal{X}^{2*}_{in_1}D^R_{n_1j})]
} $ \\ \\ $ \displaystyle{
f^{17}_2=f^{17}_1(D^L_{n_1j}\leftrightarrow D^R_{n_1j},~
\mathcal{X}^2_{in_1}\leftrightarrow \mathcal{X}^{2*}_{in_1},~
\mathcal{W}^3_{n_1j}\rightarrow \mathcal{W}^4_{n_1j})
} $ \\ \\ $ \displaystyle{
f^{18}_1=\mathcal{T}^{18}\sum^2_{c_1=1}[
c^{18}_1\mathbf{m}_{\widetilde{\chi}^-_j}\mathcal{Y}^{2*}_{jc_1}D^R_{ic_1}
+c^{18}_0\mathcal{W}^5_{c_1j}D^R_{ic_1}
+c^{18}_2(\mathbf{m}_{\widetilde{\chi}^-_j}\mathcal{Y}^{2*}_{jc_1}D^R_{ic_1}
-\mathbf{m}_{\widetilde{\chi}^0_i}\mathcal{Y}^2_{c_1j}D^L_{ic_1})]
} $ \\ \\ $ \displaystyle{
f^{18}_2=f^{18}_1(D^L_{ic_1}\leftrightarrow D^R_{ic_1},~
\mathcal{Y}^2_{c_1j}\leftrightarrow \mathcal{Y}^{2*}_{jc_1},~
\mathcal{W}^6_{c_1j}\rightarrow \mathcal{W}^6_{jc_1})
} $ \\ \\ $ \displaystyle{
f^{19}_1=\mathcal{T}^{19}\sum^4_{n_1=1}[
c^{19}_1\mathbf{m}_{\widetilde{\chi}^-_j}\mathcal{X}^3_{in_1}D^L_{n_1j}
+c^{19}_0\mathcal{X}^3_{in_1}\mathcal{W}^3_{n_1j}
+c^{19}_2(\mathbf{m}_{\widetilde{\chi}^-_j}\mathcal{X}^3_{in_1}D^L_{n_1j}
+\mathbf{m}_{\widetilde{\chi}^0_i}\mathcal{X}^{3*}_{in_1}D^R_{n_1j})]
} $ \\ \\ $ \displaystyle{
f^{19}_2=-f^{19}_1(D^L_{n_1j}\leftrightarrow D^R_{n_1j},~
\mathcal{X}^3_{in_1}\leftrightarrow \mathcal{X}^{3*}_{in_1},~
\mathcal{W}^3_{n_1j}\rightarrow \mathcal{W}^4_{n_1j})
} $ \\ \\ $ \displaystyle{
f^{20}_1=\mathcal{T}^{20}\sum^2_{c_1=1}[
c^{20}_1\mathbf{m}_{\widetilde{\chi}^-_j}\mathcal{Y}^{3*}_{jc_1}D^R_{ic_1}
+c^{20}_0\mathcal{W}^5_{c_1j}D^R_{ic_1}
+c^{20}_2(\mathbf{m}_{\widetilde{\chi}^-_j}\mathcal{Y}^{3*}_{jc_1}D^R_{ic_1}
+\mathbf{m}_{\widetilde{\chi}^0_i}\mathcal{Y}^3_{c_1j}D^L_{ic_1})]
} $ \\ \\ $ \displaystyle{
f^{20}_2=-f^{20}_1(D^L_{ic_1}\leftrightarrow D^R_{ic_1},~
\mathcal{Y}^3_{c_1j}\leftrightarrow \mathcal{Y}^{3*}_{jc_1},~
\mathcal{W}^6_{c_1j}\rightarrow \mathcal{W}^6_{jc_1})
} $ \\ \\ $ \displaystyle{
f^{21}_1=\mathcal{T}^{21}\sum^3_{g_1=1}\sum^2_{s_1=1}
(c^{21}_0+c^{21}_1+c^{22}_2)\mathbf{m}_{\widetilde{\chi}^-_j}
\mathcal{Y}^8_{js_1}\mathcal{X}^{21}_{g_1s_1}
} $ \\ \\ $ \displaystyle{
f^{21}_2=-\mathcal{T}^{21}\sum^3_{g_1=1}\sum^2_{s_1=1}
c^{21}_2\mathbf{m}_{\widetilde{\chi}^0_i}\mathcal{Y}^8_{js_1}\mathcal{X}^{21}_{g_1s_1}
} $ \\ \\ $ \displaystyle{
f^{22}_1=\mathcal{T}^{22}\sum^3_{g_1=1}\sum^2_{s_1=1}[
c^{22}_1\mathbf{m}_{\widetilde{\chi}^-_j}\mathbf{m}_{e_{g_1}}U^*_{j2}
\mathcal{X}^{21}_{g_1s_1}\mathcal{Z}^8_{g_1s_1}
+(c^{22}_0\mathbf{m}_{e_{g_1}}\mathcal{Z}^8_{g_1s_1}
+c^{22}_2\mathbf{m}_{\widetilde{\chi}^0_i}\mathcal{X}^8_{g_1s_1})
\mathcal{X}^{21}_{g_1s_1}V_{j1}\sqrt{2}c_\beta\mathbf{m}_W]
} $ \\ \\ $ \displaystyle{
f^{22}_2=\mathcal{T}^{22}\sum^3_{g_1=1}\sum^2_{s_1=1}[
c^{22}_1\mathbf{m}_{\widetilde{\chi}^-_j}V_{j1}
\mathcal{X}^{21}_{g_1s_1}\mathcal{X}^8_{g_1s_1}
\sqrt{2}c_\beta\mathbf{m}_W
+(c^{22}_0\mathbf{m}_{e_{g_1}}\mathcal{X}^8_{g_1s_1}
+c^{22}_2\mathbf{m}_{\widetilde{\chi}^0_i}\mathcal{Z}^8_{g_1s_1})
\mathbf{m}_{e_{g_1}}U^*_{j2}\mathcal{X}^{21}_{g_1s_1}]
} $ \\ \\ $ \displaystyle{
f^{23}_1=\sum^3_{g_1=1}\sum^3_{g_2=1}\sum^2_{s_1=1}\sum^2_{s_2=1}
\mathcal{T}^{23}\mathcal{X}^{23}[
c^{23}_1\mathbf{m}_{\widetilde{\chi}^-_j}\mathcal{X}^{10}_{g_1s_1}
\mathcal{Z}^{9*}_{g_2s_2}t_\beta-(c^{23}_0\mathbf{m}_{u_{g_2}}
\mathcal{Z}^{9*}_{g_2s_2}+c^{23}_2\mathbf{m}_{\widetilde{\chi}^0_i}\mathcal{Y}^{9*}_{g_2s_2})
\mathbf{m}_{u_{g_2}}R^{\widetilde{d}_{g_1}}_{s_11}V_{j2}\sqrt{2}]
} $ \\ \\ $ \displaystyle{
f^{23}_2=\sum^3_{g_1=1}\sum^3_{g_2=1}\sum^2_{s_1=1}\sum^2_{s_2=1}
\mathcal{T}^{23}\mathcal{X}^{23}[
(c^{23}_0\mathbf{m}_{u_{g_2}}\mathcal{Y}^{9*}_{g_2s_2}
+c^{23}_2\mathbf{m}_{\widetilde{\chi}^0_i}\mathcal{Z}^{9*}_{g_2s_2})
\mathcal{X}^{10}_{g_1s_1}t_\beta
-c^{23}_1\mathbf{m}_{\widetilde{\chi}^-_j}\mathbf{m}_{u_{g_2}}R^{\widetilde{d}_{g_1}}_{s_11}
V_{j2}\mathcal{Y}^{9*}_{g_2s_2}\sqrt{2}
]
} $ \\ \\ $ \displaystyle{
f^{24}_1=\sum^3_{g_1=1}\sum^3_{g_2=1}\sum^2_{s_1=1}\sum^2_{s_2=1}
\mathcal{T}^{24}\mathcal{X}^{23}[
c^{24}_1\mathbf{m}_{\widetilde{\chi}^-_j}\mathbf{m}_{d_{g_2}}
U^*_{j2}R^{\widetilde{u}_{g_1}*}_{s_11}\sqrt{2}t_\beta\mathcal{Y}^{10*}_{g_2s_2}
+(c^{24}_0\mathbf{m}_{d_{g_2}}\mathcal{Y}^{10*}_{g_2s_2}
-c^{24}_2\mathbf{m}_{\widetilde{\chi}^0_i}\mathcal{Z}^{10*}_{g_2s_2})
\mathcal{X}^9_{g_1s_1}]
} $ \\ \\ $ \displaystyle{
f^{24}_2=\sum^3_{g_1=1}\sum^3_{g_2=1}\sum^2_{s_1=1}\sum^2_{s_2=1}
\mathcal{T}^{24}\mathcal{X}^{23}[
\mathbf{m}_{d_{g_2}}U^*_{j2}R^{\widetilde{u}_{g_1}*}_{s_11}\sqrt{2}t_\beta
(c^{24}_2\mathbf{m}_{\widetilde{\chi}^0_i}\mathcal{Y}^{10*}_{g_2s_2}
-c^{24}_0\mathbf{m}_{d_{g_2}}\mathcal{Z}^{10*}_{g_2s_2})
+c^{24}_1\mathbf{m}_{\widetilde{\chi}^-_j}\mathcal{X}^9_{g_1s_1}\mathcal{Z}^{10*}_{g_2s_2}]
} $ \\ \\ $ \displaystyle{
f^{25}_1=\mathcal{T}^{25}\sum^2_{c_1=1}\sum^4_{n_1=1}[
-2b^{25}_0\mathcal{Y}^{25}_{c_1j}C^R_{n_1c_1}\mathcal{Z}^{25}_{in_1}
-c^{25}_2(\mathbf{m}_{\widetilde{\chi}^0_i}\mathcal{Y}^{25}_{c_1j}
\mathcal{Z}^{25*}_{in_1}\mathcal{W}^1_{n_1c_1}
+\mathbf{m}_{\widetilde{\chi}^-_j}\mathcal{X}^{25}_{jc_1}
\mathcal{Z}^{25}_{in_1}\mathcal{W}^2_{n_1c_1}
} $ \\ \\ $ \displaystyle{
\indent
+2\mathbf{m}^2_{H^-}\mathcal{Y}^{25}_{c_1j}C^R_{n_1c_1}\mathcal{Z}^{25}_{in_1})
-c^{25}_1(\mathbf{m}_{\widetilde{\chi}^-_j}\mathcal{X}^{25}_{jc_1}
\mathcal{Z}^{25}_{in_1}\mathcal{W}^2_{n_1c_1}
+(\mathbf{m}_{H^-}^2+\mathbf{m}_{\widetilde{\chi}^-_j}^2-\mathbf{m}_{\widetilde{\chi}^0_i}^2)
\mathcal{Y}^{25}_{c_1j}C^R_{n_1c_1}\mathcal{Z}^{25}_{in_1})
} $ \\ \\ $ \displaystyle{
\indent
-c^{25}_0(\mathbf{m}_{\widetilde{\chi}^-_{c_1}}
\mathbf{m}_{\widetilde{\chi}^0_i}\mathcal{Y}^{25}_{c_1j}\mathcal{Z}^{25*}_{in_1}C^L_{n_1c_1}
+\mathbf{m}_{\widetilde{\chi}^-_{c_1}}\mathbf{m}_{\widetilde{\chi}^-_j}
\mathcal{X}^{25}_{jc_1}C^R_{n_1c_1}\mathcal{Z}^{25}_{in_1}
+2\mathbf{m}_{\widetilde{\chi}^-_{c_1}}\mathcal{Y}^{25}_{c_1j}
\mathcal{Z}^{25}_{in_1}\mathcal{W}^2_{n_1c_1})]
} $ \\ \\ $ \displaystyle{
f^{25}_2=-f^{25}_1(L\leftrightarrow R,~
\mathcal{Y}^{25}_{c_1j}\leftrightarrow \mathcal{X}^{25}_{jc_1},~
\mathcal{Z}^{25}_{in_1}\leftrightarrow \mathcal{Z}^{25*}_{in_1},~
\mathcal{W}^1_{n_1c_1}\leftrightarrow \mathcal{W}^2_{n_1c_1},~)
} $ \\ \\ $ \displaystyle{
f^{26}_1=\mathcal{T}^{26}\sum^2_{c_1=1}\sum^4_{n_1=1}[
2b^{26}_0\mathcal{Y}^{26}_{c_1i}C^R_{n_1c_1}\mathcal{Y}^{26*}_{jn_1}
+c^{26}_1(\mathbf{m}_{\widetilde{\chi}^-_j}\mathcal{Y}^{26}_{c_1i}
\mathcal{X}^{26*}_{c_1i}\mathcal{W}^1_{n_1c_1}
+(\mathbf{m}_{H^-}^2
+\mathbf{m}_{\widetilde{\chi}^-_j}^2-\mathbf{m}_{\widetilde{\chi}^0_i}^2)
} $ \\ \\ $ \displaystyle{
\indent
\times
\mathcal{Y}^{26}_{c_1i}\mathcal{Y}^{26*}_{jn_1}C^R_{n_1c_1})
+c^{26}_2(\mathbf{m}_{\widetilde{\chi}^-_j}\mathcal{Y}^{26}_{c_1i}
\mathcal{X}^{26*}_{c_1i}\mathcal{W}^1_{n_1c_1}
+2\mathbf{m}_{H^-}^2\mathcal{Y}^{26}_{c_1i}C^R_{n_1c_1}\mathcal{Y}^{26*}_{jn_1}
-\mathbf{m}_{\widetilde{\chi}^0_i}\mathcal{X}^{26}_{jn_1}
\mathcal{Y}^{26*}_{jn_1}\mathcal{W}^2_{n_1c_1})
} $ \\ \\ $ \displaystyle{
\indent
+c^{26}_0(2\mathbf{m}_{\widetilde{\chi}^0_{n_1}}\mathcal{Y}^{26}_{c_1i}\mathcal{Y}^{26*}_{jn_1}\mathcal{W}^1_{n_1c_1}
+\mathbf{m}_{\widetilde{\chi}^-_j}\mathbf{m}_{\widetilde{\chi}^0_{n_1}}\mathcal{Y}^{26}_{c_1i}
\mathcal{X}^{26*}_{c_1i}C^R_{n_1c_1}
-\mathbf{m}_{\widetilde{\chi}^0_i}\mathbf{m}_{\widetilde{\chi}^0_{n_1}}\mathcal{X}^{26}_{jn_1}
\mathcal{Y}^{26*}_{jn_1}C^L_{n_1c_1})
]
} $ \\ \\ $ \displaystyle{
f^{26}_2=f^{26}_1(L\leftrightarrow R,~
\mathcal{W}^1_{n_1c_1}\leftrightarrow \mathcal{W}^2_{n_1c_1},~
\mathcal{X}^{26}_{jn_1}\leftrightarrow \mathcal{Y}^{26}_{c_1i},~
\mathcal{X}^{26*}_{c_1i}\leftrightarrow \mathcal{Y}^{26*}_{jn_1}
)
} $ \\ \\ $ \displaystyle{
f^{27}_1=\mathcal{T}^{27}\sum^4_{n_1=1}[
-b^{27}_0C^R_{n_1j}\mathcal{Z}^{25}_{in_1}+
c^{27}_1\mathbf{m}_{\widetilde{\chi}^-_j}C^L_{n_1j}
(\mathbf{m}_{\widetilde{\chi}^0_i}\mathcal{Z}^{25*}_{in_1}-
\mathbf{m}_{\widetilde{\chi}^0_{n_1}}\mathcal{Z}^{25}_{in_1})
} $ \\ \\ $ \displaystyle{
\indent
+c^{27}_2(\mathbf{m}_{H^-}^2C^R_{n_1j}\mathcal{Z}^{25}_{in_1}
-\mathbf{m}_{\widetilde{\chi}^0_i}\mathcal{Z}^{25*}_{in_1}\mathcal{W}^1_{n_1j}
-\mathbf{m}_{\widetilde{\chi}^-_j}\mathcal{Z}^{25}_{in_1}\mathcal{W}^2_{n_1j})
} $ \\ \\ $ \displaystyle{
\indent
+c^{27}_0(\mathbf{m}_{\widetilde{\chi}^0_i}\mathcal{Z}^{25*}_{in_1}\mathcal{W}^1_{n_1j}
+\mathbf{m}_{\widetilde{\chi}^-_j}\mathcal{Z}^{25}_{in_1}\mathcal{W}^2_{n_1j}
-\mathbf{m}_{H^-}^2C^R_{n_1j}\mathcal{Z}^{25}_{in_1})
]
} $ \\ \\ $ \displaystyle{
f^{27}_2=f^{27}_2(L\leftrightarrow R,~
\mathcal{Z}^{25}_{in_1}\leftrightarrow -\mathcal{Z}^{25*}_{in_1},~
\mathcal{W}^1_{n_1j}\leftrightarrow \mathcal{W}^2_{n_1j}
)
} $ \\ \\ $ \displaystyle{
f^{28}_1=\mathcal{T}^{28}\sum^2_{c_1=1}[
(b^{28}_0+c^{28}_0\mathbf{m}_{h^0}^2-c^{28}_2\mathbf{m}_{H^-}^2)
\mathcal{Y}^1_{c_1j}\mathcal{X}^{26}_{c_1i}
+(c^{28}_2-c^{28}_0)(\mathbf{m}_{\widetilde{\chi}^-_j}
\mathcal{X}^{26}_{c_1i}\mathcal{W}^5_{jc_1}
} $ \\ \\ $ \displaystyle{
\indent
-\mathbf{m}_{\widetilde{\chi}^0_i}
\mathcal{Y}^{26}_{c_1i}\mathcal{W}^5_{c_1j})
+c^{28}_1\mathbf{m}_{\widetilde{\chi}^-_j}\mathcal{Y}^{1*}_{jc_1}]
(\mathbf{m}_{\widetilde{\chi}^-_{c_1}}\mathcal{X}^{26}_{c_1i}
+\mathbf{m}_{\widetilde{\chi}^0_i}\mathcal{Y}^{26}_{c_1i})
} $ \\ \\ $ \displaystyle{
f^{28}_2=f^{28}_1(
\mathcal{Y}^1_{c_1j}\leftrightarrow\mathcal{Y}^{1*}_{jc_1},~
\mathcal{X}^{26}_{c_1i}\leftrightarrow\mathcal{Y}^{26}_{c_1i})
} $ \\ \\ $ \displaystyle{
f^{29}_1=f^{28}_1(
h^0\rightarrow H^0,~
\mathcal{T}^{28}\rightarrow\mathcal{T}^{29},~
b^{28}_0\rightarrow b^{29}_0,~
c^{28}_{(0,1,2)}\rightarrow c^{29}_{(0,1,2)},~
\mathcal{Y}^1_{c_1j}\rightarrow\mathcal{Y}^2_{c_1j},~
} $ \\ \\ $ \displaystyle{
\indent
\mathcal{Y}^{1*}_{c_1j}\rightarrow\mathcal{Y}^{2*}_{c_1j},~
\mathcal{W}^5\rightarrow\mathcal{W}^6
)
} $ \\ \\ $ \displaystyle{
f^{29}_2=f^{28}_2(
h^0\rightarrow H^0,~
\mathcal{T}^{28}\rightarrow\mathcal{T}^{29},~
b^{28}_0\rightarrow b^{29}_0,~
c^{28}_{(0,1,2)}\rightarrow c^{29}_{(0,1,2)},~
\mathcal{Y}^1_{c_1j}\rightarrow\mathcal{Y}^2_{c_1j},~
} $ \\ \\ $ \displaystyle{
\indent
\mathcal{Y}^{1*}_{c_1j}\rightarrow\mathcal{Y}^{2*}_{c_1j},~
\mathcal{W}^5\rightarrow\mathcal{W}^6
)
} $ \\ \\ $ \displaystyle{
f^{30}_1=-f^{29}_1(
H^0\rightarrow A^0,~
\mathcal{T}^{29}\rightarrow\mathcal{T}^{30},~
b^{29}_0\rightarrow b^{30}_0,~
c^{29}_{(0,1,2)}\rightarrow c^{30}_{(0,1,2)},~
\mathcal{Y}^1_{c_1j}\rightarrow\mathcal{Y}^2_{c_1j},~
} $ \\ \\ $ \displaystyle{
\indent
\mathcal{Y}^{1*}_{c_1j}\rightarrow-\mathcal{Y}^{2*}_{c_1j},~
\mathcal{W}^5\rightarrow\mathcal{W}^6
)
} $ \\ \\ $ \displaystyle{
f^{30}_2=-f^{29}_2(
H^0\rightarrow A^0,~
\mathcal{T}^{29}\rightarrow\mathcal{T}^{30},~
b^{29}_0\rightarrow b^{30}_0,~
c^{29}_{(0,1,2)}\rightarrow c^{30}_{(0,1,2)},~
\mathcal{Y}^1_{c_1j}\rightarrow\mathcal{Y}^2_{c_1j},~
} $ \\ \\ $ \displaystyle{
\indent
\mathcal{Y}^{1*}_{c_1j}\rightarrow-\mathcal{Y}^{2*}_{c_1j},~
\mathcal{W}^5\rightarrow\mathcal{W}^6
)
} $ \\ \\ $ \displaystyle{
f^{31}_1=\mathcal{T}^{31}[
(c^{31}_2+2c^{31}_0+2c^{31}_1)
(\mathbf{m}_{H^-}^2+\mathbf{m}_{\widetilde{\chi}^-_j}^2-
\mathbf{m}_{\widetilde{\chi}^0_i}^2)C^R_{ij}
+(b^{31}_0+2c^{31}_2\mathbf{m}_{H^-}^2)C^R_{ij}-2c^{31}_1\mathbf{m}_{\widetilde{\chi}^-_j}
\mathbf{m}_{\widetilde{\chi}^0_i}C^L_{ij}]
} $ \\ \\ $ \displaystyle{
f^{31}_2=f^{31}_1(L\leftrightarrow R)
} $ \\ \\ $ \displaystyle{
f^{32}_1=\mathcal{T}^{32}\sum^2_{c_1=1}[
b^{32}_0\mathcal{Y}^{25}_{c_1j}C^R_{ic_1}+
c^{32}_1(\mathbf{m}_{\widetilde{\chi}^-_j}\mathcal{X}^{25}_{c_1j}\mathcal{W}^2_{ic_1}+
(2\mathbf{m}_{H^-}^2+\mathbf{m}_{\widetilde{\chi}^-_j}^2
-2\mathbf{m}_{\widetilde{\chi}^0_i}^2)\mathcal{Y}^{25}_{c_1j}C^R_{ic_1}
+\mathbf{m}_{\widetilde{\chi}^-_j}\mathbf{m}_{\widetilde{\chi}^0_i}
\mathcal{X}^{25}_{c_1j}C^L_{ic_1})
} $ \\ \\ $ \displaystyle{
\indent
+c^{32}_0(2\mathbf{m}_{\widetilde{\chi}^-_j}\mathcal{X}^{25}_{c_1j}\mathcal{W}^2_{ic_1}
-2\mathbf{m}_{\widetilde{\chi}^0_i}\mathcal{Y}^{25}_{c_1j}\mathcal{W}^1_{ic_1}
+(2\mathbf{m}_{H^-}^2+\mathbf{m}_Z^2)\mathcal{Y}^{25}_{c_1j}C^R_{ic_1})
} $ \\ \\ $ \displaystyle{
\indent
+c^{32}_2(\mathbf{m}_{\widetilde{\chi}^-_j}\mathcal{X}^{25}_{c_1j}
\mathcal{W}^2_{ic_1}
+3\mathbf{m}_{H^-}^2\mathcal{Y}^{25}_{c_1j}C^R_{ic_1}
-\mathbf{m}_{\widetilde{\chi}^0_i}\mathcal{Y}^{25}_{c_1j}\mathcal{W}^1_{ic_1})
} $ \\ \\ $ \displaystyle{
f^{32}_2=f^{32}_1(L\leftrightarrow R,~
\mathcal{W}^1_{ic_1}\leftrightarrow \mathcal{W}^2_{ic_1},~
\mathcal{X}^{25}_{c_1j}\leftrightarrow\mathcal{Y}^{25}_{c_1j}
)
} $ \\ \\ $ \displaystyle{
f^{33}_1=-f^{32}_1(
\mathcal{T}^{32}\rightarrow\mathcal{T}^{33},~
b^{32}_0\rightarrow b^{33}_0,
c^{32}_{(0,1,2)}\rightarrow c^{33}_{(0,1,2)},~
\mathcal{W}^1_{ic_1}\rightarrow\mathcal{W}^8_{in_1},~
\mathcal{W}^2_{ic_1}\rightarrow\mathcal{W}^8_{n_1i},~
} $ \\ \\ $ \displaystyle{
\indent
\mathcal{Y}^{25}_{c_1j}\rightarrow\mathcal{Y}^{26*}_{jn_1},~
\mathcal{X}^{25}_{c_1j}\rightarrow\mathcal{X}^{26*}_{jn_1},
C^L_{ic_1}\rightarrow\mathcal{X}^{1*}_{in_1},~
C^R_{ic_1}\rightarrow\mathcal{X}^1_{in_1}
)
} $ \\ \\ $ \displaystyle{
f^{33}_2=f^{33}_1(
\mathcal{X}^1_{in_1}\leftrightarrow\mathcal{X}^{1*}_{in_1},~
\mathcal{X}^{26*}_{jn_1}\leftrightarrow\mathcal{Y}^{26*}_{jn_1}
)
} $ \\ \\ $ \displaystyle{
f^{34}_1=f^{33}_1(
\mathcal{T}^{33}\rightarrow\mathcal{T}^{34},~
b^{33}_0\rightarrow b^{34}_0,
c^{33}_{(0,1,2)}\rightarrow c^{34}_{(0,1,2)},~
\mathcal{X}^1_{in_1}\rightarrow\mathcal{X}^2_{in_1},~
\mathcal{X}^{1*}_{in_1}\rightarrow\mathcal{X}^{2*}_{in_1}
)
} $ \\ \\ $ \displaystyle{
f^{34}_1=f^{33}_1(
\mathcal{T}^{33}\rightarrow\mathcal{T}^{34},~
b^{33}_0\rightarrow b^{34}_0,
c^{33}_{(0,1,2)}\rightarrow c^{34}_{(0,1,2)},~
\mathcal{X}^1_{in_1}\rightarrow\mathcal{X}^2_{in_1},~
\mathcal{X}^{1*}_{in_1}\rightarrow\mathcal{X}^{2*}_{in_1}
)
} $ \\ \\ $ \displaystyle{
f^{35}_1=f^{34}_1(
\mathcal{T}^{34}\rightarrow\mathcal{T}^{35},~
b^{34}_0\rightarrow b^{35}_0,
c^{34}_{(0,1,2)}\rightarrow c^{35}_{(0,1,2)},~
\mathcal{X}^2_{in_1}\rightarrow\mathcal{X}^3_{in_1},~
\mathcal{X}^{2*}_{in_1}\rightarrow-\mathcal{X}^{3}_{in_1}
)
} $ \\ \\ $ \displaystyle{
f^{35}_1=f^{34}_1(
\mathcal{T}^{34}\rightarrow\mathcal{T}^{35},~
b^{34}_0\rightarrow b^{35}_0,
c^{34}_{(0,1,2)}\rightarrow c^{35}_{(0,1,2)},~
\mathcal{X}^2_{in_1}\rightarrow\mathcal{X}^3_{in_1},~
\mathcal{X}^{2*}_{in_1}\rightarrow-\mathcal{X}^{3*}_{in_1}
)
}
$
\newpage

\newpage

\newpage

\pagebreak

\newpage

\newpage

\newpage

\end{document}